\documentclass[preprint,showpacs,preprintnumbers,amsmath,amssymb]{revtex4}

\usepackage{amsfonts}%
\usepackage{amsmath}%  
                                      
\usepackage{amsthm}
\usepackage{mathrsfs}
\usepackage{epsfig}

\def\baselskip{19pt}

\def\cvar{\mathcal{E}}

\def\Hw{K}

\def\rf{\rho^-}
\def\rfp{\rho^+}

\newcommand{\change}%     % Puts Hash Mark # on the margin to indicate a %
{{\marginpar{\#}}}        % change or mistake in the manuscript.         %
\newcommand{\comma}{\: ,}              % Comma and Period to be used to   %
\newcommand{\Om}{\Omega}                %%%%%%%%%%%%%%%%%%%%%%%%%%%%%%%%%%%

\newcommand{\one}{{\bf 1}}
\newcommand{\cA}{{\mathcal{A}}}
\newcommand{\cB}{{\mathcal{B}}}

\newcommand{\cF}{{\mathcal{F}}}

\newcommand{\cH}{{\mathcal{H}}}

\newcommand{\cL}{{\mathcal{L}}}         %%%%%%%%%%%%%%%%%%%%%%%%
\newcommand{\cO}{{\mathcal{O}}}         %                      %
\newcommand{\cP}{{\mathcal{P}}}         %%%%%%%%%%%%%%%%%%%%%%%%

\newcommand{\cS}{{\mathcal{S}}}

\newcommand{\RR}{\mathbb{R}}            %%%%%%%%%%%%%%%%%%%%%%%%%%%
\newcommand{\ZZ}{\mathbb{Z}}            %                         %
\newcommand{\NN}{\mathbb{N}}            % Blackboard Bold Letters %
\newcommand{\CC}{\mathbb{C}}            %                         %
\newcommand{\fh}{\mathfrak{h}}         %%% Gothic Letters %%%%%%%%
\newcommand{\vA}{{\vec{A}}}             %%%%%%%%%%%%%%%%%%%%%%%%%%%%%
\newcommand{\vnabla}{{\vec{\nabla}}}
\newcommand{\veps}{{\vec{\varepsilon}}}

\newcommand{\vq}{{\vec{q}}}
\newcommand{\vk}{{\vec{k}}}

\newcommand{\vp}{{\vec{p}}}
\newcommand{\vP}{{\vec{P}}}

\newcommand{\vx}{{\vec{x}}}

\newcommand{\vy}{{\vec{y}}}
\newcommand{\cirS}{\mathop{\bigcirc\kern -.73em {\scriptstyle{\rm S}}}}

\newcommand{\at}{{el}}

\def\Bra{\Big\langle}
\def\Ket{\Big\rangle}
\def\bra{\big\langle}
\def\ket{\big\rangle}

\def\eqn{\begin{eqnarray}}
\def\eeqn{\end{eqnarray}}
\newcommand{\Proof}{\noindent{\em Proof.}}
\newcommand{\QED}{\phantom{blablabla}\hfill\qed\newline}  % End of a Proof %
\renewcommand{\thesection}%            %%%%%%%%%%%%%%%%%%%%%%%%%%%%%%%%%%%%
{\Roman{section}}                      % To begin a section, write \secct %
\renewcommand{\theequation}%           % instead of \section. The section %
{\thesection.\arabic{equation}}        % number will then be displayed    %
\newtheorem{theorem}{Theorem}[section]         %%%%%%%%%%%%%%%%%%%%%%%%%%%%%%
\newtheorem{lemma}[theorem]{Lemma}             % These Theoremlike environm.%
\theoremstyle{plain}
\renewcommand{\theequation}{\thesection.\arabic{equation}}
\newcommand{\resetequ}{\setcounter{equation}{0}}
\begin{document}
\bibliographystyle{plain}
%%%%%%%%%%%%%%%%%%%%%%%%%%%%%%%%%%%%%%%%%%%%%%%%%%%%%%%%%%%%%%%%%%%%%%%%%%%%

\baselineskip=\baselskip
\thispagestyle{empty}

\title{Infraparticle Scattering States in Non-Relativistic QED:
II. Mass Shell Properties}

\author{Thomas Chen}
\address{Department of Mathematics, University of Texas at Austin, 1 University Station C1200,
Austin, TX 78712, USA}
\email{tc@math.utexas.edu}
\author{J\"{u}rg Fr\"{o}hlich} 
\address{Institut f\"ur Theoretische Physik, ETH H\"{o}nggerberg,  
CH-8093 Z\"{u}rich, Switzerland, and IH\'ES, Bures sur Yvette, France}
\email{juerg@itp.phys.ethz.ch}
\author{Alessandro Pizzo}
\address{Department of Mathematics, One Shields Avenue,
University of California Davis, Davis, CA 95616, USA}
\email{pizzo@math.ucdavis.edu}

%\date{\DATUM}

\begin{abstract}
We study the infrared problem in the usual model of QED with non-relativistic
matter. 
We prove spectral and regularity properties characterizing the mass shell
of an electron
and one-electron infraparticle states of this model.
Our results are crucial
for the construction of infraparticle scattering states, 
which are treated in a separate paper. 
\end{abstract}


\maketitle

\section{Introduction}

We study the dynamics of an electron interacting with the quantized electromagnetic field
in the framework of non-relativistic Quantum Electrodynamics (QED).
%We establish some key spectral and regularity results that characterize 
%the generalized mass shell and the infraparticle states of this model.
%The properties inquired
%here address the mass shell structure of a non-relativistic electron
%interacting with the quantized radiation field. In particular, we prove certain
%regularity results for the one-particle states of the electron 
%under the removal of an
%infrared cutoff used to regularize the model, 
%Our results are crucial for the construction of infraparticle scattering states,
%which we present in a companion paper, \cite{ChFrPi2}. 
%
%
%Here, we briefly outline the background in literature, and explain the scheme of
%our analysis.  
In a theory describing a massive particle (the electron) interacting
with a field of {\em massless} bosons (the photons), 
massive one-particle states do, in general, not exist
in the physical Hilbert space of the theory.
This fact was first observed by Schroer \cite{Schroer}, 
who also coined the term ``infraparticle'',
a notion that generalizes that of a particle.
In relativistic QED,
charged infraparticles were shown to occur, 
using arguments from general quantum field
theory; see \cite{FrMoSt,Buchholz2}.  
For  the spectrum of $(H,\vP)$ 
in Nelson's model, a simplified variant of non-relativistic QED,
with $H$ denoting the Hamiltonian, and  
$\vP$ the  total, conserved momentum of the massive particle and
the massless bosons,
it was proven in \cite{Fr73,Fr74}
that the bottom of the spectrum of the fiber
Hamiltonian $H_{\vP}$ at a fixed total momentum $\vP\in\RR^3$ is
not an eigenvalue of $H_\vP$, for any  value of $\vP$ with $\frac{|\vP|}{m} <\rho_{0}(\lambda)<1$ where 
$\lambda$ is the coupling constant and $m$ is the electron mass. 
To prove this result, one introduces an infrared
cutoff $\sigma>0$ in the Hamiltonian $H_{\vP}$ turning off all interactions of the
non-relativistic, massive particle with the soft
modes (with frequencies $<\sigma$) of the relativistic, massless boson field.
One then aims to establish  
spectral properties of the model in the limit $\sigma\to0$. 

Extending results of \cite{Fr73,Fr74}, an iterative algorithm for constructing the
ground state vector $\Psi_{\vP}^\sigma$ of the infrared regularized Hamiltonian 
$H_{\vP}^\sigma$ in Nelson's model has been developed in  \cite{Pizzo2003} 
using a novel multiscale analysis technique. In \cite{Pizzo2003}, important regularity 
properties have been derived,
which are crucial for the analysis of the asymptotic dynamics of the
electron. Similarly as in
\cite{Fr74}, the strategy in \cite{Pizzo2003} is to apply a specific
{\em Bogoliubov transformation} to the photon variables in $H_{\vP}^\sigma$, in order to obtain a
Hamiltonian $\Hw_{\vP}^\sigma$ whose ground state 
$\Phi_{\vP}^\sigma$ remains in   Fock space, as $\sigma\to0$. 
Subsequently, one derives
properties of the ground state vector of the physical Hamiltonian $H_{\vP}$ 
in the singular limit $\sigma\to0$ by inverting the Bogoliubov
transformation. In the limit $\sigma\to0$, the latter  
gives rise to a coherent representation of the 
observable algebra of the boson field {\em unitarily inequivalent} 
to the Fock representation 
and to the coherent representations associated to different values of
the total momentum.

The identification of the correct Bogoliubov transformation  is crucial for the 
constructions in \cite{Pizzo2003,Pizzo2005}. 
For Nelson's model, this Bogoliubov transformation has
been found in \cite{Fr73} by a method that exploits the  linearity
of the interaction in the Nelson Hamiltonian with respect to the
creation- and annihilation operators. 
Due to the more complicated   structure of the interaction
Hamiltonian in non-relativistic
QED, this argument cannot be applied, and the correct
Bogoliubov transformation  for non-relativistic QED
has only recently been identified in \cite{ChFr}, based on  
uniform bounds on the renormalized electron mass established in \cite{Chen}.
This makes it possible to extend the constructions and methods of 
\cite{Pizzo2003,Pizzo2005} to non-relativistic QED.

By a generalization of the multiscale methods based on  
\emph{recursive analytic perturbation theory} introduced in
\cite{Pizzo2003}, we present a new construction of the correct Bogoliubov transformation,
and we    prove the following main results:
\begin{itemize}
\item
The ground state vectors $\Phi_{\vP}^\sigma$ of the 
Bogoliubov-transformed Hamiltonians $\Hw_{\vP}^\sigma$ converge strongly
to a vector in Fock space,
in the limit $\sigma\to0$. The convergence rate is estimated by
$\cO(\sigma^{\eta})$, for some 
explicit $\eta>0$.
\item 
The vectors $\Phi_{\vP}^\sigma$ in Fock space
are H\"older continuous in $\vP$, 
uniformly in $\sigma$.
\end{itemize}
These properties are key ingredients for the construction of
infraparticle scattering states, which we present in \cite{ChFrPi2}.
A key difficulty in this analysis is the fact that
the infrared behavior of the interaction in QED is,
in the terminology of renormalization group theory, 
of \emph{marginal} type (see also \cite{Chen}).

\newpage

%%%%%%%%%%%%%%%%%%%%%%%%%%%%%%%%%%%%%%%%%%%%%%%%%%%%%%%%%%%%%%%%%%%
%%%%%%%%%%%%%%%%%%%%%%%%%%%%%%%%%%%%%%%%%%%%%%%%%%%%%%%%%%%%%%%%%%%
\section{Definition of the Model }
\label{sec-II.1} 
\label{sec-I.2}
%%%%%%%%%%%%%%%%%%%%%%%%%%%%%%%%%%%%%%%%%%%%%%%%%%%%%%%%%%%%%%%%%%%
 The Hilbert space of pure state vectors of the system
consisting of one non-relativistic electron interacting with the quantized
electromagnetic field is given by
\begin{equation} \label{eq-I-1}
\cH \; := \; \cH_\at \, \otimes \, \cF  \, ,
\end{equation}
where $\cH_\at = L^2(\RR^3)$ is the Hilbert space for a
single Schr\"odinger electron (for expository convenience, we neglect
the spin of the electron).
The Fock space used to describe the states of the transverse
modes of the quantized electromagnetic field (the \emph{photons}) in
the Coulomb gauge is given by
\begin{equation} \label{eq-I-2}
\cF \ := \ \bigoplus_{N=0}^\infty \cF^{(N)} \comma \hspace{6mm}
\cF^{(0)} = \CC \, \Om \comma
\end{equation}
where $\Om$ is the vacuum vector (the state of the electromagnetic
field without any excited modes), and
\begin{equation} \label{eq-I-3}
\cF^{(N)} \ := \ \cS_N \, \bigotimes_{j=1}^N \fh \comma \hspace{6mm}
N \geq 1 \comma
\end{equation}
where the Hilbert space $\fh$ of a single photon is
\begin{equation} \label{eq-I-4}
\fh \ := \ L^2( \RR^3 \times \ZZ_2 ) \,.
\end{equation}
Here, $\RR^3$ is momentum space, and $\ZZ_2$ accounts for the
two independent transverse polarizations (or helicities) of a
photon.  In (\ref{eq-I-3}), $\cS_N$ denotes the orthogonal
projection onto the subspace of $\bigotimes_{j=1}^N \fh$ of totally
symmetric $N$-photon wave functions, to account for the fact
that photons satisfy Bose-Einstein statistics. Thus, $\cF^{(N)}$ is
the subspace of $\cF$ of state vectors for configurations of exactly
$N$ photons. 
%It is convenient to represent the Hilbert space $\cH$
%as the space of square-integrable wave functions on the electron
%position space $\RR^3$ with values in the photon Fock space $\cF$,
%i.e.,
%%
%\begin{equation} \label{eq-I-5}
%\cH \ \cong \  L^2 ( \RR^3 \, ; \, \cF ) \period
%\end{equation}
%%

In this paper, we use units such that Planck's constant $\hbar$,
the speed of light $c$, and the
mass of the electron are equal to unity.
The dynamics of the system is generated by the Hamiltonian
\begin{equation} \label{eq-I-6}
H \, := \, \frac{\big(-i\vnabla_{\vx} \, + \, \alpha^{1/2} \vA(\vx)
\, \big)^2}{2} \, + \, H^{f}\,.
\end{equation}
The multiplication operator $\vx\in\RR^3$ accounts for the position of
the electron. The electron momentum operator is given by $\vp=-i\vnabla_\vx$.
$\alpha \cong 1/137$ is the finestructure constant (which, in this paper, plays
the r\^ole of a small parameter), $\vA(\vx)$ denotes the vector
potential of the transverse modes of the quantized electromagnetic
field in the \emph{Coulomb gauge},
\begin{equation} \label{eq-I-7}
\vnabla_\vx \cdot \vA (\vx) \ = \ 0 \, .
\end{equation}

The operator $H^f$ is the Hamiltonian of the quantized, free
electromagnetic field,  
\begin{equation} \label{eq-I-10}
    H^f \; := \;  \sum_{\lambda = \pm} \int d^3k \; |\vk| \,
    a^*_{\vk,\lambda} \,  a_{\vk, \lambda} \comma
\end{equation}
where $a^*_{\vk, \lambda}$ and $a_{\vk, \lambda}$ are the usual
photon creation- and annihilation operators,   satisfying the
canonical commutation relations
\begin{eqnarray}
\label{eq-I-12}
    [a_{\vk, \lambda} \, , \, a^*_{\vk', \lambda'}] & = &
    \delta_{\lambda \lambda'} \, \delta (\vk - \vk') \comma
    \\
    \label{eq-I-11}
    [a^\#_{\vk, \lambda} \, , \, a^\#_{\vk', \lambda'}] & = & 0
\end{eqnarray}
(where $a^\#=a$ or $a^*$). The vacuum vector $\Om$ is characterized by
the condition
\begin{equation}
    a_{\vk, \lambda} \, \Om \; = \; 0 \comma  \label{eq-I-13}
\end{equation}
for all $\vk  \in \RR^3$ and $\lambda  \in \ZZ_2
\equiv \{\pm\}$.

The quantized electromagnetic vector potential is given by
\begin{eqnarray} \label{eq-I-14}
    \vA(\vx) \; := \;
    \sum_{\lambda = \pm} \int_{\cB_{\Lambda}} \frac{d^3k}{\sqrt{ |\vk| \,}}\,
    \big\{ \veps_{\vk, \lambda} e^{-i\vk \cdot \vx}
    a^*_{\vk, \lambda} \, + \, \veps_{\vk, \lambda}^{\;*}
    e^{i\vk \cdot \vx} a_{\vk, \lambda} \big\} \comma
\end{eqnarray}
where $\veps_{\vk, -}$, $\veps_{\vk, +}$ are photon polarization
vectors, i.e., two unit vectors in $\RR^3 \otimes\CC$ satisfying
\begin{equation} \label{eq-I-15}
    \veps_{\vk, \lambda}^{\;*} \cdot \veps_{\vk, \mu} \; = \;
    \delta_{\lambda \mu} \comma \hspace{8mm} \vk \cdot \veps_{\vk,
    \lambda} \; = \; 0 \comma
\end{equation}
for $\lambda, \mu = \pm$. The equation $\vk \cdot \veps_{\vk,
\lambda} = 0$ expresses the Coulomb gauge condition. Moreover,
$\cB_{\Lambda}$ is a ball of radius $\Lambda$ centered at the origin
in momentum space. $\Lambda$ represents an \emph{ultraviolet cutoff}
that will be kept fixed throughout our analysis. The vector
potential defined in (\ref{eq-I-14}) is thus cut off in the
ultraviolet.

Throughout this paper, it will be assumed that $\Lambda\approx 1$
(the rest energy of an electron), and that $\alpha$ is sufficiently small.
Under these assumptions, the Hamitonian $H$ is selfadjoint on $D(H^0)$, i.e.,
on the domain of definition of the operator
\begin{equation}
    H^0 \; := \; \frac{(-i\vnabla_{\vx})^2}{2} \, + \, H^f \;.
\end{equation}
The perturbation $H-H^0$ is small in the sense of Kato; see, e.g., \cite{Simon}.

The operator measuring the total momentum of the system consisting of the
electron and the electromagnetic radiation field is given by
\begin{equation}
    \vP\,:=\,\vp+\vP^f \, ,
\end{equation}
where $\vp=-i\vnabla_{\vx}$ is the momentum operator for the electron, and
\begin{equation}
    \vP^f\,:=\, \sum_{\lambda = \pm} \int  d^3k \; \vk \,
    a^*_{\vk, \lambda} \, a_{\vk, \lambda}
\end{equation}
is the momentum operator associated with the photon field.

The operators $H$ and $\vP$ are essentially selfadjoint on the 
domain $D(H_0)$, and since the dynamics is invariant under
translations, they commute, $[H,\vP]=\vec 0$.
The Hilbert space $\cH$ can be decomposed on the joint
spectrum, $\RR^3$,  of the component-operators of $\vP$.
Their spectral measure is
absolutely continuous with respect to Lebesque measure.
Thus,
\begin{equation}
\cH\,:=\,\int^{\oplus}\cH_{\vP} \, d^3P \, ,
\end{equation}
where each fiber space $\cH_{\vP}$ is a copy of Fock space $\cF$.
\\

\noindent {\bf{Remark}} \emph{Throughout
this paper, the symbol $\vP$ stands both for a variable in $\RR^3$
and for a vector operator in $\cH$, depending on the context.
Similarly, a double meaning is also
associated with functions  of the total
momentum operator.}
\\

We recall that vectors $\Psi\in\cH$ are given by sequences
\begin{equation}
\{\Psi^{(m)}(\vx; \vk_1,\lambda_1;\dots; \vk_m,\lambda_m)\}_{m=0}^{\infty}\,,
\end{equation}
of functions, $\Psi^{(m)}$, where $\Psi^{(0)}(\vx)\in L^2(\RR^3)$, of the electron position $\vx$ and of $m$ photon momenta
$\vk_1,\dots,\vk_m $ and helicities $\lambda_1, \dots, \lambda_m$, with the following properties:
\begin{itemize}
\item[(i)]
$\Psi^{(m)}(\vx; \vk_1,\lambda_1;\dots; \vk_m,\lambda_m)$
is totally symmetric in its $m$ arguments $(\vk_j,\lambda_j)_{j=1,\dots,m}$.
\item[(ii)]
$\Psi^{(m)}$ is square-integrable, for all $m$.
\item[(iii)]
If $\Psi$ and $\Phi$ are two vectors in $\cH$ then 
\begin{eqnarray}
\lefteqn{(\Psi\,,\,\Phi) }\\
&=&\sum_{m=0}^{\infty}\big(\sum_{\lambda_j=\pm}\,\int\,d^3x\,\prod_{j=1}^{m}\,d^3k_j\,\overline{\Psi^{(m)}(\vx;\vk_1,\lambda_1;\dots; \vk_m,\lambda_m)}\,\Phi^{(m)}(\vx;\vk_1,\lambda_1;\dots; \vk_m,\lambda_m)\big)\,.\nonumber
\end{eqnarray}
\end{itemize}
We identify a square integrable function $g(\vx)$ with the sequence 
\begin{equation}
\{\Psi^{(m)}(\vx; \vk_1,\lambda_1;\dots; \vk_m,\lambda_m)\}_{m=0}^{\infty}\,,
\end{equation}
where $\Psi^{(0)}(\vx)\equiv g(\vx)$, and $\Psi^{(m)}(\vx; \vk_1,\lambda_1;\dots; \vk_m,\lambda_m)\equiv 0$ for all $m>0$;  analogously, a square integrable function $g^{(m)}(\vx; \vk_1,\lambda_1;\dots; \vk_m,\lambda_m)$, $m\geq 1$,  is identified with the sequence
\begin{equation}
\{\Psi^{(m)}(\vx; \vk_1,\lambda_1;\dots; \vk_m,\lambda_m)\}_{m=0}^{\infty}\,,
\end{equation}
where $\Psi^{(m)}(\vx; \vk_1,\lambda_1;\dots; \vk_m,\lambda_m)\equiv g^{(m)} $, and  $\Psi^{(m')}(\vx; \vk_1,\lambda_1;\dots; \vk_m,\lambda_m)\equiv 0$ for all $m'\neq m$. From now on, a sequence describing a quantum state with a fixed number of photons is identified with  its nonzero component wave function; vice versa, a wave function corresponds to a sequence according to the previous identification.
The elements of the fiber space $\cH_{\vP^*}$ are obtained by linear combinations of the (improper) eigenvectors  of the total momentum operator $\vP$ with eigenvalue $\vP^*$, e.g., the plane wave $e^{i\vP^*\cdot\vx}$ is the eigenvector describing a state with an electron and no photon. 
Given any $\vP\in\RR^3$, there is an isomorphism, $I_{\vP}$,
\begin{equation}
    I_{\vP}\,:\,\cH_{\vP}\,\longrightarrow\,\cF^{b}\,,
\end{equation}
from the fiber space $\cH_{\vP}$ to the Fock space $\cF^{b}$, acted upon by the annihilation- and
creation operators $b_{\vk,\lambda}$, $b^*_{\vk,\lambda}$,
where
$b_{\vk,\lambda}$ corresponds to $e^{i\vk\cdot\vx}  a_{\vk,\lambda}$, and
$b_{\vk,\lambda}^*$ to $e^{-i\vk\cdot\vx} a_{\vk,\lambda}^* $, and
with vacuum $\Omega_{f}:=I_{\vP}(e^{i\vP\cdot\vx})$.
To define $I_{\vP}$ more precisely, we consider a vector
$\psi_{(f^{(n)};\vP)}\in \cH_{\vP}$
with a definite total momentum, $\vP$, describing an electron and $n$ photons. Its wave function in the variables
$(\vx;\vk_1,\lambda_1;\dots,\vk_n,\lambda_n)$ is given by
\begin{equation}
 e^{i(\vP-\vk_1-\cdots-\vk_n)\cdot\vx}f^{(n)}(\vk_1,\lambda_1;\dots;\vk_n,\lambda_n)
\end{equation}
where $f^{(n)}$ is totally symmetric in its $n$ arguments.
The isomorphism $I_{\vP}$ acts by way of
\begin{eqnarray}
    \lefteqn{I_{\vP}\big( e^{i(\vP-\vk_1-\cdots-\vk_n)\cdot\vx}f^{(n)}(\vk_1,\lambda_1;\dots;\vk_n,\lambda_n)\big)}
    \\
    &= &\frac{1}{\sqrt{n!}}\sum_{\lambda_1,\dots,\lambda_n}\int \, d^3k_1\dots d^3k_n \,f^{(n)}(\vk_1,\lambda_1;\dots;\vk_n,\lambda_n)\,
    b_{\vk_1,\lambda_1}^* \cdots
    b_{\vk_n,\lambda_n}^*  \, \Omega_f \,.\nonumber
\end{eqnarray}

The Hamiltonian $H$ maps each $\cH_{\vP}$ into itself, i.e.,
it can be written as
\begin{equation}
    H\,=\,\int H_{\vP}\,d^3P\,,
\end{equation}
where
\begin{equation}
    H_{\vP}\,:\,\cH_{\vP}\longrightarrow\cH_{\vP}\,.
\end{equation}
Written in terms of the operators $b_{\vk,\lambda}$, $b^*_{\vk,\lambda}$, and of the
variable $\vP$, the fiber Hamiltonian $H_{\vP}$ has the form
\begin{equation}\label{eq-fibHam}
    H_{\vP} \; := \; \frac{\big(\vP-\vP^f +\alpha^{1/2} \vA \big)^2}{2} \;
    + \;H^{f}\,,
\end{equation}
where
\begin{eqnarray}
    \vP^f & = & \sum_{\lambda}\, \int d^3k \, \vk \,  b^*_{\vk,\lambda} \,  b_{\vk,\lambda} \, ,
    \\
    H^f & = & \sum_{\lambda}\, \int d^3k \, |\vk| \, b^*_{\vk,\lambda} \, b_{\vk,\lambda} \,,
\end{eqnarray}
and
\begin{equation}
    \vA \, := \,\sum_{\lambda}\, \int_{\cB_{\Lambda}}\,
    \frac{d^3k}{\sqrt{|\vk| \,}} \, \big\{  b^*_{\vk,\lambda} \veps_{\vk,\lambda}
    \, + \, \veps^{\;*}_{\vk,\lambda} b_{\vk,\lambda} \big\} \,.
\end{equation}
In the following, we will only construct infraparticle states of momentum $\vP\in\mathcal{S}$, 
where 
\begin{equation}
    \mathcal{S}:=\lbrace\,
    \vP\in\RR^3\,:\,|\vP|<\frac{1}{3}\,\rbrace\,.
\end{equation}
(Our results can be extended to a region $ \mathcal{S}$ (inside the unit ball) of radius larger than $1/3$.)

In order to give a well-defined meaning to the operations we use in the sequel,
we introduce an infrared cut-off at energy $\sigma>0$ in the interaction term
\begin{equation}
	H_{I,\vP} \, := \, \alpha^{1/2} \, \vA \cdot (\vP - \vP^f) 
	\, + \, \alpha \, \frac{\vA^2}{2}
\end{equation} 
of the  Hamiltonian $H_\vP$, which is imposed on the vector potential $\vA$.
Its removal is the main problem solved in this paper.
Our results are crucial ingredients for infraparticle scattering theory; see \cite{ChFrPi2}.
We will start by studying the regularized fiber Hamiltonian
\begin{equation}\label{eq:H-fiber}
    H_{\vP}^\sigma \, := \, \frac{\big(\vec{P}-\vec{P}^{f}
    +\alpha^{1/2} \vA^{\sigma} \big)^2}{2} \,  + \, H^{f}\,
\end{equation}
acting on the
fiber space $\mathcal{H}_{\vec{P}}$, for $\vec{P}\in \mathcal{S}$,
where
\begin{eqnarray}
    \vA^{\sigma} \, := 
    \,\sum_{\lambda}\, \int_{\mathcal{B}_{\Lambda}\setminus \mathcal{B}_{\sigma}}\,
    \frac{d^3k}{\sqrt{ |\vk| \,}} \, \big\{  b^*_{\vk,\lambda} \veps_{\vk,\lambda}\, + \,
    \veps^*_{\vk,\lambda} b_{\vk,\lambda} \big\}
\end{eqnarray}
and where $\mathcal{B}_{\sigma }$ is a ball of radius $\sigma$.
We will consider a sequence $(\sigma_j)_{j=0}^\infty$
of infrared cutoffs given by $\sigma_j \, := \, \Lambda\epsilon^j$,
with $0<\epsilon<1$ and $j\in\mathbb{N}_0:=\NN\cup \{0\}$.

In Section {\ref{sec-IV}}, we construct the ground state vector
$(\Psi_{\vP}^{\sigma_j})$ of the Hamiltionan $(H_{\vP}^{\sigma_j})$, and
we   compare ground state vectors $\Psi_{\vP}^{\sigma_j}$,
$\Psi_{\vP'}^{\sigma_{j'}}$ corresponding to different fiber Hamiltonians
$H_{\vP}^{\sigma_j}$, $H_{\vP'}^{\sigma_{j'}}$ with
$\vP\not=\vP'$. We compare the vectors  $\Psi_{\vP}^{\sigma_j}$,
$\Psi_{\vP'}^{\sigma_{j'}}$ as elements of the space $\cF^b$.
More precisely, we use the expression
\begin{equation}
    \|\Psi_{\vP}^{\sigma_j}-\Psi_{\vP'}^{\sigma_j}\|_{\cF}
\end{equation}
as an abbreviation for
\begin{equation}
    \|I_{\vP}(\Psi_{\vP}^{\sigma_j})-I_{\vP'}(\Psi_{\vP'}^{\sigma_j})\|_{\cF}\,.
\end{equation}

\subsection{Background}

In a companion paper \cite{ChFrPi2}, we construct a vector
$\psi_{h,\Lambda_1}(t)$ converging to a scattering state
$\psi_{h,\Lambda_1}^{out/in}$, as time $t$ tends to infinity, applying and
extending 
mathematical techniques developed in \cite{Pizzo2005} for Nelson's model.
%Convergence of $\psi_{h,\Lambda_1}(t)$
%to a non-zero vector $(\psi_{h,\Lambda_1}^{out/in})$ in the Hilbert space $\cH$
%as $t\rightarrow\pm\infty$ is shown.
The vector $\psi_{h,\Lambda_1}^{out/in}$ represents an electron with a wave
function $h$ in the momentum variable with support contained
in $\cS=\{\vP : |\vP|<\frac13\}$, accompanied
by a cloud of real photons described by a Bloch-Nordsieck factor, and with
an upper photon frequency cutoff $\Lambda_1$.

In \cite{ChFrPi2} we also construct the scattering subspaces $\cH^{out/in}$,
starting from certain subspaces,  $\cH^{1\;out/in}$, and applying "hard" asymptotic
photon creation operators. These spaces carry representations of the  
algebras, $\cA_{ph}^{out/in}$ and $\cA_{el}^{out/in}$, of asymptotic photon-
and electron observables, respectively, and the fact that their actions
commute proves, mathematically,   asymptotic decoupling of the electron 
and photon dynamics, as time $t\rightarrow\pm\infty$.
Properties of the representations of $\cA_{ph}^{out/in}$
in the infrared expected on the basis of
the Bloch-Nordsieck paradigm are rigorously established; see \cite{ChFrPi2}.

\newpage

%%%%%%%%%%%%%%%%%%%%%%%%%%%%%%%%%%%%%%%%%%%%%%%%%%%%%%%%
\section{Statement of the Main Results} \label{sec-II}
%%%%%%%%%%%%%%%%%%%%%%%%%%%%%%%%%%%%%%%%%%%%%%%%%%%%%%%%
 
\resetequ

The main results of our paper are summarized in Theorem \ref{thm-cfp-2} below.
They are fundamental for the construction of scattering states in \cite{ChFrPi2}
and are very similar to those used in the analysis of Nelson's model
in \cite{Pizzo2005}.

We define the energy of a dressed one-electron state of momentum $\vP$ by
\begin{equation}
    E^\sigma_{\vP} \,  = \, \inf {\rm spec} H_{\vP}^\sigma
    \quad , \quad\quad
    E_{\vP}  \,  = \, \inf {\rm spec} H_{\vP} \, = \, E^{\sigma=0}_{\vP} \,.
\end{equation}
We   refer to $E_{\vP}^\sigma$ as the {\em ground state energy} of the fiber
Hamiltonian $H_{\vP}^\sigma$.
If it exists the corresponding ground state is denoted by $\Psi_\vP^\sigma$.
We   always assume that $\vP \in \mathcal{S}:=\{\vP\in \RR^3 \, : \, |\vP|<\frac13 \,\}$ 
and that $\alpha$ is so small that, for all $\vP\in\mathcal{S}$,
\begin{equation}
    |\vnabla E_{\vP}^\sigma| \, < \, \nu_{max} \, < \, 1
\end{equation}
for some constant $\nu_{max}$, uniformly in $\sigma$.
The existence of $\vnabla E_{\vP}^\sigma$ will be proven in Section \ref{section-V-1}.

Let $\delta_\vP^\sigma(\widehat k)$ be given by
\begin{equation}
	\delta_\vP^\sigma(\widehat k) \, := \, 1 \, - \, \frac{\vk\cdot\vnabla E_\vP^\sigma}{|\vk|} \,.
\end{equation}
We introduce an operator  
\begin{equation}\label{eq-II-2}
    W_{\sigma}(\vnabla E_{\vP}^{\sigma}) \, := \, \exp\Big(\, \alpha^{\frac{1}{2}}
    \sum_{\lambda} \int_{\mathcal{B}_{\Lambda}\setminus \mathcal{B}_{\sigma}} d^3 k
    \frac{\vnabla E_{\vP}^{\sigma}}{|\vec{k}|^{\frac{3}{2}}\delta_{\vP}^\sigma(\widehat{k})} \cdot
    (\veps_{\vk,\lambda}b_{\vk,\lambda}^{*} - h.c.)\Big) \, ,
\end{equation}
on $\cH_{\vP}$, which is unitary for $\sigma>0$, and consider the transformed
fiber Hamiltonian 
\begin{equation}\label{eq-II-1}
    \Hw_{\vP}^\sigma\,:=\,W_{\sigma}(\vnabla
    E_{\vP}^{\sigma})H_{\vP}^\sigma W_{\sigma}^{*}(\vnabla E_{\vP}^{\sigma})\,.
\end{equation}
Conjugation by $W_{\sigma}(\vnabla E_{\vP}^{\sigma}) $ acts on the 
creation- and annhilation operators by a (Bogoliubov) translation
\begin{equation}
	W_{\sigma}(\vnabla
    E_{\vP}^{\sigma}) \, b^\#_{\vk,\lambda} \, W_{\sigma}^{*}(\vnabla E_{\vP}^{\sigma})
    \, = \, b^\#_{\vk,\lambda} \, - \,\alpha^{1/2} 
    \frac{\one_{\sigma,\Lambda}(\vk)}{|\vec{k}|^{\frac{3}{2}}\delta_{\vP}^\sigma(\widehat{k})} 
    \vnabla E_{\vP}^{\sigma}\cdot\veps_{\vk,\lambda}^{\;\#} \,,
\end{equation}
where $\one_{\sigma,\Lambda}(\vk)$ stands for the characteristic function of the set
$\mathcal{B}_{\Lambda}\setminus\mathcal{B}_{\sigma}$.
Our methods exploit regularity properties in $\sigma$
and $\vec{P}$ of the ground state vector, $\Phi_{\vec{P}}^{\sigma}$,
and of the ground state energy, $E^\sigma_{\vP}$, of $\Hw_{\vP}^\sigma$.
These properties are formulated in the following theorem, which is the main result of this paper.

\begin{theorem}
\label{thm-cfp-2}
For $\vP\in\mathcal{S}$ and for $\alpha>0$ sufficiently small, the following statements hold.

\begin{itemize}
\item[($\mathscr{I}1$)]
The energy $E^{\sigma}_{\vP}$ is a simple eigenvalue of the operator $\Hw_\vP^\sigma$
on $\cF^b$.
Let $\cB_{\sigma}:=\{ \vk \in \RR^3 \, | \, |\vk|\leq\sigma \}$, and
let $\cF_{\sigma}$ denote the Fock space over $L^2((\RR^3\setminus \cB_\sigma)\times\ZZ_2)$. 
%in  $b_{\vk,\lambda}$ and $b^*_{\vk,\lambda}$. Likewise,
Likewise, we define $\cF_0^{\sigma}$ to be the Fock space over $L^2(\cB_\sigma\times\ZZ_2)$;
hence $\cF^b=\cF_\sigma\otimes\cF_0^\sigma$.
On $\cF_{\sigma}$, the operator $\Hw_{\vP}^\sigma$
has a spectral gap of size $\rf\sigma$ or larger,
separating $E^{\sigma}_{\vP}$ from the rest of its spectrum,
for some constant 
$\rf$ (depending on $\alpha$), with $0<\rf<1$.

The contour
\begin{equation}
    \gamma\,:=\lbrace z\in\CC\,|
    |z-E_{\vec{P}}^{\sigma}|=\frac{\rf\sigma}{2}\rbrace \; \;  , \; \sigma>0\,  
\end{equation}
bounds a disc which intersects the spectrum of $\Hw_{\vP}^\sigma|_{\cF_\sigma}$
in only one point, $\{E_{\vec{P}}^{\sigma}\}$. The normalized ground state vectors
%,
%\begin{equation}
%	\Phi_\vP^\sigma \, = \, \zeta \, W_\sigma(\vnabla E_\vP^\sigma) \,
%	\frac{\Psi_\vP^\sigma}{\|\Psi_\vP^\sigma\|_{\cF}}
%	\; \; \; , \; \; \zeta\in\CC \; \; , \; \; |\zeta|\, = \, 1  \,,
%\end{equation}
of the operators $\Hw_\vP^\sigma$ are given by
\begin{equation}\label{eq-II-3}
    \Phi_{\vec{P}}^{\sigma} \, := \, \frac{\frac{1}{2\pi\,i}\int_{\gamma}
    \frac{1}{\Hw_{\vP}^\sigma-z} \,dz\,\Omega_f}
    {\|\frac{1}{2\pi\,i}\int_{\gamma}\frac{1}{\Hw_{\vP}^\sigma-z}  
    \,dz\,\Omega_f\|_{\cF} } 
    %\, \otimes \, \Omega_f
\end{equation}
and converge 
strongly   to a
non-zero vector $\Phi_{\vec{P}}\in \cF^b$, in the limit $\sigma\to 0$.
The rate of convergence
is, at least, of order $\sigma^{\frac{1}{2}(1-\delta)}$, for any
$0<\delta<1$.  
Formula (\ref{eq-II-3}) fixes the phase of $\Phi_\vP^\sigma$ such that 
$|( \Phi_{\vP}^\sigma \, , \, \Omega_f)|>0$.

The dependence of the ground state energies $E_\vP^\sigma$ of the fiber Hamiltonians
$\Hw_\vP^\sigma$ on the infrared cutoff $\sigma$
is characterized by the following estimates.
%
%The ground state energies $E_{\vP}^\sigma$ are Lipschitz in $\sigma$,
\begin{equation}\label{eq-II-4}
    | \, E_{\vec{P}}^{\sigma} - E_{\vec{P}}^{\sigma'} \, | \, \leq \,  \cO(\sigma)\,,
\end{equation}
and 
%their gradients are well-defined and H\"older continuous with
\begin{equation}
	| \, \vnabla E_{\vec{P}}^{\sigma} - \vnabla E_{\vec{P}}^{\sigma'} \, | 
	\, \leq \, \cO(\sigma^{\frac{1}{2}(1-\delta)}) \,,
\end{equation}
for any $0<\delta<1$, with $\sigma>\sigma'>0$.

\item[($\mathscr{I}2$)]
The following H\"older regularity properties in  $\vec{P}\in\mathcal{S}$ hold uniformly
in $\sigma \geq 0$:
\begin{equation}\label{eq-II-5}
    \|\Phi_{\vec{P}}^{\sigma}-\Phi_{\vec{P}+\Delta\vec{P}}^{\sigma}\|_{\cF} \leq
    C_{\delta'}|\Delta \vec{P}|^{\frac{1}{4}-\delta'}
\end{equation}
and
\begin{equation}\label{eq-II-6}
    |\vnabla E_{\vec{P}}^{\sigma}-\vnabla E_{\vec{P}+\Delta\vec{P}}^{\sigma}|\leq
    C_{\delta''}|\Delta \vec{P}|^{\frac{1}{4}-\delta''} \, ,
\end{equation}
for $0<\delta''<\delta'<\frac{1}{4}$, with $\vec{P}\,,\,\vec{P}+\Delta\vec{P} \in
\mathcal{S}$, where $C_{\delta'}$ and $C_{\delta''}$ are
finite constants depending on $\delta'$ and $\delta''$, respectively.

\item[($\mathscr{I}3$)]
Given a positive number $0<\nu_{min}<1$, there are numbers
$\nu_{max}$ independent of $\nu_{min}$
as long as $0<\nu_{min}<\nu_{max}<1$, and $r_\alpha = \nu_{min}+\cO(\alpha)>0$,
such that, for $\vP\in\mathcal{S}\setminus\cB_{r_{\alpha}}$ and for $\alpha$ sufficiently small,
\begin{equation}\label{eq-II-7}
    1>\nu_{max}\geq|\vnabla E_{\vec{P}}^{\sigma}|\geq\nu_{min}>0 \, ,
\end{equation}
uniformly in $\sigma$.  

\item[($\mathscr{I}4$)]
For $\vP\in\mathcal{S}$ and for any $\vk\not=0$, the following inequality holds
uniformly in $\sigma$, for $\alpha$ small enough:
\begin{equation}\label{eq-II-8}
    E^{\sigma}_{\vP-\vk}>E^{\sigma}_{\vP}-C_\alpha|\vk|\,,
\end{equation}
where $E^{\sigma}_{\vP-\vk}:=\inf {\rm{spec}} \, H_{\vP-\vk}^\sigma$ and $\frac13< C_\alpha<1$,
with $C_\alpha\rightarrow\frac13$ as $\alpha\rightarrow0$.  

\item[($\mathscr{I}5$)] 
Let $\Psi_{\vP}^\sigma \in \cF$ denote the ground state vector of the fiber Hamiltonian
$H_{\vP}^\sigma$, so that
\begin{equation}\label{eq-Phi-Psi-phase-1}
	\Phi_\vP^\sigma \, = \, \zeta \, W_\sigma(\vnabla E_\vP^\sigma) \,
	\frac{\Psi_\vP^\sigma}{\|\Psi_\vP^\sigma\|_{\cF}}
	\; \; \; , \; \; \zeta\in\CC \; \; , \; \; |\zeta|\, = \, 1  \,.
\end{equation}
For $\vP\in\mathcal{S}$, one has that
\begin{equation}\label{eq-II-9}
  	\| \, b_{\vk,\lambda}\frac{\Psi_\vP^\sigma}{\|\Psi_\vP^\sigma\|_{\cF}} \, \|_{\cF} \, \leq C \, \alpha^{1/2}
  	\, \frac{\one_{\sigma,\Lambda}(\vk)}{|\vk|^{3/2}}\,,
\end{equation} 
see Lemma 6.1 of \cite{ChFr} which can be extended to $\vk\in\RR^3$ using ($\mathscr{I}4$).
\end{itemize}

\end{theorem}

%The rest of this paper is dedicated to the proof of this theorem. Establishing
%the statement ($\mathscr{I}1$) will be the most demanding task, and will 
%occupy most of our efforts.
%\\

\begin{figure*}
%\centerline{\epsffile{cfp-fig2.pdf} }
%\centerline{\epsffile{cfp-fig2.eps} }
\includegraphics{cfp-fig2.epsf}

\caption{The condition ($\mathscr{I}4$).}
\end{figure*}

The proof of statement ($\mathscr{I}1$) is given in Section {\ref{sec-Jot-1}};
the proofs of statements ($\mathscr{I}2$) and ($\mathscr{I}3$)
are presented in Section \ref{sec-Jot-2-3}.
Statement ($\mathscr{I}4$) is proven in Section \ref{sec-Jot-4}.
We note that condition ($\mathscr{I}4$) plays an important r\^ole also in 
atomic and molecular bound
state problems, see for instance \cite{lomisp}. 

We note that in Section \ref{subsect-4.2} below, we will, by a slight abuse
of notation, use the same symbol $\Phi_{\vP}^\sigma$ for the ground state vector
of $\Hw_{\vP}^\sigma$ without normalization.

\subsection{Remark about infrared representations}

The statement ($\mathscr{I}5$), which states that
\begin{equation}\label{eq-II-9bisbis}
  	\| \, b_{\vk,\lambda}\Psi_{\vP}^{\sigma} \, \|_{\cF} \, \leq C \, \alpha^{1/2}
  	\, \frac{\one_{\sigma,\Lambda}(\vk)}{|\vk|^{3/2}}\,,
\end{equation}
follows from the
identity
\begin{equation}\label{eq-II-9bis}
  b_{\vk,\lambda}\Psi_{\vP}^{\sigma}\,=\,-\,
  \alpha^{\frac{1}{2}}\,\frac{\one_{\sigma,\Lambda}(\vk)}{|\vk|^{\frac{1}{2}}}
  \, \frac{1}{H_{\vP-\vk,\sigma}+|\vk|-E_{\vP}^{\sigma}}
  \, \veps_{\vk,\lambda}\cdot\vnabla_{\vP}
  H_{\vP}^\sigma\,\Psi_{\vP}^{\sigma}\,
\end{equation}
which is derived by using a standard 
"pull-through argument". Combined with
the {\em uniform bounds on the renormalized mass} of the electron established in \cite{Chen},
it is used in \cite{ChFr} to prove the bound
\begin{equation}\label{eq-II-10}
    \bra \, \Psi_{\vP}^{\sigma} \, , \, N^f \, \Psi_{\vP}^{\sigma} \, \ket \, := \, 
    \int d^3k \, \bra \, \Psi_{\vP}^{\sigma}\,,\,b^*_{\vk, \lambda}
    b_{\vk,\lambda}\Psi_{\vP}^{\sigma} \, \ket 
    \; \leq \; C \alpha (  1   +  |\vP|^2   |\ln(\sigma)|  ) \,
\end{equation}
on the expected number of photons in the ground state $\Psi_{\vP}^{\sigma}$.
Without using the uniform bounds on the renormalized mass, one obtains the weaker
upper bound (\ref{eq-II-9bisbis}).
Important implications of this result, analyzed in \cite{ChFr} 
and used in \cite{ChFrPi2}, can be summarized as follows.

Let $\mathfrak{A}_\rho$ denote the $C^*$-algebra of
bounded operators on the Fock space $\mathcal{F}(L^2((\RR^3\setminus B_\rho)\times\ZZ_2))$,
where $B_\rho=\{ \vk \in \RR^3 \, | \, |\vk|\leq\rho \}$, and let $\mathfrak{A}$ denote the $C^*$-algebra
$\mathfrak{A}:=\overline{\bigvee_{\rho>0}\mathfrak{A}_\rho}^{\| \, \cdot \, \|_{op}}$,
where the closure is taken in the operator norm.
%denotes the Weyl algebra generated by
%$\lbrace b_{\vk,\lambda}(f)\,,\,b^*_{\vk, \lambda}(g) \, , \, {\bf 1}\rbrace$,
%with $f,g\in L^2(\RR^3)$, and
We define the
state $\omega_{\vP}^\sigma:=\langle \, \Psi_{\vP}^{\sigma}\,,
\,( \; \cdot \; )\,\Psi_{\vP}^{\sigma} \, \rangle$ on $\mathfrak{A}$.
We will show that the weak-* limit of the family of states $\omega_{\vP}^\sigma$,
as $\sigma\rightarrow0$, exists and defines a state $\omega_{\vP}$ on $\mathfrak{A}$.
A somewhat weaker result of this kind (convergence of a subsequence) has
been proven in \cite{ChFr}.
An important ingredient in \cite{ChFr} are the uniform 
bounds on the renormalized electron mass established in \cite{Chen}.

The representation of $\mathfrak{A}$ determined by $\omega_{\vP}$ through 
the GNS construction can be characterized as follows.
Let $\alpha_{\vP}:\mathfrak{A}\rightarrow\mathfrak{A}$ denote the {\em Bogoliubov
automorphism} defined by
\begin{equation}
    \alpha_{\vP}(A) \, = \, \lim_{\sigma\rightarrow0}W_{\sigma}(\vnabla E_{\vP}^{\sigma}) \, A \,
    W_{\sigma}^*(\vnabla E_{\vP}^{\sigma})
\end{equation}
with $W_{\sigma}(\vnabla E_{\vP}^{\sigma})$ defined in (\ref{eq-II-2}), and $A\in\mathfrak{A}$.
Then the GNS representation $\pi_{\vP}$ of $\mathfrak{A}$ is equivalent
to $\pi_{Fock}\circ\alpha_{\vP}$,
where $\pi_{Fock}$ denotes the Fock representation.
In particular, $\pi_{\vP}$ is a {\em coherent} infrared representation unitarily
{\em inequivalent} to $\pi_{Fock}$, for $\vP\neq\vec 0$, and identical to
$\pi_{Fock}$ if $\vP=\vec 0$; see also \cite{ChFr}.

%%%%%%%%%%%%%%%%%%%%%%%%%%%%%%%%%%%%%%%%%%%%%%%%%%%%%%%%%%%%%%%%%%%%%%%%%%%%%%%%%%%%%%%%%%%%%%%%%%%%%%%%%%%%%%%%%%%%%%%%%%%%%%%%%%%%%%%%%%%%%%%%%%%%%%%%%%%%%%%%%%%%%%%%%%%%%%%%%%%%%%%%%%%%%%%%%%%%%%%%%%%%%%%%%%%%%%%%%%%%%%%%%%%%%%%%%%%%%%%%%%%%%%%%%%%%%%SPECTRAL     RESULTS%%%%%%%%%%%%%%%%%%%%%%%%%%%%%%%%%%%%%%%%%%%%%%%%%%%%%%%%%%%%%%%%%%%%%%%%%%%%%%%%%%%%%%%%%%%%%%%%%%%%%%%%%%%%%%%%%%%%%%%%%%%%%%%%%%%%%%%%%%%%%%%%%%%%%%%%%%%%%%%%%%%%%%%%%%%%%%%%%%%%%%%%%%%%%%%%%%%%%%%%%%%%%%%%%%%%%%%%%%%%%%%%%%%%%%%%%%%%%%%%%%%%%%%%%%%%%%%%%%%%%%%%%%%%%%%%%%%%%%%%%%%%%%%%%%%%%%%%%%%%%%%%%%%%%%%%%%%%%%%%%%%%%%%%%%%%%%%%%%%%%%%%%%%%%%%%%%%%%%%%%%%%%%
\newpage

%%%%%%%%%%%%%%%%%%%%%%%%%%%%%%%%%%%%%%%%%%%%%%%%%%%%%%%%
\section{Proof of ($\mathscr{I}1$) in  the main Theorem} \label{sec-IV}
\label{sec-Jot-1}
%%%%%%%%%%%%%%%%%%%%%%%%%%%%%%%%%%%%%%%%%%%%%%%%%%%%%%%%
\resetequ
In this section, we prove the statements ($\mathscr{I}1$)  in Theorem {\ref{thm-cfp-2}}.
This is the most involved part of our analysis. 

In the following, we write $\|\psi\|$, instead of $\|\psi\|_{\cF}$, 
for the norm of a vector $\psi\in\cF^{b}\cong \cH_{\vP}$.
We also use the notation $\|A\|_{\cH}=\|A|_{\cH}\|$ for the
norm of a bounded operator $A$ acting on a Hilbert space $\cH$.
Typically, $\cH$ will be some subspace of $\cF^{b}$.

\subsection{Construction of the sequence  $\{\Psi_{\vP}^{\sigma_j}\}$ of ground states}
\label{section-V-1}
We recall the definition of the fiber Hamiltonian from (\ref{eq-fibHam}),
\begin{equation}
H_{\vP}^{\sigma_j} \, = \,
\frac{\big(\vec{P}-\vec{P}^{f}
+\alpha^{1/2} \vec{A}^{\sigma_j} \big)^2}{2}
\; + \;H^{f} \,.
\end{equation} 
It acts on a fixed fiber space $\mathcal{H}_{\vec{P}}$, with $\vec{P}\in \mathcal{S}$,
where
\begin{eqnarray}
  \vec{A}^{\sigma_j} \, = \,
\sum_{\lambda = \pm} \int_{\mathcal{B}_{\Lambda}\setminus \mathcal{B}_{\sigma_j}}\frac{d^3k}{\sqrt{|\vk| \,}} \,
\big\{ \veps_{\vk, \lambda} b^*_{\vk, \lambda} \, + \,
\veps_{\vk, \lambda}^{\;*} b_{\vk, \lambda} \big\} \,
\end{eqnarray}
contains an infrared cutoff at 
\begin{equation}
	\sigma_j \, := \, \Lambda \, \epsilon^{j}  
	\; \; \; , \; \; \;
	j\in\mathbb{N}_0 \,,
\end{equation} 
with $0<\epsilon<1$ to be fixed later (we recall that $\Lambda\approx1$).
As we will see, the Hamiltonian $H_{\vP}^{\sigma_j}$ has a {\em unique} ground state
$\Psi_{\vec{P}}^{\sigma_j}$, which we construct below
using an approach developed in \cite{Pizzo2003}.

We define the Fock spaces
$$
    \cF_{\sigma_j}:=\cF^b(L^2((\RR^3\setminus \mathcal{B}_{\sigma_j})\times\ZZ_2))
    \; \; \; \; {\rm and} \; \; \; \;
    \cF_{\sigma_{j+1}}^{\sigma_{j}}:=\cF^b( L^2((\mathcal{B}_{\sigma_j}\setminus
    \mathcal{B}_{\sigma_{j+1}})\times\ZZ_2)) \, .
$$
It is clear that
\begin{equation}
     \cF_{\sigma_{j+1}} \; = \; \cF_{\sigma_j} \,\otimes\,
    \cF_{\sigma_{j+1}}^{\sigma_j}\, ,
\end{equation}
and that the Hamiltonians $\lbrace H_{\vP}^{\sigma_j}\,|\,j\in\NN_0 \rbrace$
are related to one another by
\begin{equation}
     H_{\vP}^{\sigma_{j+1}} \; = \; H_{\vP}^{\sigma_j} \, + \, \Delta H_{\vec{P}}|^{\sigma_j}_{\sigma_{j+1}} \;,
\end{equation}
where
\begin{equation}
    \label{eq:dH-def-1}
    \Delta H_{\vec{P}}|^{\sigma_j}_{\sigma_{j+1}} \; := \; \, \alpha^{\frac{1}{2}} \,
    \vnabla_{\vec{P}} H_{\vP}^{\sigma_j} \cdot
    \vec{A}|^{\sigma_j}_{\sigma_{j+1}} \, + \, \frac{\alpha}{2}  \, (\vec{A}|^{\sigma_j}_{\sigma_{j+1}})^2  \;
\end{equation}
and
\begin{equation}
\vec{A}|_{\sigma_{j+1}}^{\sigma_j} \; := \;
\sum_{\lambda = \pm} \int_{\mathcal{B}_{\sigma_j}\setminus \mathcal{B}_{\sigma_{j+1}}}\frac{d^3k}{\sqrt{|\vk| \,}} \,
\big\{ \veps_{\vk,\lambda} b^*_{\vk, \lambda} \, + \,
\veps_{\vk, \lambda}^{\;*} b_{\vk, \lambda} \big\} \,.
\end{equation}

For $\alpha$ sufficiently small and $\vec{P}\in\mathcal{S}$, we construct ground state
vectors $\lbrace \Psi_{\vec{P}}^{ \sigma_{j}}\rbrace$ of the
Hamiltonians $\lbrace H_{\vP}^{\sigma_j}\rbrace$, $j\in\NN$.
%We note that at this stage of our construction,
%the vectors $\Psi_{\vec{P}}^{ \sigma_{j}}$ are {\em not} normalized.
We will prove the following results, adapting recursive arguments developed in \cite{Pizzo2003}.

We introduce four parameters $\epsilon$, $\rfp$, $\rf$, $\mu$ with the properties that
\eqn\label{eq-rfmurfp-1}
	0 \, < \, \rf \, < \,\mu \, < \, \rfp \, < \, 1 \, - \, C_\alpha \, < \, \frac23
	\\
	\label{eq-rfmurfp-2}
	0 \, < \, \epsilon \, < \, \frac{\rf}{\rfp}
\eeqn
where $C_\alpha$ is defined in (\ref{eq-II-8}). 
Then, for $\alpha$ small enough depending on
$\Lambda$, $\epsilon$, $\rf$, $\mu$, $\rfp$, we prove:

\begin{itemize}
\item
The infimum of the spectrum of $H_{\vP}^{\sigma_j}$ on $\cF_{\sigma_j}$,
which we denote by $E^{\sigma_j}_{\vec{P}}$, is
an isolated, simple eigenvalue which is
separated from the rest of the spectrum by a gap
$\rf\sigma_j$ or larger.
\item
$E_{\vec{P}}^{\sigma_{j}}$ is also the ground state energy of the operators
$H_{\vP}^{\sigma_j}$ and $H_{\vP}^{\sigma_j} - (1-C_\alpha)H^{f}|^{\sigma_j}_{\sigma_{j+1}}$ on
$\cF_{\sigma_{j+1}}$, where $H^{f}|^{\sigma_j}_{\sigma_{j+1}}$ is
defined in Eq~(\ref{eq:II.51}).
Note that $E_{\vP}^{\sigma_{j}}=\inf{\rm spec} H_{\vP}^{\sigma_j}|_{\cF_\sigma}$,
for any $\sigma\leq\sigma_j$, and that $E_{\vP}^{\sigma_j}$ 
is a simple eigenvalue of $H_{\vP}^{\sigma_j}|_{\cF_{\sigma_{j+1}}}$
separated by a gap $\geq\rfp\sigma_{j+1}$ from the rest of the spectrum.

\item
The ground state energies $E_{\vec{P}}^{\sigma_{j}}$ and
$E_{\vec{P}}^{\sigma_{j+1}}$ of the Hamiltonians 
$H_{\vP}^{\sigma_j}$ and $H_{\vP}^{\sigma_{j+1}}$, respectively,
acting on the same space $\cF_{\sigma_{j+1}}$  satisfy
\begin{equation}\label{eq:gsen-diff-est-1}
0 \; \leq \; E_{\vec{P}}^{\sigma_{j+1}} \; \leq \; E_{\vec{P}}^{\sigma_{j}} \, + \, c \, \alpha \, \sigma_{j}^2 \;,
\end{equation}
where $c$ is a constant independent of $j$ and $\alpha$ but $\Lambda$-dependent.

\end{itemize}

We   recursively construct the ground state vector, $\Psi_\vP^{\sigma_j}$ (which, at this stage, is not normalized),
of $H_\vP^{\sigma_j}$ on $\cF_{\sigma_j}$, as follows. In the initial step, we set $\Psi_\vP^{\sigma_0}=\Omega_f$.

Let $\Psi_{\vec{P}}^{\sigma_j}$ denote the ground state of
the Hamiltonian $H_{\vP}^{\sigma_j}$ on $\cF_{\sigma_j}$
with non-degenerate eigenvalue $E_{\vec{P}}^{\sigma_{j}}$ and a spectral gap 
at least as large as $\rf\sigma_j$.
We note that $E_\vP^{\sigma_0}\equiv \frac{\vP^2}{2}$ is a non-degenerate eigenvalue of $H_\vP^{\sigma_0}$
on $\cF_{\sigma_0}$, and that 
\eqn
	{\rm gap}(H_\vP^{\sigma_0}|_{\cF_{\sigma_{0}}}) 
 	\, \geq \, \frac23  \, \sigma_0 \, \geq \, \rf\sigma_{0} \,,
\eeqn
where
\eqn
	{\rm gap}(H) \, := \, \inf\{ \, {\rm spec}(H) \, \setminus \,
	\{ \, \inf{\rm spec}( H) \, \} \, \} \, - \, \inf{\rm spec}(H) \,.
\eeqn
We observe that
\begin{equation}
    \Psi_{\vP}^{\sigma_j} \, \otimes \, \Omega_f \; \; \in \; \cF_{\sigma_{j+1}} \; = \;
    \cF_{\sigma_j} \otimes \cF_{\sigma_{j+1}}^{\sigma_j} \, ,
    \label{eq:II.45}
\end{equation}
where
\eqn
	\| \, \Psi_\vP^{\sigma_j} \otimes \Om_f \, \| \, = \, \| \, \Psi_\vP^{\sigma_j} \, \| \,,
\eeqn
is an eigenvector of $H_{\vP}^{\sigma_j}|_{\cF_{\sigma_{j+1}}}$.
In (\ref{eq:II.45}), $\Omega_f$ stands for the vacuum state in
$\cF_{\sigma_{j+1}}^{\sigma_j}$ (if not further specified otherwise, $\Omega_f$ denotes the
vacuum state in any of the photon Fock spaces). 
Moreover, we note that (\ref{eq:II.45}) is   the ground state of 
$H_{\vP}^{\sigma_j}$ restricted to $\cF_{\sigma_{j+1}}$, because
\eqn
	\lefteqn{
	\inf{\rm spec} \Big( \, H_\vP^{\sigma_j} \big|_{\cF_{\sigma_{j+1}}\ominus 
	\{ \CC \Psi_\vP^{\sigma_j}\otimes\Omega_f \} } \, - \, E_\vP^{\sigma_j}\, \Big)
	}
	\nonumber\\
	& \geq & \min\Big\{ \, \rf \sigma_j \, , \, \inf_{\vk\in\RR^3\setminus\cB_{\sigma_{j+1}}}
	\{ \, E_{\vP+\vk}^{\sigma_j} + |\vk| - E_\vP^{\sigma_j} \, \} \, , \, \sigma_{j+1} \, \Big\}
	\nonumber\\
	& \geq & \min \Big\{ \, \rf \sigma_j \, , \, (1-C_\alpha) \, \sigma_{j+1} \, \Big\}
	\nonumber\\
	& \geq & \rfp \sigma_{j+1} \, > \, 0 \,, 
	\label{eq-spec-gap-1}
\eeqn
where $\cF_{\sigma_{j+1}}\ominus\{ \CC \Psi_\vP^{\sigma_j}\otimes\Omega_f \}$ is
the orthogonal complement in $\cF_{\sigma_{j+1}}$ of the one-dimensional
subspace $\{ \CC \Psi_\vP^{\sigma_j}\otimes\Omega_f \}$.
%We remark that, since $E_\vP^{\sigma_j}$ is assumed to be an isolated eigenvalue and 
%$H_\vP^{\sigma_j}|_{\cF_{\sigma_j}}$ is an analytic family of type A, 
%one can use 
We use property ($\mathscr{I}4$),
which holds for $\inf{\rm spec}(H_{\vP}^{\sigma_j}|_{\cF_{\sigma}})$,
$0\leq\sigma\leq\sigma_j$, with the same $C_\alpha$, to pass from the first to the second line,
and from the second to the third
line in (\ref{eq-spec-gap-1}); for a proof of property ($\mathscr{I}4$)
see Section \ref{sect-proof-I4}.

Consequently, the spectral gap of $H_{\vP}^{\sigma_j}$ restricted to $\cF_{\sigma_{j+1}}$
is bounded from below by 
\eqn
	{\rm gap}(H_\vP^{\sigma_j}|_{\cF_{\sigma_{j+1}}}) \, \geq \, \rfp\sigma_{j+1} \,.
\eeqn
%
%The state ,
%\begin{equation}
%    \| \,  \Psi_{\vec{P}}^{\sigma_j} \, \otimes \, \Omega_f \, \|_{\cF}\; =
%    \; \| \,  \Psi_{\vec{P}}^{\sigma_j} \, \|_{\cF} \, .
%\end{equation}
We define the contour $\gamma_{\sigma_{j+1}}:=\{z_{j+1}\in\CC \, \big| \,
|z_{j+1}-E_{\vec{P}}^{\sigma_j}| \, =\,\mu\sigma_{j+1}\}$ which is the boundary
of a closed disc that contains the non-degenerate
ground state eigenvalue $E_{\vec{P}}^{\sigma_j}$ of $H_{\vec{P}}^{\sigma_j}$, but no
other elements of the spectrum of $H_{\vec{P}}^{\sigma_j}|_{\cF_{\sigma_{j+1}}}$;
see also Figure 2 below.

%We can construct the ground state vector $\Psi_{\vec{P}}^{\sigma_{j}}$
%(which is at this stage not normalized) of $H_{\vec{P}}^{\sigma_{j}}$
%on $\cF_{\sigma_{j+1}}$ recursively as follows. In the initial step, we set
%$\Psi_{\vec{P}}^{\sigma_{0}}\equiv\Omega_f$, $\|\Omega_f\|_{\cF}=1$. 

Then we define
\begin{eqnarray}
	\Psi_{\vec{P}}^{\sigma_{j+1}} &:=& \frac{1}{2\pi i}
	\oint_{\gamma_{j+1}} dz_{j+1} \, \frac{1}{H_{\vP}^{\sigma_{j+1}} - z_{j+1} } 
	\Psi_{\vec{P}}^{\sigma_j}\otimes\Omega_f
    \nonumber\\
    &=& \sum_{n\geq0 } \frac{1}{2\pi i} \oint_{\gamma_{j+1}} dz_{j+1} \, 
    \frac{1}{H_{\vP}^{\sigma_j}- z_{j+1} }
   	\label{eq:II.47} \\
    &&\quad\quad\quad
    \Big(- \Delta H_{\vec{P}}|^{\sigma_j}_{\sigma_{j+1}}
	\frac{1}{H_{\vP}^{\sigma_j} - z_{j+1} }\Big)^n \Psi_{\vec{P}}^{\sigma_j}\otimes\Omega_f\,,
    \nonumber
\end{eqnarray}
which is, by construction, the ground state eigenvector
of $H_\vP^{\sigma_{j+1}}|_{\cF_{\sigma_{j+1}}}$.
The associated ground state eigenvalue $E_\vP^{\sigma_{j+1}}$,
with $H_\vP^{\sigma_{j+1}}\Psi_{\vec{P}}^{\sigma_{j+1}}
=E_\vP^{\sigma_{j+1}}\Psi_{\vec{P}}^{\sigma_{j+1}}$, is non-degenerate by Kato's theorem.
To control the expansion in (\ref{eq:II.47}) for sufficiently small $\alpha$, we show that,
for $z_{j+1}\in\gamma_{j+1}$,
\begin{eqnarray}
	\lefteqn{
	\sup_{z_{j+1}\in\gamma_{j+1}}\Big\| \,
	\Big(\frac{1}{H_{\vP}^{\sigma_j}- z_{j+1}}\Big)^{\frac12} \, \Delta
	H_{\vec{P}}|^{\sigma_j}_{\sigma_{j+1}} \,\Big(\frac{1}{H_{\vP}^{\sigma_j}-
  	z_{j+1}}\Big)^{\frac12} \, \Big\|_{\cF_{\sigma_{j+1}}}
	}
	\nonumber\\
	&&\quad \quad \quad
	\leq \,  C \, \frac{\alpha^{1/2}}{\epsilon^{1/2} \, 
	[\min\{(\rfp-\mu),\mu\}]^{1/2}} \, ,
\quad
\label{eq:II.48}
\end{eqnarray}
where the   constant on the r.h.s. depends on $\vP$ and $\Lambda$.
The largest value of $\alpha$ such that $(\ref{eq:II.48})<1$
may depend on $\epsilon$ and $\mu$.
The estimate (\ref{eq:II.48}) is obtained from the following bounds, 
which depend critically
on the spectral gap
(as in the model treated in \cite{Pizzo2003}):

\begin{itemize}
\item[i)] For $z_{j+1}\in\gamma_{j+1}$,
\begin{eqnarray} 
	\sup_{z_{j+1}\in\gamma_{j+1}}\Big\| \,
	\Big(\frac{1}{H_{\vP}^{\sigma_j} -z_{j+1} }\Big)^{\frac{1}{2}} \,
	(\vnabla_{\vec{P}}H_{\vP}^{\sigma_j})^2 \, \Big(\frac{1}{H_{\vP}^{\sigma_j} -z_{j+1}}\Big)^{\frac{1}{2}}
	\, \Big\|_{\cF_{\sigma_{j+1}}}
	\nonumber\\
	\leq\mathcal{O}\Big( \, \frac{1}{ \epsilon^{j+1} \, 
	\min\{(\rfp-\mu),\mu\}} \, \Big)\, \quad \quad\quad
\end{eqnarray}
where the implicit constant depends on $\vP$ and $\Lambda$.

\item[ii)]
Writing $(\vec{A}|_{\sigma_{j+1}}^{\sigma_j})^{-}$ and
$(\vec{A}|_{\sigma_{j+1}}^{\sigma_j})^{+}$ for the parts in
$\vec{A}|_{\sigma_{j+1}}^{\sigma_j}$ which contain
annihilation-  and   creation operators, respectively, we have that
\begin{eqnarray}
   \| \, (\vec{A}|_{\sigma_{j+1}}^{\sigma_j})^- \psi\|  & \leq & 
   \Big( \, 2 \, \int_{\mathcal{B}_{\sigma_{j}}\setminus \mathcal{B}_{\sigma_{j+1}}}
    \frac{d^3k}{|\vk|^2}\Big)^{1/2} \, \|( H^{f}|^{\sigma_j}_{\sigma_{j+1}})^{1/2}\psi\|
    \nonumber\\
    & \leq & c \, \epsilon^{\frac{j}{2}}\|( H^{f}|^{\sigma_j}_{\sigma_{j+1}})^{1/2}\psi\| \;,
    \quad\quad\quad\quad
\end{eqnarray}
where
\begin{equation}
    H^{f}|^{\sigma_j}_{\sigma_{j+1}}\, := \, \sum_{\lambda} \int_{\mathcal{B}_{\sigma_{j}}
    \setminus \mathcal{B}_{\sigma_{j+1}}} d^3k \, | \vec{k}| \,
    b_{\vk,\lambda}^* \, b_{\vk,\lambda} \; ,
    \label{eq:II.51}\,
\end{equation}
with $\psi$ in the domain of $(H^{f}|^{\sigma_j}_{\sigma_{j+1}})^{1/2}$.
%The constant $c$ is independent of $j$, $\alpha$, and $\epsilon$, 
Moreover,
\begin{eqnarray}
     0 \; < \; [(\vec{A}|_{\sigma_{j+1}}^{\sigma_j})^- \, , \,
     (\vec{A}|_{\sigma_{j+1}}^{\sigma_j})^+] 
     %& \leq & \, c\,
     %\int_{\mathcal{B}_{\sigma_j}\setminus \mathcal{B}_{\sigma_{j+1}}}\frac{d\vec{k}}{|\vec{k}|}
     \;\leq \; c' \,\epsilon^{2j} \;,
     \quad\quad\quad
\end{eqnarray}
where the constants $c$, $c'$ are proportional to $\Lambda^{1/2}$
and $\Lambda$, respectively.  

\item[iii)]
For $z_{j+1}\in\gamma_{j+1}$,
\begin{equation}\label{eq:II.16}
 \sup_{z_{j+1}\in\gamma_{j+1}}\Big\| \, \Big(\frac{1}{H_{\vP}^{\sigma_j}- z_{j+1}}\Big)^{\frac{1}{2}}
 \; H^{f}|^{\sigma_j}_{\sigma_{j+1}} \;
 \Big( \frac{1}{H_{\vP}^{\sigma_j}- z_{j+1}}\Big)^{\frac12} \, \Big\|_{\cF_{\sigma_{j+1}}} \leq \, 
 \cO(\frac{1}{\rfp-\mu}) \; ,
\end{equation}
%where the implicit constant is uniform in  $\epsilon$,
%$\rfp$, $\rf$, and $\mu$, for $\alpha\Lambda$ small enough.
which follows from the spectral theorem for the commuting operators $H^{f}|^{\sigma_j}_{\sigma_{j+1}}$
and $H_{\vP}^{\sigma_j}$ (one can for instance see this by adding and subtracting a suitable
fraction of $H^{f}|^{\sigma_j}_{\sigma_{j+1}}$ in the denominator).
\end{itemize}

\noindent
Using (\ref{eq:II.48}), one concludes that
\begin{equation}
    \| \Psi_{\vec{P}}^{\sigma_{j+1}} \, - \, \Psi_{\vec{P}}^{\sigma_{j}}\| \,
    \leq \, C \, \alpha^{\frac{1}{2}}\,\|\Psi_{\vec{P}}^{\sigma_{j}} \| \, ,
\end{equation}
with $C$ uniform in $j$, such that, for $\alpha$ small enough,
\begin{equation}
    \| \, \Psi_{\vec{P}}^{\sigma_{j+1}}\| \,
     \geq \, C' \, \| \, \Psi_{\vec{P}}^{\sigma_{j}}\| \,,
\end{equation}
for a constant $C'>0$ independent of $j$. In particular, the vector constructed in  (\ref{eq:II.47})
is indeed non-zero.

Because of (\ref{eq:gsen-diff-est-1}), which follows from a variational argument, we find that,
for $\alpha$ small enough and $\Lambda$-dependent, but independent of $j$,
\eqn
	{\rm gap}(H_\vP^{\sigma_{j+1}}|_{\cF_{\sigma_{j+1}}}) \,
	\geq \, \mu \sigma_{j+1} \, - \, c \, \alpha \, \sigma_j^2
	\, \geq \, \rf \sigma_{j+1} \,.
\eeqn
This estimate allows us to proceed to the next scale.
\\

\begin{figure*}
%\centerline{\epsffile{cfp-1-fig-2.pdf} }
%\centerline{\epsffile{cfp-1-fig-2.eps} }
\includegraphics{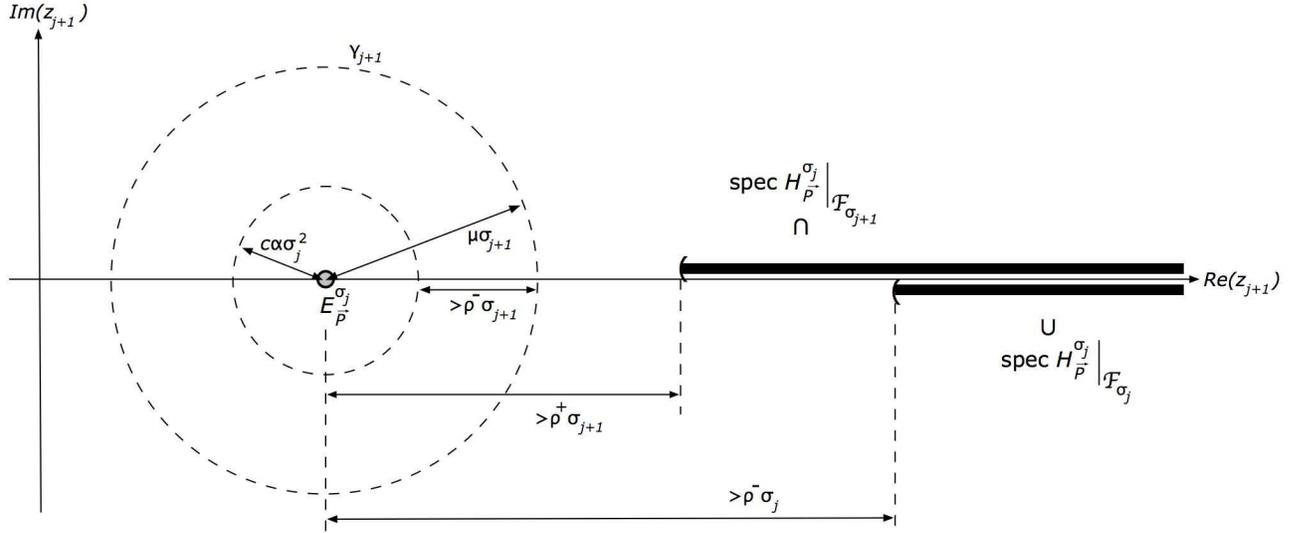}
\caption{The contour integral in the energy plane.}
\end{figure*}

It easily follows from the previous results that $E_\vP^{\sigma_j}$
is simple and isolated, and $(H_\vP^{\sigma_j})_{\vP\in\mathcal{S}}$ is an analytic family of type A. 
In particular, this allows us to express $\vnabla E_\vP^{\sigma_j}$,
as a function of $\vP$, by using the Feynman-Hellman formula; see (\ref{eq-FH-appl-1}) below.
 
%%%%%%%%%%%%%%%%%%%%%%%%%%%%%%%%%%%%%%%%%%%%%%%%%%%%%%%%%%%%%%%%%%%%%%%%%%%%%%%%%%%%%%%%%%%%%%%%%%%%%%%%%%%%%%%%%%%%%%%%%%%%%%%%%%%%%%%%%%%%%%%%%%%%%%%%%%%%%%%%%%%%%%%%%%%%%%%%%%%%%%%%%%%%%%%%%%%%%%%%%%%%%%%%%%%%%%%%%%%%%%%%%%%%%%%%%%%%%%%%%%%%%%%%%%%%%%%%%%%%%%%%%%%%%%%%%%%%%%%%%%%%%%%%%%%%%%%%%%%%%%%%%%%%%%%%%%%%%%%

\subsection{Transformed Hamiltonians and the sequence of ground states $\lbrace
  \Phi_{\vec{P}}^{\sigma_j} \rbrace $}
  \label{subsect-4.2}
In this section, we consider the Hamiltonians obtained from
$\{H_{\vP}^{\sigma_j}\}$ after a $j-$dependent Bogoliubov
transformation of the photon variables. In the limit
$j\to\infty$, this transformation coincides with the one identified in \cite{ChFr},
which provides the correct representation of the photon degrees of freedom for which the Hamiltonian
$H_{\vec{P}}$ has a ground state.
%%%%%%%%%%%%%%%%%%%%%%%%%%%%%%%%%%%%%%%%%%%%%%%%%%%%%%%%%%%%%%%%%%%%%%%%%%%%%%%%%%%%%%%%%%%%%%%%%%%%%%%%%%%%%%%%%%%%%%%%%%%%%%%%%%%%%%%%%%%%%%%%%%%%%%%%%%%%%%%%%%%%%%%%%%%%%%%%%%%%%%%%%%%%%%%%%%%%%%%%%%%%%%%%%%%%%%%%%%%%%%%%%%%%%%%%%%%%%%%%%%%%%%%%%%%%%%%%%%%%%%%%%%%%%%%%%%%%%%%%%%%%%%%%%%%%%%%%%%%%%%%%%%%%%%%%%%%%%%
\subsubsection{Bogoliubov transformation and canonical form of the Hamiltonian}
\label{sec:iv-2-1}

The Feynman-Hellman formula yields
\begin{equation}\label{eq-FH-appl-1}
\vnabla E_{\vec{P}}^{\sigma_j} \; = \; \vec{P} \, - \, \bra \, \vec{P}^{f}
\, - \, \alpha^{1/2} \vec{A}^{\sigma_j} \, \ket_{\Psi^{\sigma_j}_{\vec{P}}} \;,
\end{equation}
where
\begin{equation}
\bra \, \vec{P}^{f}
-\alpha^{1/2} \vec{A}^{\sigma_j} \, \ket_{\Psi^{\sigma_j}_{\vec{P}}} \, := \,
\frac{\bra \, \Psi_{\vec{P}}^{\sigma_j}\,,\,(\vec{P}^{f}
-\alpha^{1/2}
\vec{A}^{\sigma_j}) \, \Psi_{\vec{P}}^{\sigma_j} \, \ket}
{\bra \, \Psi_{\vec{P}}^{\sigma_j}\,,\,\Psi_{\vec{P}}^{\sigma_j} \, \ket }\, .
\end{equation}
We define
\begin{eqnarray}
\vec{\beta}^{\sigma_j}&:=&\vec{P}^{f}
-\alpha^{1/2} \vec{A}^{\sigma_j}
\nonumber\\
\delta_{\vP}^{\sigma_j}(\widehat{k})&:=&1-\widehat{k}\cdot\vnabla E_{\vec{P}}^{\sigma_j}
\nonumber\\
    c^{*}_{\vk, \lambda}&:=&b^{*}_{\vk, \lambda}+\alpha^{\frac{1}{2}}\frac{\vnabla
    E_{\vec{P}}^{\sigma_j}\cdot \veps_{\vk,
  \lambda}^{\;*}}{|\vec{k}|^{\frac{3}{2}}\delta_{\vP}^{\sigma_j}(\widehat{k})}
  \nonumber\\
    c_{\vk, \lambda}&:=&b_{\vk, \lambda}+\alpha^{\frac{1}{2}}\frac{\vnabla
  E_{\vec{P}}^{\sigma_j}\cdot \veps_{\vk,
  \lambda}}{|\vec{k}|^{\frac{3}{2}}\delta_{\vP}^{\sigma_j}(\widehat{k})}\, .
\end{eqnarray}
We then rewrite $H_{\vP}^{\sigma_j}$ as
\begin{equation}
H_{\vP}^{\sigma_j} \, = \,
\frac{\big(\vec{P}-\vec{\beta}^{\sigma_j} \big)^2}{2}
\; + \;H^{f}\, ,
\end{equation}
and
\begin{equation}
\vec{P}\, =\,\vnabla
E_{\vec{P}}^{\sigma_j}+\langle\vec{\beta}^{\sigma_j}\rangle_{\Psi^{\sigma_j}_{\vec{P}}} \;,
\end{equation}
thus obtaining
\begin{eqnarray}
H_{\vP}^{\sigma_j}&=&\frac{\vec{P}^2}{2}-(\vnabla
E_{\vec{P}}^{\sigma_j}+\langle\vec{\beta}^{\sigma_j}\rangle_{\Psi^{\sigma_j}_{\vec{P}}})
\cdot\vec{\beta}^{\sigma_j}+\frac{(\vec{\beta}^{\sigma_j})^2}{2}+H^{f}\quad
\nonumber\\
&=& \frac{\vec{P}^2}{2}+\frac{(\vec{\beta}^{\sigma_j})^2}{2}-\langle\vec{\beta}^{\sigma_j}
\rangle_{\Psi^{\sigma_j}_{\vec{P}}}\cdot\vec{\beta}^{\sigma_j}
\nonumber\\
& &+\sum_{\lambda}\int_{\RR^3\setminus (\mathcal{B}_{\Lambda}\setminus
\mathcal{B}_{\sigma_j})}|\vec{k}|\delta_{\vP}^{\sigma_j}(\widehat{k})\,b^*_{\vk, \lambda}b_{\vk,\lambda} d^3k
\nonumber\\
& &+\sum_{\lambda}\int_{\mathcal{B}_{\Lambda}\setminus
  \mathcal{B}_{\sigma_j}}|\vec{k}|\delta_{\vP}^{\sigma_j}(\widehat{k})\,c^*_{\vk,
  \lambda}c_{\vk,\lambda} d^3k
\\
& &-\alpha\, \sum_{\lambda}\int_{\mathcal{B}_{\Lambda}\setminus
  \mathcal{B}_{\sigma_j}}|\vec{k}|\delta_{\vP}^{\sigma_j}(\widehat{k})\,
  \frac{\vnabla E_{\vec{P}}^{\sigma_j}\cdot \veps_{\vk, \lambda}^{\;*}}{|\vec{k}|^{\frac{3}{2}}
  \delta_{\vP}^{\sigma_j}(\widehat{k})}\frac{\vnabla E_{\vec{P}}^{\sigma_j}\cdot
  \veps_{\vk, \lambda}}{|\vec{k}|^{\frac{3}{2}}\delta_{\vP}^{\sigma_j}(\widehat{k})}\,d^3k\,.\quad\quad
  \nonumber
\end{eqnarray}
Adding and subtracting
$\frac{1}{2}\langle\vec{\beta}^{\sigma_j}\rangle_{\Psi^{\sigma_j}_{\vec{P}}}^2$,
one gets
\begin{eqnarray}
H_{\vP}^{\sigma_j}&=&\frac{\vec{P}^2}{2}-\frac{\langle\vec{\beta}^{\sigma_j}
\rangle_{\Psi^{\sigma_j}_{\vec{P}}}^2}{2}+
\frac{\big(\vec{\beta}^{\sigma_j}-\langle\vec{\beta}^{\sigma_j}\rangle_{\Psi^{\sigma_j}_{\vec{P}}}\big)^2}{2}
\nonumber\\
& &+\sum_{\lambda}\int_{\RR^3\setminus (\mathcal{B}_{\Lambda}\setminus \mathcal{B}_{\sigma_j})}
|\vec{k}|\delta_{\vP}^{\sigma_j}(\widehat{k})\,b^*_{\vk, \lambda}b_{\vk,\lambda} d^3k
\nonumber\\
& &+\sum_{\lambda}\int_{\mathcal{B}_{\Lambda}\setminus  \mathcal{B}_{\sigma_j}}|\vec{k}|
\delta_{\vP}^{\sigma_j}(\widehat{k})\,c^*_{\vk,\lambda}c_{\vk,\lambda} d^3k
\\
& &-\alpha\, \sum_{\lambda}\int_{\mathcal{B}_{\Lambda}\setminus
  \mathcal{B}_{\sigma_j}}|\vec{k}|\delta_{\vP}^{\sigma_j}(\widehat{k})
  \frac{\vnabla E_{\vec{P}}^{\sigma_j}\cdot \veps_{\vk, \lambda}^{\;*}}{|\vec{k}|^{\frac{3}{2}}
  \delta_{\vP}^{\sigma_j}(\widehat{k})}\frac{\vnabla E_{\vec{P}}^{\sigma_j}\cdot
  \veps_{\vk, \lambda}}{|\vec{k}|^{\frac{3}{2}}\delta_{\vP}^{\sigma_j}(\widehat{k})}\,d^3k\,.\quad\quad
  \nonumber
\end{eqnarray}
Next, we apply the Bogoliubov transformation
\begin{eqnarray}
b^{*}_{\vk, \lambda}&\longrightarrow \,&W_{\sigma_j}(\vnabla
E_{\vec{P}}^{\sigma_j}) b^{*}_{\vk, \lambda}W_{\sigma_j}^*(\vnabla
E_{\vec{P}}^{\sigma_j})=b^{*}_{\vk, \lambda}-\alpha^{\frac{1}{2}}\frac{\vnabla
  E_{\vec{P}}^{\sigma_j}\cdot \veps_{\vk,
  \lambda}^{\;*}}{|\vec{k}|^{\frac{3}{2}}\delta_{\vP}^{\sigma_j}(\widehat{k})}
  \nonumber\\
b_{\vk, \lambda}&\longrightarrow \,&W_{\sigma_j}(\vnabla
E_{\vec{P}}^{\sigma_j})b_{\vk, \lambda}W_{\sigma_j}^*(\vnabla
E_{\vec{P}}^{\sigma_j})=b_{\vk, \lambda}-\alpha^{\frac{1}{2}}\frac{\vnabla
  E_{\vec{P}}^{\sigma_j}\cdot \veps_{\vk,
  \lambda}}{|\vec{k}|^{\frac{3}{2}}\delta_{\vP}^{\sigma_j}(\widehat{k})}
  \quad\quad\quad
\end{eqnarray}
for $\vec{k}\in\mathcal{B}_{\Lambda}\setminus\mathcal{B}_{\sigma_j}$, where
\begin{equation}
W_{\sigma_j}(\vnabla
E_{\vec{P}}^{\sigma_j}) \, := \, \exp\Big(\, \alpha^{\frac{1}{2}}\sum_{\lambda}
\int_{\mathcal{B}_{\Lambda}\setminus \mathcal{B}_{\sigma_j}} d^3k \,
    \frac{\vnabla E_{\vec{P}}^{\sigma_j}}{|\vec{k}|^{\frac{3}{2}}
    \delta_{\vP}^{\sigma_j}(\widehat{k})} \cdot (\veps_{\vk,
        \lambda}b_{\vk,\lambda}^{*} - h.c.)\Big) \;.
\end{equation}
It is evident that $W_{\sigma_j}$ acts
as the identity on $\cF^b(L^2(\cB_{\sigma_j}\times\ZZ_2))$ and on 
$\cF^b(L^2((\RR^3\setminus \cB_{\Lambda})\times\ZZ_2))$.
Moreover, we define the vector operators
\begin{eqnarray}
\vec{\Pi}_{\vP}^{\sigma_j} & := & W_{\sigma_j}(\vnabla
E_{\vec{P}}^{\sigma_j})\,\vec{\beta}^{\sigma_j} \, W_{\sigma_j}^*(\vnabla
E_{\vec{P}}^{\sigma_j})
\nonumber\\
&&\quad\quad- \, \langle W_{\sigma_j}(\vnabla
E_{\vec{P}}^{\sigma_j})\,\vec{\beta}^{\sigma_j} \, W_{\sigma_j}^*(\vnabla
E_{\vec{P}}^{\sigma_j})\rangle_{\Omega_f}\; ,
\end{eqnarray}
noting that
\begin{eqnarray}\label{eq-iv-37-final}
\langle\vec{\beta}^{\sigma_j}\rangle_{\Psi^{\sigma_j}_{\vec{P}}}&=&\vec{P}-\vnabla
E_{\vec{P}}^{\sigma_j}\label{eq:IV.29}\\
&= &\frac{\bra  \, \Phi_{\vec{P}}^{\sigma_j}\,,\,\vec{\Pi}_{\vP}^{\sigma_j}
\Phi_{\vec{P}}^{\sigma_j} \, \ket}{\bra \, \Phi_{\vec{P}}^{\sigma_j}\,,\,\Phi_{\vec{P}}^{\sigma_j} \, \ket}
+\langle W_{\sigma_j}(\vnabla
E_{\vec{P}}^{\sigma_j})\,\vec{\beta}^{\sigma_j} \, W_{\sigma_{j}}^*(\vnabla  E_{\vec{P}}^{\sigma_{j}})\rangle_{\Omega_f} \;,
\nonumber
\end{eqnarray}
where $\Phi_{\vec{P}}^{\sigma_j}$ is the ground state of the
Bogoliubov-transformed Hamiltonian
\begin{equation}\label{eq:iv-30}
\Hw_{\vP}^{\sigma_j} \, := \, W_{\sigma_j}(\vnabla
E_{\vec{P}}^{\sigma_j})H_{\vP}^{\sigma_j} W_{\sigma_{j}}^*(\vnabla  E_{\vec{P}}^{\sigma_{j}})\; .
\end{equation}
We remark that although we have not specified the phase $\zeta$ in (\ref{eq-Phi-Psi-phase-1}) yet, 
the expression
in (\ref{eq-iv-37-final}) is uniquely defined, since it does not depend on $\zeta$ and on the normalization of $\Phi_{\vec{P}}^{\sigma_j} $.

It thus follows that
\begin{equation}\label{eq:iv-30bis}
    W_{\sigma_j}(\vnabla E_{\vec{P}}^{\sigma_j})\,\vec{\beta}^{\sigma_j} \,
    W_{\sigma_{j}}^*(\vnabla  E_{\vec{P}}^{\sigma_{j}})-\langle\vec{\beta}^{\sigma_j}
    \rangle_{\Psi^{\sigma_j}_{\vec{P}}}
    \, = \, \vec{\Pi}_{\vP}^{\sigma_j}-\langle\vec{\Pi}_{\vP}^{\sigma_j}
    \rangle_{\Phi^{\sigma_j}_{\vec{P}}}\,.
\end{equation}
As in  \cite{Pizzo2003}, it is convenient to write
$\Hw_{\vP}^{\sigma_j}$ in the "canonical" form
\begin{equation}\label{eq-IV.36}
    \Hw_{\vP}^{\sigma_j} \; = \; \frac{\big(\vec{\Gamma}_{\vP}^{\sigma_j})^2}{2}+
    \sum_{\lambda}\int_{\RR^3}|\vec{k}|\delta_{\vP}^{\sigma_j}(\widehat{k})\, b^{*}_{\vk, \lambda}
    b_{\vk,\lambda} d^3k+\cvar_{\vec{P}}^{\sigma_j} \, ,
\end{equation}
where
\begin{equation}\label{eq-IV-33}
    \vec{\Gamma}_{\vP}^{\sigma_j} \, := \, \vec{\Pi}_{\vP}^{\sigma_j}-
    \bra \, \vec{\Pi}_{\vP}^{\sigma_j} \, \ket_{\Phi^{\sigma_j}_{\vec{P}}} \, ,
\end{equation}
so that
\begin{equation}
    \bra \, \vec{\Gamma}_{\vP}^{\sigma_j} \, \ket_{\Phi_{\vec{P}}^{\sigma_j}} \, = \, 0\,,
    \label{eq:II.87}
\end{equation}
and
\begin{eqnarray} 
    \cvar_{\vec{P}}^{\sigma_j}&:=&\frac{\vec{P}^2}{2}
    \, - \, \frac{(\vec{P}-\vnabla E_{\vec{P}}^{\sigma_j})^2}{2}
    \\
    & &- \, \alpha\,
    \sum_{\lambda}\int_{\mathcal{B}_{\Lambda}\setminus \mathcal{B}_{\sigma_j}}|\vec{k}|
    \delta_{\vP}^{\sigma_j}(\widehat{k})\frac{\vnabla
    E_{\vec{P}}^{\sigma_j}\cdot \veps_{\vk,
    \lambda}^{\;*}}{|\vec{k}|^{\frac{3}{2}}\delta_{\vP}^{\sigma_j}(\widehat{k})}\,
    \frac{\vnabla E_{\vec{P}}^{\sigma_j}\cdot \veps_{\vk, \lambda}}{|\vec{k}|^{\frac{3}{2}}
    \delta_{\vP}^{\sigma_j}(\widehat{k})}\,d^3k\,. \quad\nonumber
\end{eqnarray}
One   arrives at (\ref{eq-IV.36})   using
\begin{eqnarray}
W_{\sigma_j}(\vnabla
E_{\vec{P}}^{\sigma_j})\,c^{*}_{\vk, \lambda}\,W_{\sigma_{j}}^*(\vnabla  E_{\vec{P}}^{\sigma_{j}})
\,&=&\,b^{*}_{\vk, \lambda} \,,
\nonumber\\
W_{\sigma_j}(\vnabla
E_{\vec{P}}^{\sigma_j})\,c_{\vk, \lambda}\,W_{\sigma_{j}}^*(\vnabla  E_{\vec{P}}^{\sigma_{j}})
\,&=&\,b_{\vk, \lambda}\,,
\end{eqnarray}
for $\vec{k}\in\mathcal{B}_{\Lambda}\setminus\mathcal{B}_{\sigma_j}$.
The Hamiltonian $\Hw_{\vP}^{\sigma_j}$ has a structure
similar to the Bogoliubov-transformed Nelson Hamiltonian in \cite{Pizzo2003}.

Following ideas of \cite{Pizzo2003}, we define the intermediate Hamiltonian
\begin{equation}
    \widehat{\Hw}_{\vP}^{\sigma_{j+1}}\,:=\,W_{\sigma_{j+1}}(\vnabla
    E_{\vec{P}}^{\sigma_j})H_{\vP}^{\sigma_{j+1}} W_{\sigma_{j+1}}^*(\vnabla
    E_{\vec{P}}^{\sigma_{j}}) \,,
\end{equation}
where
\begin{equation}
    W_{\sigma_{j+1}}(\vnabla E_{\vec{P}}^{\sigma_j}) \, := \,
    \exp\Big(\, \alpha^{\frac{1}{2}}\sum_{\lambda}
    \int_{\mathcal{B}_{\Lambda}\setminus \mathcal{B}_{\sigma_{j+1}}} d^3k \,
    \frac{\vnabla E_{\vec{P}}^{\sigma_j}}{|\vec{k}|^{\frac{3}{2}}\delta_{\vP}^{\sigma_j}(\widehat{k})}
    \cdot (\veps_{\vk,\lambda}b_{\lambda}^{*}(\vk) - h.c.)\Big) \; ,
\end{equation}
and split it into different terms similarly as for $\Hw_{\vP}^{\sigma_j}$. We write
\begin{equation}
    H_{\vP}^{\sigma_{j+1}} \; = \;
    \frac{\vec{P}^2}{2}-\vec{P}\cdot\vec{\beta}^{\sigma_{j+1}}+\frac{(\vec{\beta}^{\sigma_{j+1}})^2}{2}+H^{f}\,,
\end{equation}
and replace $\vec{P}$ by $\vnabla
E_{\vec{P}}^{\sigma_j}+\langle\vec{\beta}^{\sigma_j}\rangle_{\Psi^{\sigma_j}_{\vec{P}}}$,
thus obtaining
\begin{eqnarray}
    H_{\vP}^{\sigma_{j+1}}&=&\frac{\vec{P}^2}{2}-\big(\vnabla
    E_{\vec{P}}^{\sigma_j}+\langle\vec{\beta}^{\sigma_j}\rangle_{\Psi^{\sigma_j}_{\vec{P}}}\big)
    \cdot\vec{\beta}^{\sigma_{j+1}}+\frac{(\vec{\beta}^{\sigma_{j+1}})^2}{2}+H^{f}\quad
    \nonumber\\
    &=& \frac{\vec{P}^2}{2}+\frac{(\vec{\beta}^{\sigma_{j+1}})^2}{2}-\langle\vec{\beta}^{\sigma_j}
    \rangle_{\Psi^{\sigma_j}_{\vec{P}}}\cdot\vec{\beta}^{\sigma_{j+1}}
    \nonumber\\
    & &+\sum_{\lambda}\int_{\RR^3\setminus (\mathcal{B}_{\Lambda}\setminus\mathcal{B}_{\sigma_{j+1}})}
    |\vec{k}|\delta_{\vP}^{\sigma_j}(\widehat{k})\,b^*_{\vk, \lambda}b_{\vk,\lambda} d^3k
    \nonumber\\
    & &+\sum_{\lambda}\int_{\mathcal{B}_{\Lambda}\setminus
  \mathcal{B}_{\sigma_{j+1}}}|\vec{k}|\delta_{\vP}^{\sigma_j}(\widehat{k})\,c^*_{\vk,
  \lambda}c_{\vk,\lambda} d^3k
  \\
    & &-\alpha\, \sum_{\lambda}\int_{\mathcal{B}_{\Lambda}\setminus
  \mathcal{B}_{\sigma_{j+1}}}|\vec{k}|\delta_{\vP}^{\sigma_j}(\widehat{k})
  \frac{\vnabla E_{\vec{P}}^{\sigma_j}\cdot \veps_{\vk, \lambda}^{\;*}}{|\vec{k}|^{\frac{3}{2}}
  \delta_{\vP}^{\sigma_j}(\widehat{k})}\frac{\vnabla E_{\vec{P}}^{\sigma_j}\cdot \veps_{\vk, \lambda}}
  {|\vec{k}|^{\frac{3}{2}}\delta_{\vP}^{\sigma_j}(\widehat{k})}\,d^3k\, .\quad\quad
  \nonumber
\end{eqnarray}
We add and subtract $\frac{1}{2}\langle\vec{\beta}^{\sigma_j}\rangle_{\Psi^{\sigma_j}_{\vec{P}}}^2$,
and apply a Bogoliubov transformation by conjugating with
the unitary operator $W_{\sigma_{j+1}}(\vnabla E_{\vec{P}}^{\sigma_j})$. 
Formally, we find that
\begin{eqnarray}\label{eq-BogHam}
\widehat{\Hw}_{\vP}^{\sigma_{j+1}}
& = &\frac{\big(\vec{\Gamma}_{\vP}^{\sigma_j}+\vec{\mathcal{L}}_{\sigma_{j+1}}^{\sigma_j}+
\vec{\mathcal{I}}_{\sigma_{j+1}}^{\sigma_j}\big)^2}{2}
\quad\label{eq-IV.42}\\
& &+ \, \sum_{\lambda}
\int_{\RR^3}|\vec{k}|\delta_{\vP}^{\sigma_j}(\widehat{k})\,
b^{*}_{\vk, \lambda}b_{\vk,\lambda} d^3k
\, + \, \widehat{\cvar}_{\vP}^{\sigma_{j+1}}\nonumber
\end{eqnarray}
where
\begin{eqnarray}
    \vec{\mathcal{L}}_{\sigma_{j+1}}^{\sigma_j}
    &:=&- \, \alpha^{\frac{1}{2}}\sum_{\lambda}\int_{\mathcal{B}_{\sigma_j}\setminus
    \mathcal{B}_{\sigma_{j+1}}}\vec{k}
    \frac{\vnabla E_{\vec{P}}^{\sigma_j}\cdot\veps_{\vk,\lambda}^{*}b_{\vec{k},\lambda}+h.c.}
    {|\vec{k}|^{\frac{3}{2}}\delta_{\vP}^{\sigma_j}(\widehat{k})}\,d^3k
    \nonumber\\
    &&- \, \alpha^{\frac{1}{2}}\vec{A}|_{\sigma_{j+1}}^{\sigma_j}\, \quad\quad
    \label{eq-III.42.1}\\
    \vec{\mathcal{I}}_{\sigma_{j+1}}^{\sigma_j}
    &:= &\alpha\,
    \sum_{\lambda}\int_{\mathcal{B}_{\sigma_j}\setminus \mathcal{B}_{\sigma_{j+1}}}\,\vec{k}\,\frac{\vnabla
    E_{\vec{P}}^{\sigma_j}\cdot \veps_{\vk,\lambda}^{\;*}\,
    \vnabla E_{\vec{P}}^{\sigma_j}\cdot \veps_{\vk,\lambda}}
    {|\vec{k}|^{3}(\delta_{\vP}^{\sigma_j}(\widehat{k}))^2} \,d^3k\,
    \label{eq-III.42.2}\\
    & &+\alpha\,
    \sum_{\lambda}\int_{\mathcal{B}_{\sigma_j}\setminus \mathcal{B}_{\sigma_{j+1}}}\,\big[\veps_{\vk,
    \lambda}\,\frac{\vnabla  E_{\vec{P}}^{\sigma_j}\cdot \veps_{\vk,
    \lambda}^{*} }{|\vec{k}|^{\frac{3}{2}}\delta_{\vP}^{\sigma_j}(\widehat{k})}+h.c.\big]
    \,\frac{d^3k}{\sqrt{|\vec{k}|}}
    \nonumber\\
    \widehat{\cvar}_{\vP}^{\sigma_{j+1}}
    &:= &\frac{\vec{P}^2}{2}-\frac{(\vec{P}-\vnabla
    E_{\vec{P}}^{\sigma_j})^2}{2}\label{eq-III.43}
    \\
    & &-\alpha\,
    \sum_{\lambda}\int_{\mathcal{B}_{\Lambda}\setminus \mathcal{B}_{\sigma_{j+1}}}|\vec{k}|
    \delta_{\vP}^{\sigma_j}(\widehat{k})\frac{\vnabla
    E_{\vec{P}}^{\sigma_j}\cdot \veps_{\vk,
    \lambda}^{\;*}}{|\vec{k}|^{\frac{3}{2}}\delta_{\vP}^{\sigma_j}(\widehat{k})}
    \frac{\vnabla E_{\vec{P}}^{\sigma_j}\cdot \veps_{\vk, \lambda}}{|\vec{k}|^{\frac{3}{2}}
    \delta_{\vP}^{\sigma_j}(\widehat{k})}\,d^3k \, .
    \nonumber
\end{eqnarray}
For details on the derivation of (\ref{eq-IV.42}) and for
the proof that (\ref{eq-IV.36}) and (\ref{eq-IV.42}) hold in the operator sense
(and not only formally),
we refer to Lemmata {\ref{lm-A-1}} and {\ref{lm:B-2}} in the Appendix.

We also define the operators ($j\geq1$)
\begin{eqnarray}\label{eq-IV-46bis}
\widehat{\vec{\Pi}}_{\vP}^{\sigma_j} \; := \;
W_{\sigma_{j}}(\vnabla E_{\vec{P}}^{\sigma_{j-1}})
W_{\sigma_{j}}^*(\vnabla  E_{\vec{P}}^{\sigma_{j}})\vec{\Pi}_{\vP}^{\sigma_j}
W_{\sigma_{j}}(\vnabla E_{\vec{P}}^{\sigma_{j}})W_{\sigma_{j}}^*(\vnabla  E_{\vec{P}}^{\sigma_{j-1}})\quad
\end{eqnarray}
and
\begin{eqnarray}\label{eq:IV.46}
\widehat{\vec{\Gamma}}_{\vP}^{\sigma_j} \; := \; \widehat{\vec{\Pi}}_{\vP}^{\sigma_j}
\, - \, \langle\widehat{\vec{\Pi}}_{\vP}^{\sigma_j} \rangle_{\widehat{\Phi}_{\vec{P}}^{\sigma_j}}\, ,
\end{eqnarray}
which are used in the proofs in the next section.
Here, $\widehat{\Phi}_{\vec{P}}^{\sigma_j}$ denotes the ground state vector of the
Hamiltonian $\widehat{\Hw}_{\vP}^{\sigma_j}:=W_{\sigma_{j}}(\vnabla E_{\vec{P}}^{\sigma_{j-1}})
H_\vP^{\sigma_j} W_{\sigma_{j}}^*(\vnabla  E_{\vec{P}}^{\sigma_{j-1}})$.

%%%%%%%%%%%%%%%%%%%%%%%%%%%%%%%%%%%%%%%%%%%%%%%%%%%%%%%%%%%%%%%%%%%%%%%%%%%%%%%%%%%%%%%%%%%%%%%%%%%%%%%%%%%%%%%%%%%%%%%%%%%%%%%%%%%%%%%%%%%%%%%%%%%%%%%%%
\subsection{Construction and convergence of $\lbrace
  \Phi_{\vec{P}}^{\sigma_j} \rbrace $}
%%%%%%%%%%%%%%%%%%%%%%%%%%%%%%%%%%%%%%%%%%%%%%%%%%%%%%%%%%%%%%%%%%%%%%%%%%%%%%%%%%%%%%%%%%%%%%%%%%%%%%%%%%%%%%%%%%%%%%%%%%%%%%%%%%%%%%%%%%%%%%%%%%%%%%%%%

In this section, we construct a sequence
$\lbrace \Phi_{\vec{P}}^{\sigma_j}\,|\,j\in\NN \rbrace$ of unnormalized ground state
vectors of the (Bogoliubov-transformed) Hamiltonians $\Hw_{\vP}^{\sigma_j}$,
and establish the existence of
\begin{equation}
    %\Phi_{\vec{P}} \, := \, 
    s-\lim_{j\to\infty}\Phi_{\vec{P}}^{\sigma_j}\,.
\end{equation}
(We warn the reader that, with an abuse of notation, we use the same symbol introduced for the normalized ground state  vector in (\ref{eq-II-3}).)
%%by estimating the rate of convergence, and by
%We estimate the rate of convergence, and establish regularity
%properties with respect to $\vec{P}$.
%Our results are similarly to those in \cite{Pizzo2003} for the Nelson model.

%In the following, we outline the method, and
%point out all the counterparts of the estimates used in \cite{Pizzo2003} for Nelson's model.

%, in an iterative procedure starting from the
%vacuum state,  $\Phi_{\vec{P}}^{\sigma_0}\equiv \Omega_f$,
%where we exploit all the spectral results stated in the
%previous section. We recall that in our discussion, the
%ultraviolet cut-off is fixed, $\Lambda=1$.
%Our results hold for all sufficiently small $\alpha$.

%Using the orthogonality property in (\ref{eq:II.87}), and exploiting the
%structure of $\vec{\mathcal{L}}_{\sigma_{j+1}}^{\sigma_j}$ and
%$\vec{\mathcal{I}}_{\sigma_{j+1}}^{\sigma_j}$
%(linearity in $b_{\vec{k},\lambda}\,,\,b_{\vec{k},\lambda}^{*}$ of
%$\vec{\mathcal{L}}_{\sigma_{j+1}}^{\sigma_j}$, restriction of the
%support of $\vec{k}$ to dyadic momentum shells, and power
%counting properties of both $\vec{\mathcal{L}}_{\sigma_{j+1}}^{\sigma_j}$ and
%$\vec{\mathcal{I}}_{\sigma_{j+1}}^{\sigma_j}$), we can apply the
%same techniques as in \cite{Pizzo2003}.

In the initial step of the construction corresponding to $j=0$, we define
$\Phi_{\vec{P}}^{\sigma_0}:=\Omega_f$, with $\|\Omega_f\|=1$.

To pass from scale $j$ to $j+1$, we proceed in two steps.
First, we construct an intermediate vector $\widehat{\Phi}_{\vec{P}}^{\sigma_{j+1}}$
\begin{eqnarray}
\widehat{\Phi}_{\vec{P}}^{\sigma_{j+1}}
\, = \,  \sum_{n=0}^{ \infty}\frac{1}{2\pi i}
\int_{\gamma_{j+1}}dz_{j+1} \, \frac{1}{\Hw_{\vP}^{\sigma_j}-z_{j+1}}\big[-\Delta
\Hw_{\vec{P}}|^{\sigma_j}_{\sigma_{j+1}}\frac{1}{\Hw_{\vP}^{\sigma_j}-z_{j+1}}\big]^{n}
\Phi_{\vec{P}}^{\sigma_{j}} \;,
\nonumber\\
\label{eq:II.122}
\end{eqnarray}
where
\begin{eqnarray}\label{eq:II.123}
    \Delta \Hw_{\vec{P}}|^{\sigma_j}_{\sigma_{j+1}}\,&:=&\,
    \widehat{\Hw}_{\vP}^{\sigma_{j+1}}-\widehat{\cvar}_{\vP}^{\sigma_{j+1}}+
    \cvar_{\vec{P}}^{\sigma_{j}}-\Hw_{\vP}^{\sigma_j}
    \nonumber\\
    &=&\frac12\Big[ \, \vec{\Gamma}_{\vP}^{\sigma_j}\cdot\big(\vec{\mathcal{L}}_{\sigma_{j+1}}^{\sigma_j}+
    \vec{\mathcal{I}}_{\sigma_{j+1}}^{\sigma_j}\big) \, + \, h.c. \, \Big] \, + \,
    \big(\vec{\mathcal{L}}_{\sigma_{j+1}}^{\sigma_j}+\vec{\mathcal{I}}_{\sigma_{j+1}}^{\sigma_j}\big)^2 \;.
\end{eqnarray}
Then, we define
\begin{equation}
    \Phi_{\vec{P}}^{\sigma_{j+1}}\,:=\,W_{\sigma_{j+1}}(\vnabla E_{\vec{P}}^{\sigma_{j+1}})
    W_{\sigma_{j+1}}^*(\vnabla E_{\vec{P}}^{\sigma_j}) \widehat{\Phi}_{\vec{P}}^{\sigma_{j+1}}\,.
\end{equation}
The series in (\ref{eq:II.122}) is termwise well-defined 
and converges strongly to a non-zero vector, provided $\alpha$ is small enough
(independently of $j$). This
follows from operator-norm estimates of the type used for (\ref{eq:II.48}).

To prove the convergence of the sequence
$\lbrace \Phi_{\vec{P}}^{\sigma_{j}}\rbrace$,
we proceed as follows. The key point is to show that the term
\begin{equation}
    \frac12\Big[ \, \vec{\Gamma}_{\vP}^{\sigma_j}\cdot\big(\vec{\mathcal{L}}_{\sigma_{j+1}}^{\sigma_j}+
    \vec{\mathcal{I}}_{\sigma_{j+1}}^{\sigma_j}\big) \, + \, h.c. \, \Big]
\end{equation}
contained in (\ref{eq:II.123}), which
is superficially marginal in the infrared 
by power counting (using the terminology of renormalization group theory),
is in fact irrelevant. This is a consequence of the orthogonality relation
\begin{equation}\label{eq:II.126}
	 \bra \, \Phi_{\vec{P}}^{\sigma_j}\,,
	 \,\vec{\Gamma}_{\vP}^{\sigma_j}\Phi_{\vec{P}}^{\sigma_j} \, \ket \, = \, 0\,,
\end{equation}
as we will show.
We then proceed to showing that terms like
\begin{equation}\label{eq:II.128}
    \big\|\big(\frac{1}{\Hw_{\vP}^{\sigma_j}-z_{j+1}}\big)^{\frac{1}{2}}
    \big[\vec{\Gamma}_{\vP}^{\sigma_j}\cdot
    \big(\vec{\mathcal{L}}_{\sigma_{j+1}}^{\sigma_j\,(+)}+\vec{\mathcal{I}}_{\sigma_{j+1}}^{\sigma_j}\big)\big]
    \big(\frac{1}{\Hw_{\vP}^{\sigma_j}-z_{j+1}}\big)^{\frac{1}{2}}\Phi_{\vec{P}}^{\sigma_j}\big\| \,
\end{equation}
(where $\vec{\mathcal{L}}_{\sigma_{j+1}}^{\sigma_j\,(+)}$ stands for the part
which contains only photon creation operators) are of order $\cO(\epsilon^{\eta j})$, for some
$\eta>0$, and we consequently deduce that
\begin{equation}
	\|\widehat{\Phi}_{\vec{P}}^{\sigma_{j+1}}-\Phi_{\vec{P}}^{\sigma_j}\| 
\end{equation}
tends to 0, as $j\to\infty$.

\begin{theorem}
\label{thm:III-1}
The strong limit
\begin{equation}
    %\Phi_{\vec{P}}  \, = \, 
    s-\lim_{j\to\infty}\Phi_{\vec{P}}^{\sigma_j}\,
\end{equation}
exists and is non-zero, and the rate of convergence
is, at least,  $\cO(\sigma_j^{\frac{1}{2}(1-\delta)})$, for any
$0<\delta<1$.
\end{theorem} 

In the proof, we can import results from \cite{Pizzo2003} at various
places. Thus, we will be sketchy in part of our presentation. 
\\
 
%%%%%%%%%%%%%%%%%%%%%%%%%%%%%%%%%%%%%%%%%%%%%%%%%%%
\subsection{Key ingredients of the proof of Theorem {\ref{thm:III-1}}}
\label{ssec-keyingred}
%%%%%%%%%%%%%%%%%%%%%%%%%%%%%%%%%%%%%%%%%%%%%%%%%%%

%The main elements of the 
%are given in the following list ($\mathscr{A}1$) -- ($\mathscr{A}4$).
$\;$

\noindent$\bullet\;$
\emph{Constraints on $\epsilon$, $\mu$ and $\alpha$}

In addition to the conditions on $\alpha$, $\epsilon$ and $\mu$
imposed in our discussion so far, the analysis in this part will require an upper
bound on $\mu$ 
%(see Lemma {\ref{lm:B-3}}) 
and an upper bound on $\epsilon$ strictly smaller than
%1  in (\ref{eq:II.166}) 
the ones imposed by the inequalities (\ref{eq-rfmurfp-1}), (\ref{eq-rfmurfp-2});
see Lemma {\ref{lm:B-3}} and  (\ref{eq:II.166}) below.
We note that the more restrictive conditions on $\mu$ and $\epsilon$
imply new bounds on $\rf$, $\rfp$.
Moreover,
$\epsilon$ must satisfy a lower bound $\epsilon>C \, \alpha^{\frac{1}{2}}$,
with $C>0$ sufficiently large.
We will point out below where these constraints are needed.
\\

\noindent$\bullet\;$
\emph{Estimates on the shift of the ground state energy and its gradient}

There are constants $C_1$, $C_2$ such that the following hold.

\begin{itemize}
\item[($\mathscr{A}1$)]
\begin{equation}
    | \, E^{\sigma_j}_{\vec{P}} \, - \, E^{\sigma_{j+1}}_{\vec{P}} \, |
    \, \leq \, C_{1}\,\alpha\,\epsilon^{j}
\end{equation}
This estimate can be proved as inequality (II.19) in \cite{BachFroehlichPizzo2005}.

\item[($\mathscr{A}2$)]
\begin{equation}\label{eq:II.132}
    | \, \vnabla E^{\sigma_{j+1}}_{\vec{P}}-\vnabla E^{\sigma_{j}}_{\vec{P}} \, |
    \, \leq \,
    C_{2}\Big(\Big\|\frac{\widehat{\Phi}_{\vec{P}}^{\sigma_{j+1}}}{\|\widehat{\Phi}_{\vec{P}}^{\sigma_{j+1}}\| }
    -\frac{\Phi_{\vec{P}}^{\sigma_{j}}}{\|\Phi_{\vec{P}}^{\sigma_{j}}\| }
    \Big\| +\epsilon^{\frac{j+1}{2}}\Big)
\end{equation}
%This estimate follows by using the Feynman-Hellman formula,  where $\vnabla E^{\sigma_j}_{\vec{P}}$
%is expressed in terms of the operator
%$\widehat{\Pi}_{\vP}^{\sigma_j}$, and
%$\vnabla E^{\sigma_{j-1}}_{\vec{P}}$ is expressed in terms of the operator
%$\Pi_{\vP}^{\sigma_{j-1}}$. For the comfort of the reader, an outline of the proof is provided in Lemma B.2..
For the proof, see Lemma  {\ref{lm:B-2}} in the Appendix.
\end{itemize}

\noindent$\bullet\;$\emph{Bounds relating expectations of operators to those of their absolute values}

There are constants $C_{3}$, $C_{4}>1$ such that the following hold.
\begin{itemize}

\item[($\mathscr{A}3$)]
For $z_{j+1}\in\gamma_{j+1}$,
\begin{eqnarray}
    \lefteqn{\Bra \, (\Gamma_{\vP}^{\sigma_j})^{i}\Phi_{\vec{P}}^{\sigma_j}\,,\,
    \big|\frac{1}{\Hw_{\vP}^{\sigma_j}-z_{j+1}}\big|\,(\Gamma_{\vP}^{\sigma_j})^i
    \Phi_{\vec{P}}^{\sigma_j}\, \Ket
    }
    \nonumber\\
    &\leq &C_{3}\Big|\Bra \, (\Gamma_{\vP}^{\sigma_j})^{i}\Phi_{\vec{P}}^{\sigma_j}\,,\,
    \frac{1}{\Hw_{\vP}^{\sigma_j}-z_{j+1}}\, (\Gamma_{\vP}^{\sigma_j})^i
    \Phi_{\vec{P}}^{\sigma_j}\, \Ket\Big|\, .\label{eq:IV.59}
\end{eqnarray}

\item[($\mathscr{A}4$)]
For $z_{j+1}\in\gamma_{j+1}$,
\begin{eqnarray}
    \lefteqn{
    \Bra \, \vec{\mathcal{L}}_{\sigma_{j+1}}^{\sigma_j\,(+)}(\Gamma_{\vP}^{\sigma_j})^{i}
    \Phi_{\vec{P}}^{\sigma_j}\,,\,\big|\frac{1}{\Hw_{\vP}^{\sigma_j}-z_{j+1}}
    \big|\,\vec{\mathcal{L}}_{\sigma_{j+1}}^{\sigma_j\,(+)}
    (\Gamma_{\vP}^{\sigma_j})^i\Phi_{\vec{P}}^{\sigma_j}\, \Ket
    }
    \nonumber\\
    &\leq&C_{4}\Big|\Bra \, \vec{\mathcal{L}}_{\sigma_{j+1}}^{\sigma_j\,(+)}
    (\Gamma_{\vP}^{\sigma_j})^{i}\Phi_{\vec{P}}^{\sigma_j}\,,\,
    \frac{1}{\Hw_{\vP}^{\sigma_j}-z_{j+1}}\,\vec{\mathcal{L}}_{\sigma_{j+1}}^{\sigma_j\,(+)}
    (\Gamma_{\vP}^{\sigma_j})^i\Phi_{\vec{P}}^{\sigma_j}\,\Ket\Big|\,.\quad\quad\quad\label{eq:IV.60}
\end{eqnarray}
To obtain these two bounds, it suffices to exploit the fact that the spectral support
(with respect to $\Hw_{\vP}^{\sigma_j}$) of the two vectors
$(\Gamma_{\vP}^{\sigma_j})^i\Phi_{\vec{P}}^{\sigma_j}$ and
$\vec{\mathcal{L}}_{\sigma_{j+1}}^{\sigma_j\,(+)}(\Gamma_{\vP}^{\sigma_j})^i\Phi_{\vec{P}}^{\sigma_j}$
is strictly above the ground state energy, since they are both orthogonal 
to the ground state $\Phi_{\vec{P}}^{\sigma_j}$.
\end{itemize}

\noindent
\emph{\bf{Remark:}}
The constants $C_1,\dots\,, C_4$ are independent of $\alpha$, $\epsilon$, $\mu$,
and $j\in \NN_0$, provided that
$\alpha,\,\epsilon$, and $\mu$ are sufficiently small.
\\

\newpage
 
\subsection{Proof of the convergence of $(\Phi_\vP^{\sigma_j})_{j=0}^\infty$}

The proof of Theorem {\ref{thm:III-1}} consists of four main steps.

\noindent
%%%%%%%%%%%%%%%%%%%%%%%%%%%%%%%%%%%%%%%%%%%%%%%%%%%
\subsubsection*{\underline{Step (1)}}
%%%%%%%%%%%%%%%%%%%%%%%%%%%%%%%%%%%%%%%%%%%%%%%%%%
%{{\emph{In the first step, we verify the following:}}}
$\;$\\
\\
{{\emph{(i) Assuming the bound
\begin{equation}\label{eq:II.137}
\Big|\Bra \, (\Gamma_{\vP}^{\sigma_j})^{i}\Phi_{\vec{P}}^{\sigma_j}\,,\,
\frac{1}{\Hw_{\vP}^{\sigma_j}-z_{j+1}}\,(\Gamma_{\vP}^{\sigma_j})^i
\Phi_{\vec{P}}^{\sigma_j}\, \Ket \Big| \, \leq \,
\frac{R_0}{\alpha\,\epsilon^{j\delta}}\quad\quad 1>\delta>0\,,
\end{equation}
where $R_0$ is a constant uniform in
$j\in \NN$, for
$\alpha$, $\epsilon$, $\mu$ sufficiently small,
we prove that
\begin{equation}\label{eq:II.138}
\big\|\big(\frac{1}{\Hw_{\vP}^{\sigma_j}-z_{j+1}}\big)^{\frac{1}{2}}
\big[\vec{\Gamma}_{\vP}^{\sigma_j}\cdot\big(\vec{\mathcal{L}}_{\sigma_{j+1}}^{\sigma_j\,(+)}+
\vec{\mathcal{I}}_{\sigma_{j+1}}^{\sigma_j}\big)\big]
\big(\frac{1}{\Hw_{\vP}^{\sigma_j}-z_{j+1}}\big)^{\frac{1}{2}}\Phi_{\vec{P}}^{\sigma_j}\big\| 
\, \leq \, \cO(\epsilon^{\frac{j}{2}(1-\delta)})\, ;
\end{equation}
(see  (\ref{eq:II.128})).
(ii) For $\alpha$ and $R_0$ small enough
independently of $j$, we prove that}}}
\begin{equation}\label{eq:II.139}
	\|\widehat{\Phi}_{\vec{P}}^{\sigma_{j+1}}-\Phi_{\vec{P}}^{\sigma_{j}}\| 
	\, \leq \, \epsilon^{\frac{j+1}{2}(1-\delta)}.
\end{equation}
%{\em (by combining (\ref{eq:II.138}) with an estimate analogous to (\ref{eq:II.16})).}
%\\

For the term on the l.h.s. of (\ref{eq:II.138}) proportional to
to $\vec{\mathcal{I}}_{\sigma_{j+1}}^{\sigma_j}$,  the asserted
upper bound is readily obtained from   estimate $(\mathscr{A}3)$ 
combined with (\ref{eq:II.137}). For the term proportional to
$\vec{\mathcal{L}}_{\sigma_{j+1}}^{\sigma_j\,(+)}$, we prove (\ref{eq:II.138})
following arguments developed in \cite{Pizzo2003}; see Lemma {\ref{lm:B-3}} of the Appendix for details.
This involves the application of a "pull-through formula", a
resolvent expansion, and the bounds $(\mathscr{A}3)$, $(\mathscr{A}4)$.
\\

\noindent
%%%%%%%%%%%%%%%%%%%%%%%%%%%%%%%%%%%%%%%%%%%%%%%%%%%
\subsubsection*{\underline{Step (2)}}
%%%%%%%%%%%%%%%%%%%%%%%%%%%%%%%%%%%%%%%%%%%%%%%%%%
$\;$\\
\\
{{\emph{We relate the l.h.s.
of (\ref{eq:II.137}) to the corresponding quantity with $j$
replaced by $j-1$, and to the norm difference
\begin{equation}
    \|\widehat{\Phi}_{\vec{P}}^{\sigma_{j}}-\Phi_{\vec{P}}^{\sigma_{j-1}}\|\,
\end{equation}
(see  (\ref{eq:IV.72}) -- (\ref{eq:IV.75}) below).}}}
\\

By unitarity of $W_{\sigma_{j}}(\vnabla E_{\vec{P}}^{\sigma_{j-1}})
W_{\sigma_{j}}^*(\vnabla E_{\vec{P}}^{\sigma_{j}}) $,
the l.h.s. of (\ref{eq:II.137}) equals
\begin{equation}
    \label{eq:IV-99}
    \Big|\Bra \, (\widehat{\Gamma}_{\vP}^{\sigma_j})^{i}\widehat{\Phi}_{\vec{P}}^{\sigma_j}\,,\,
    \frac{1}{\widehat{\Hw}_{\vP}^{\sigma_j}-z_{j+1}}\,(\widehat{\Gamma}_{\vP}^{\sigma_j})^i
    \widehat{\Phi}_{\vec{P}}^{\sigma_j}\, \Ket\Big|\,.
\end{equation}
Assuming that $\alpha$ is small enough and $\epsilon>C \, \alpha^{\frac{1}{2}}$,
with $C>0$ sufficiently large, we may use $(\mathscr{A}1)$ to re-expand the
resolvent and find
\begin{eqnarray}
    \lefteqn{
    \Big|\Bra \, (\widehat{\Gamma}_{\vP}^{\sigma_j})^{i}\widehat{\Phi}_{\vec{P}}^{\sigma_j}\,,\,
    \frac{1}{\widehat{\Hw}_{\vP}^{\sigma_j}-z_{j+1}}\,(\widehat{\Gamma}_{\vP}^{\sigma_j})^i
    \widehat{\Phi}_{\vec{P}}^{\sigma_j}\, \Ket\Big|
    }
    \nonumber\\
    &\leq&2 \, \Big|\Bra \, (\widehat{\Gamma}_{\vP}^{\sigma_j})^{i}
    \widehat{\Phi}_{\vec{P}}^{\sigma_j}\,,\,\Big|\frac{1}{\Hw_{\vP}^{\sigma_{j-1}}-z_{j+1}}
    \Big|\,(\widehat{\Gamma}_{\vP}^{\sigma_j})^i\widehat{\Phi}_{\vec{P}}^{\sigma_j}\, \Ket\Big|\,.
\end{eqnarray}
We then readily obtain that
\begin{eqnarray}
\lefteqn{
2 \, \Big|\Bra \, (\widehat{\Gamma}_{\vP}^{\sigma_j})^{i}\widehat{\Phi}_{\vec{P}}^{\sigma_j}\,,\,
\Big|\frac{1}{\Hw_{\vP}^{\sigma_{j-1}}-z_{j+1}}
\Big| \, (\widehat{\Gamma}_{\vP}^{\sigma_j})^i\widehat{\Phi}_{\vec{P}}^{\sigma_j}\, \Ket\Big|
}
\nonumber\\
&\leq&4 \, \Big\|\Big(\frac{1}{\Hw_{\vP}^{\sigma_{j-1}}-z_{j+1}}\Big)^{\frac{1}{2}}\,
( (\widehat{\Gamma}_{\vP}^{\sigma_j})^{i}\widehat{\Phi}_{\vec{P}}^{\sigma_j}-
\, (\Gamma_{\vP}^{\sigma_{j-1}})^{i}\Phi_{\vec{P}}^{\sigma_{j-1}})\Big\|^2
\label{eq:II.145}\\
& &+ \, 4 \, \Big|\Bra \, (\Gamma_{\vP}^{\sigma_{j-1}})^{i}
\Phi_{\vec{P}}^{\sigma_{j-1}}\,,\,\Big|\frac{1}{\Hw_{\vP}^{\sigma_{j-1}}-z_{j+1}}
\Big| \, (\Gamma_{\vP}^{\sigma_{j-1}})^{i}\Phi_{\vec{P}}^{\sigma_{j-1}}\, \Ket\Big|\,.
\label{eq:II.146}
\end{eqnarray}
Our strategy is to construct a recursion that relates (\ref{eq:II.146}) to the initial
expression (\ref{eq:IV-99}) with $j$ replaced by $j-1$,
while (\ref{eq:II.145}) is a remainder term.

We bound the remainder term (\ref{eq:II.145}) by
\begin{eqnarray}
    \lefteqn{
    4 \, \Big\|\Big(\frac{1}{\Hw_{\vP}^{\sigma_{j-1}}-z_{j+1}}\Big)^{\frac{1}{2}}\,
    ( (\widehat{\Gamma}_{\vP}^{\sigma_j})^{i}\widehat{\Phi}_{\vec{P}}^{\sigma_j}-
    (\Gamma_{\vP}^{\sigma_{j-1}})^{i}\Phi_{\vec{P}}^{\sigma_{j-1}})\Big\|^2
    }
    \nonumber\\
    &\leq&8 \, \Big\|\Big(\frac{1}{\Hw_{\vP}^{\sigma_{j-1}}-z_{j+1}}\Big)^{\frac{1}{2}}\,
    ((\widehat{\Gamma}_{\vP}^{\sigma_j})^{i}\widehat{\Phi}_{\vec{P}}^{\sigma_j}-
    (\Gamma_{\vP}^{\sigma_{j-1}})^{i}\widehat{\Phi}_{\vec{P}}^{\sigma_j})\Big\|^2
    \label{eq:IV.69}\\
    &&+ \, 8 \, \Big\|\Big(\frac{1}{\Hw_{\vP}^{\sigma_{j-1}}-z_{j+1}}\Big)^{\frac{1}{2}}\,
    (\Gamma_{\vP}^{\sigma_{j-1}})^{i}(\widehat{\Phi}_{\vec{P}}^{\sigma_j}-\Phi_{\vec{P}}^{\sigma_{j-1}})
    \Big\|^2
    \nonumber\\
    &\leq&\frac{R_1}{\epsilon^{\frac{j}{2}}}
    \Big(\frac{\|\widehat{\Phi}_{\vec{P}}^{\sigma_{j}}-\Phi_{\vec{P}}^{\sigma_{j-1}}\| +
    \epsilon^{\frac{j}{2}}}{\epsilon^{\frac{j}{4}}}\Big)^2
    \label{eq:IV.70}\\
    & &+ \, \frac{R_2}{\epsilon^{\frac{j}{2}}}
    \Big(\frac{\big\|\frac{\widehat{\Phi}_{\vec{P}}^{\sigma_{j}}}{\|\widehat{\Phi}_{\vec{P}}^{\sigma_{j}}\|}-
    \frac{\Phi_{\vec{P}}^{\sigma_{j-1}}}{\|\Phi_{\vec{P}}^{\sigma_{j-1}}\|}\big\|+
    \epsilon^{\frac{j}{2}}}{4\epsilon^{\frac{j}{4}}}\Big)^2 \, ,\nonumber
\end{eqnarray}
where $R_1$ and $R_2$ are constants independent of $\alpha$,
%, $\epsilon$, 
$\mu$, and
$j\in \NN$, provided that $\alpha$, $\epsilon$, and $\mu$ are sufficiently small,
and $\epsilon>C \, \alpha^{\frac{1}{2}}$.
For details on the step from (\ref{eq:IV.69}) to (\ref{eq:IV.70}),
we refer to Lemma {\ref{lm:B-4}} of the Appendix.

To bound the term (\ref{eq:II.146}), we use $(\mathcal{A}3)$ and the orthogonality property
expressed in (\ref{eq:II.126}). We find that, for any $z_j\in\gamma_j$, 
\begin{eqnarray}
\lefteqn{
4 \, \Big|\Bra \, (\Gamma_{\vP}^{\sigma_{j-1}})^{i}\Phi_{\vec{P}}^{\sigma_{j-1}}\,,\,
\Big|\frac{1}{\Hw_{\vP}^{\sigma_{j-1}}-z_{j+1}}
\Big| \, (\Gamma_{\vP}^{\sigma_{j-1}})^{i}\Phi_{\vec{P}}^{\sigma_{j-1}}\, \Ket\Big|
}
\nonumber\\
&\leq &4C_3\,\Big|\Bra \, (\Gamma_{\vP}^{\sigma_{j-1}})^{i}\Phi_{\vec{P}}^{\sigma_{j-1}}\,,\,
\frac{1}{\Hw_{\vP}^{\sigma_{j-1}}-z_{j+1}} \,
(\Gamma_{\vP}^{\sigma_{j-1}})^{i}\Phi_{\vec{P}}^{\sigma_{j-1}}\, \Ket\Big|
\label{eq:II.153}\\
&\leq &8C_3^2\,\Big|\Bra \, (\Gamma_{\vP}^{\sigma_{j-1}})^{i}\Phi_{\vec{P}}^{\sigma_{j-1}}\,,\,
\frac{1}{\Hw_{\vP}^{\sigma_{j-1}}-z_{j}} \, (\Gamma_{\vP}^{\sigma_{j-1}})^{i}
\Phi_{\vec{P}}^{\sigma_{j-1}}\, \Ket\Big|\,.\label{eq:II.154}
\end{eqnarray}
In passing from (\ref{eq:II.153}) to (\ref{eq:II.154}), we have used the constraint on the spectral
support
(with respect to $\Hw_{\vP}^{\sigma_{j-1}}$) of the vector
$(\Gamma_{\vP}^{\sigma_{j-1}})^{i}\Phi_{\vec{P}}^{\sigma_{j-1}}$.

Therefore,  for sufficiently small values of the parameters
$\epsilon$ and $\alpha$, we conclude that
\begin{eqnarray}
	\lefteqn{\Big|\Bra \, (\Gamma_{\vP}^{\sigma_j})^{i}\Phi_{\vec{P}}^{\sigma_j}\,,\,
	\frac{1}{\Hw_{\vP}^{\sigma_j}-z_{j+1}} \,
	(\Gamma_{\vP}^{\sigma_j})^{i}\Phi_{\vec{P}}^{\sigma_j}\, \Ket \Big|}
	\label{eq:IV.72}
	\\
	&\leq&\frac{R_1}{\epsilon^{\frac{j}{2}}}
	\Big(\frac{\|\widehat{\Phi}_{\vec{P}}^{\sigma_{j}}-\Phi_{\vec{P}}^{\sigma_{j-1}}\|+
	\epsilon^{\frac{j}{2}}}{\epsilon^{\frac{j}{4}}}\Big)^2\label{eq:IV.73}
	\\
	& &+ \, \frac{R_2}{\epsilon^{\frac{j}{2}}}\Big(\frac{\big\|
	\frac{\widehat{\Phi}_{\vec{P}}^{\sigma_{j}}}{\|\widehat{\Phi}_{\vec{P}}^{\sigma_{j}}\|}-
	\frac{\Phi_{\vec{P}}^{\sigma_{j-1}}}{\|\Phi_{\vec{P}}^{\sigma_{j-1}}\|}\big\|+
	\epsilon^{\frac{j}{2}}}{4\epsilon^{\frac{j}{4}}}\Big)^2\label{eq:IV.74}
	\\
	& &+ \, 8C_3^2\,\Big|\Bra \, (\Gamma_{\vP}^{\sigma_{j-1}})^{i}\Phi_{\vec{P}}^{\sigma_{j-1}}\,,\,
	\frac{1}{\Hw_{\vP}^{\sigma_{j-1}}-z_{j}} \, (\Gamma_{\vP}^{\sigma_{j-1}})^{i}
	\Phi_{\vec{P}}^{\sigma_{j-1}}\, \Ket\Big|\,.\label{eq:IV.75}
\end{eqnarray}

\noindent
%%%%%%%%%%%%%%%%%%%%%%%%%%%%%%%%%%%%%%%%%%%%%%%%%%%
\subsubsection*{\underline{Step (3)}}
%%%%%%%%%%%%%%%%%%%%%%%%%%%%%%%%%%%%%%%%%%%%%%%%%%%
$\;$\\
\\
{\emph{We prove that}
\begin{equation}\label{eq:II.159}
    \|\Phi_{\vec{P}}^{\sigma_{j}}-\widehat{\Phi}_{\vec{P}}^{\sigma_{j}}\|
    \, \leq \, C_5 \, \alpha^{\frac{1}{2}}
    \,|\vnabla E_{\vec{P}}^{\sigma_{j-1}}-\vnabla E_{\vec{P}}^{\sigma_{j}}|\,|\ln(\epsilon^j)|\,,
\end{equation} 
where $C_5$ is independent of $\alpha$, $\epsilon$, $\mu$, and $j\in\NN$, provided that $\alpha$, $\epsilon$,  and $\mu$ are sufficiently small.
$\;$

>From the definition
\begin{equation}
    \Phi_{\vec{P}}^{\sigma_{j}} \, := \, W_{\sigma_{j}}(\vnabla E_{\vec{P}}^{\sigma_{j}})
    W_{\sigma_{j}}^*(\vnabla  E_{\vec{P}}^{\sigma_{j-1}})\widehat{\Phi}_{\vec{P}}^{\sigma_{j}}\,,
\end{equation}
we get that
\begin{eqnarray}
    \| \, \Phi_{\vec{P}}^{\sigma_{j}}-\widehat{\Phi}_{\vec{P}}^{\sigma_{j}} \, \|
    \, = \, \|W_{\sigma_{j}}^*(\vnabla  E_{\vec{P}}^{\sigma_{j-1}})W_{\sigma_{j}}(\vnabla
    E_{\vec{P}}^{\sigma_{j}})\Psi_{\vec{P}}^{\sigma_{j}}-\Psi_{\vec{P}}^{\sigma_{j}}\|
\end{eqnarray}
where (with an abuse of notation)  we have
denoted by $\Psi_{\vec{P}}^{\sigma_{j}}$ the ground state eigenvector
\begin{equation}
    W_{\sigma_{j}}^*(\vnabla  E_{\vec{P}}^{\sigma_{j}})\Phi_{\vec{P}}^{\sigma_{j}}\,,
\end{equation} 
$\|W_{\sigma_{j}}^*(\vnabla E_{\vec{P}}^{\sigma_{j}})\Phi_{\vec{P}}^{\sigma_{j}}\|\leq1
$, of the
Hamiltonian $H_{\vP}^{\sigma_j}$. Then, we apply formula (\ref{eq-II-9})
(which was derived in \cite{ChFr}), and obtain the logarithmic bound 
$\langle N^f\rangle_{\Psi_{\vec{P}}^{\sigma_{j}}} \leq \cO(|\ln\sigma_j|)$ for the
expectation value of the photon number operator $N^f$ in
$\Psi_{\vec{P}}^{\sigma_{j}}$, where $\sigma_j=\Lambda\epsilon^j$, and $\Lambda\approx1$.
Hence, the estimate (\ref{eq:II.159}) follows.
\\

\noindent
%%%%%%%%%%%%%%%%%%%%%%%%%%%%%%%%%%%%%%%%%%%%%%%%%%
\subsubsection*{\underline{Step (4)}}
%%%%%%%%%%%%%%%%%%%%%%%%%%%%%%%%%%%%%%%%%%%%%%%%%%
$\;$\\
\\
{\em We prove the bound (\ref{eq:II.137}) assumed in step (1)
by an inductive argument (see (\ref{eq:IV.87}) below)}.
\\

We assume $\alpha$, $\epsilon$, and $\mu$ to be sufficiently small for all our previous
results to hold, and such that:

\begin{itemize}
\item[i)]
\begin{equation}
S_{1}^{j}:=\sum_{m=1}^{ j}\Big[\epsilon^{\frac{m}{2}(1-\delta)}+
4\,C_5\,C_2\,\alpha^{\frac{1}{2}}\,\epsilon^{\frac{m}{2}(1-\delta)}|\ln(\epsilon^{m})|\Big]\leq\frac{1}{3}\;,
\end{equation}
uniformly in $j$.

\item[ii)]
\begin{equation}
	\|\widehat{\Phi}_{\vec{P}}^{\sigma_1}-\Phi_{\vec{P}}^{\sigma_0}\|
	\, \leq \, \epsilon^{\frac{1}{2}(1-\delta)}\;.
\end{equation}
\item[iii)]
The bound (\ref{eq:II.137}) holds for $j=1$, and
\begin{equation}
	0<R_1+R_2\leq\,(1-8\,C_3\,\epsilon^{\delta})\frac{R_0}{\alpha}\,.\label{eq:II.166}
\end{equation}
\end{itemize}
Notably, (\ref{eq:II.166}) imposes a more restrictive 
upper bound on the admissible values of $\epsilon$.
Then, we proceed with the   induction in $j$.
\\

\begin{itemize}
\item
\underline{\em Inductive hypotheses}
We assume that, for $j-1(\geq 1)$
\\

\begin{itemize}
\item[($\mathcal{H}$1)] we have an estimate
\begin{eqnarray*}
    \|\Phi_{\vec{P}}^{\sigma_{j-1}}-\Phi_{\vec{P}}^{\sigma_{0}}\|
    \,\leq \,
    S_{1}^{j-1}=\sum_{m=1}^{ j-1}\Big[\epsilon^{\frac{m}{2}(1-\delta)}+
    4\,C_5\,C_2\,\,\alpha^{\frac{1}{2}}\,\epsilon^{\frac{m}{2}(1-\delta)}|\ln(\epsilon^{m})|\Big] \,;
\end{eqnarray*}
\item[($\mathcal{H}$2)]
the bound (\ref{eq:II.137}) holds for $j-1(\geq 1)$.
\\
\end{itemize}

\item
\underline{\em Induction step from $j-1$ to $j$}
\\
\\
>From $(\mathcal{H}2)$, we get that
\begin{equation}\label{eq:II.139bis}
\|\widehat{\Phi}_{\vec{P}}^{\sigma_j}-\Phi_{\vec{P}}^{\sigma_{j-1}}\|
\leq\epsilon^{\frac{j}{2}(1-\delta)}\;.
\end{equation}
>From $(\mathcal{H}1)$, $(\mathcal{H}2)$ and $(\mathscr{A}2)$, we can conclude that
\begin{eqnarray}\label{eq-Phi-lowbd-1}
& &\|\Phi_{\vec{P}}^{\sigma_{j-1}}\|\geq\|\Phi_{\vec{P}}^{\sigma_{0}}\|-
\|\Phi_{\vec{P}}^{\sigma_{j-1}}-\Phi_{\vec{P}}^{\sigma_{0}}\|\geq\frac{2}{3}
\\
& &|\vnabla E^{\sigma_j}(\vec{P})-\vnabla E^{\sigma_{j-1}}(\vec{P})|\leq
4C_{2}\epsilon^{\frac{j}{2}(1-\delta)}\,.\,
\end{eqnarray}
and then, by combining (\ref{eq:II.139bis}), (\ref{eq:II.159}) and (\ref{eq:II.132}), that
\begin{equation}\label{eq:II.170}
\|\Phi_{\vec{P}}^{\sigma_{j  }}-\Phi_{\vec{P}}^{\sigma_{0}}\| \, \leq \,
  S_{1}^{j}  \; .
\end{equation}
Finally, we obtain from (\ref{eq:IV.73}) -- (\ref{eq:IV.75}) that
\begin{eqnarray}
\lefteqn{
\Big|\Bra \, (\Gamma_{\vP}^{\sigma_j})^{i}\Phi_{\vec{P}}^{\sigma_j}\,,\,
\frac{1}{\Hw_{\vP}^{\sigma_j}-z_{j+1}} \, (\Gamma_{\vP}^{\sigma_j})^{i}
\Phi_{\vec{P}}^{\sigma_j}\, \Ket \Big|
}
\nonumber\\
&\leq&\frac{R_1}{\epsilon^{\frac{j}{2}}}
\Big(\frac{\|\widehat{\Phi}_{\vec{P}}^{\sigma_{j}}-\Phi_{\vec{P}}^{\sigma_{j-1}}\|+
\epsilon^{\frac{j}{2}(1-\delta)}}{2\epsilon^{\frac{j}{4}}}\Big)^2
\nonumber\\
& &+ \, \frac{R_2}{\epsilon^{\frac{j}{2}}}
\Big(\frac{\big\|\frac{\widehat{\Phi}_{\vec{P}}^{\sigma_{j}}}{\|\widehat{\Phi}_{\vec{P}}^{\sigma_{j}}\|}-
\frac{\Phi_{\vec{P}}^{\sigma_{j-1}}}{\|\Phi_{\vec{P}}^{\sigma_{j-1}}\|}\big\|+
\epsilon^{\frac{j}{2}(1-\delta)}}{4\epsilon^{\frac{j}{4}}}\Big)^2
\nonumber\\
&
&+ \, 8C_3^2\,\Big|\Bra \, (\Gamma_{\vP}^{\sigma_{j-1}})^{i}
\Phi_{\vec{P}}^{\sigma_{j-1}}\,,\,\frac{1}{\Hw_{\vP}^{\sigma_{j-1}}-z_{j}} \,
(\Gamma_{\vP}^{\sigma_{j-1}})^{i}\Phi_{\vec{P}}^{\sigma_{j-1}}\,\Ket\Big|
\nonumber\\
&\leq&\frac{R_1}{\epsilon^{j\delta}}+\frac{R_2}{\epsilon^{j\delta}}+
8C_3^2\,\frac{R_0}{\alpha\epsilon^{(j-1)\delta}}
\; \leq \; \frac{R_0}{\alpha\epsilon^{j\delta}}\,.\label{eq:IV.87}
\end{eqnarray}
\end{itemize}
This proves (\ref{eq:II.138}) and implies 
that the sequence $\lbrace\Phi_{\vec{P}}^{\sigma_j} \rbrace $  
converges. This follows from the same argument yielding (\ref{eq:II.170}).
The limit is a non-zero vector because of (\ref{eq-Phi-lowbd-1}) which holds uniformly in $j$.

This concludes the proof of statement ($\mathscr{I}1$)  in Theorem {\ref{thm-cfp-2}}
for the sequence $\sigma_j=\Lambda\epsilon^j$ of infrared cutoffs. For general sequences
of infrared cutoffs,  ($\mathscr{I}1$) follows by arguments in \cite{Pizzo2003}.
\QED

%%%%%%%%%%%%%%%%%%%%%%%%%%%%%%%%%%%%%%%%%%%%%%%%%%%%%%%%%%%%%%%%%%%%%%

\section{Proof of Statements $(\mathscr{I}2)$ and $(\mathscr{I}3)$ in the main Theorem}
\label{sec-Jot-2-3}

\resetequ

%The following two results can be straightforwardly adopted from \cite{Pizzo2003} where we
%refer for proofs.
%First of all, the analysis given in \cite{Pizzo2003}, Section 3, implies that for the contour
%$\gamma\,:=\lbrace z\in\CC\,|
%|z-E_{\vec{P}}^{\sigma}|=\frac{\sigma}{4}\rbrace$, the vector
%\begin{equation}
%    \Phi_{\vec{P}}^{\sigma} \, := \,
%    \frac{-\frac{1}{2\pi\,i}\int_{\gamma}\frac{1}{\Hw_{\vP}^\sigma-z}\,dz\,\Omega_f}
%    {\|-\frac{1}{2\pi\,i}\int_{\gamma}\frac{1}{\Hw_{\vP}^\sigma-z}\,dz\,\Omega_f\|_{\cF}}
%\end{equation}
%converges strongly to a
%non-zero vector $\Phi_{\vec{P}}$ in the limit $\sigma\to 0$. The rate of convergence
%is at most of order $\cO(\sigma^{\frac{1}{2}(1-\delta)})$, for some
%$0<\delta<1$.

Statement  $(\mathscr{I}2)$  in Theorem {\ref{thm-cfp-2}} expresses
H\"older regularity of $\Phi_{\vec{P}}^{\sigma}$
and $\vnabla E_{\vec{P}}^{\sigma}$ with respect to  $\vec{P}\in\mathcal{S}$, uniformly
in $\sigma \geq 0$. That is,
\begin{equation}
    \| \, \Phi_{\vec{P}}^{\sigma} \, - \, \Phi_{\vec{P}+\Delta\vec{P}}^{\sigma} \, \| \, \leq \,
    C_{\delta'} \, | \, \Delta \vec{P} \, |^{\frac{1}{4}-\delta'}
\end{equation}
and
\begin{equation}
    | \, \vnabla E_{\vec{P}}^{\sigma} \, - \, \vnabla E_{\vec{P}+\Delta\vec{P}}^{\sigma} \, | \, \leq \,
    C_{\delta''} \, | \, \Delta \vec{P} \, |^{\frac{1}{4}-\delta''} \;,
\end{equation}
for any $0<\delta''<\delta'<\frac{1}{4}$, where $\vec{P}\,,\,\vec{P}+\Delta\vP \in
\mathcal{S}$. The constants $C_{\delta'}$ and $C_{\delta''}$
depend on $\delta'$ and $\delta''$, respectively.
This result can be taken over from \cite{Pizzo2003}.

Statement $(\mathscr{I}3)$ follows easily from  ($\mathscr{I}5$).
In fact, we recall from the beginning of Section {\ref{sec:iv-2-1}} that
\begin{eqnarray}
	\vP - \vnabla E_{\vP}^\sigma &=& \bra \, \vP^f-\alpha^{1/2}\vec{A}_\sigma \, \ket_{\Psi_{\vP}^\sigma} \, .
\end{eqnarray}
%Using ($\mathscr{I}5$), which we recall in (\ref{eq-II-9bisbis}) below, 
We then find that
\begin{eqnarray}\label{eq-V-93-1}
	| \, \bra \, \vP^f  \, \ket_{\Psi_{\vP}^\sigma} \, | &\leq&
	\sum_\lambda \int d^3k \, |\vk| \, \frac{\| \, b_{\vk,\lambda}\Psi_{\vP}^{\sigma} \, \|^2}{\| \, \Psi_{\vP}^{\sigma} \, \|^2}
	\nonumber\\
	&\leq& C' \, \alpha \, \int_{\cB_\Lambda}\frac{ d^3k}{|\vk|^2} 
	\; \leq \; C \, \alpha \, ,
\end{eqnarray}
and
\begin{eqnarray}\label{eq-V-93}
	| \, \bra \, \alpha^{1/2}\vec{A}_\sigma \, \ket_{\Psi_{\vP}^\sigma} \, | &\leq& 2 \, \alpha^{1/2}
	\sum_\lambda \int \frac{d^3k}{  |\vk|^{1/2}} \, \frac{\| \, b_{\vk,\lambda}\Psi_{\vP}^{\sigma} \, \|}{\| \, \Psi_{\vP}^{\sigma} \, \|}
	\nonumber\\
	&\leq& C' \, \alpha \, \int_{\cB_\Lambda}\frac{ d^3k}{|\vk|^2} 
	\; \leq \; C \, \alpha \, ,
\end{eqnarray}
where we used  ($\mathscr{I}5$) in (\ref{eq-V-93}).
Therefore, 
\begin{equation}
	| \, \vP - \vnabla E_{\vP}^\sigma \, | \, \leq \, C \, \alpha \,,
\end{equation}
for a constant $C$ independent of $\vP\in\mathcal{S}$ and $\sigma$. Statement $(\mathscr{I}3)$ then
follows immediately.

\newpage
%%%%%%%%%%%%%%%%%%%%%%%%%%%%%%%%%%%%%%%%%%%%%%%%%%%%%%%%%%%%% T H E O R E M   B.1.    %%%%%%%%%%%%%%%%%%%%%%%%%%%%%%%%%%%%%%%%%%%%%%%%%%%%%%%%%%%%%%%%%%%%%%%%%%%%%%%%%
\section{Proof of Statement $(\mathscr{I}4)$ in the main Theorem}
\label{sect-proof-I4}
\label{sec-Jot-4}

\resetequ

To prove   statement  $(\mathscr{I}4)$ in Theorem {\ref{thm-cfp-2}}, 
we must show that, for $\vP\in\mathcal{S}$, $\alpha$ small enough,
$\vk\neq0$ and $\sigma\geq0$,  
\begin{equation}\label{eq-B-15}
E^{\sigma}_{\vP-\vk} \, > \, E^{\sigma}_{\vP}-C_\alpha|\vk| 
\end{equation}
holds, where $E^{\sigma}_{\vP-\vk}:=\inf {\text{spec}}H_{\vP-\vk}^{\sigma}$,
and $\frac13< C_\alpha < 1$, with $C_\alpha\rightarrow\frac13$ as $\alpha\rightarrow0$.

To prove (\ref{eq-B-15}), we first note that
\eqn
	H_{\vP+\vk}^\sigma \, = \, H_{\vP }^\sigma \, + \, \vk\cdot \vnabla  H_{\vP }^\sigma 
	\, + \, \frac{|\vk|^2}{2}  \,,
\eeqn
and that
\eqn
	\bra\phi \, , \, H_{\vP+\vk}^\sigma \, \phi \ket
	&\geq& \bra\phi \, , \, H_{\vP }^\sigma \, \phi \ket \, - \, |\vk| \,
	\bra\phi \, , \, (\vnabla H_{\vP }^\sigma)^2 \, \phi \ket^{1/2}
	\, + \, \frac{|\vk|^2}{2}  
	\nonumber\\
	&\geq&\bra\phi \, , \, H_{\vP }^\sigma \, \phi \ket \, - \, \sqrt2 \, |\vk| \,
	\bra\phi \, , \, H_{\vP }^\sigma \, \phi \ket^{1/2}
	\, + \, \frac{|\vk|^2}{2}  
\eeqn
for $\phi\in D(H_{\vP+\vk}^\sigma)$, with $\|\phi\|=1$. Thus, we 
obtain the inequality
\eqn\label{eq-EPk-EP-lowbd-1}
	\lefteqn{
	\bra\phi \, , \, H_{\vP+\vk}^\sigma \, \phi\ket 
	%\, + \, |k| 
	\, - \, E_\vP^\sigma
	}
	\nonumber\\
	& \geq &\inf_{z\geq0}\big\{ \,
	(z+E_\vP^\sigma) \, - \, \sqrt2 |\vk| \, (z+E_\vP^\sigma)^{1/2} 
	%\, + \, |k| 
	\, + \, \frac{|\vk|^2}{2} 
	\, - \, E_\vP^\sigma \, \big\}
	\nonumber\\
	&=&\inf_{x\geq   (E_\vP^\sigma)^{1/2}}
	\big\{ \, x^2 \, - \, \sqrt2 |\vk| \, x 
	%\, + \, |k| 
	\, + \, \frac{|\vk|^2}{2} \, - \, E_\vP^\sigma \, \big\} \,,
\eeqn
where $z:=\bra\phi \,, \, H_\vP^\sigma \, \phi \ket - E_\vP^\sigma\geq0$ in the
expression on the second line.

Setting $\partial_x(\cdots)=0$ in the expression on the last line of (\ref{eq-EPk-EP-lowbd-1}), we find
\eqn 
	2x - \sqrt2 |\vk| \, = \, 0 \,.
\eeqn
The minimum is therefore attained at 
$x=\frac{\sqrt2}{2}|\vk|$, if $\frac{\sqrt2}{2}|\vk|\geq (E_\vP^\sigma)^{1/2}$, and 
at $x=(E_\vP^\sigma)^{1/2}$, corresponding to $z=0$, otherwise. That is,
\eqn
	x_{min} \, = \, \max\{\frac{\sqrt2}{2}|\vk| \, , \, (E_\vP^\sigma)^{1/2} \} \,.
\eeqn
Now, for  $\frac{\sqrt2}{2}|\vk|\geq (E_\vP^\sigma)^{1/2}$, so that $x_{min}=\frac{\sqrt2}{2}|\vk|$, 
we evaluate the lower bound and get
\eqn
	\frac{|\vk|^2}{2} \, - \, |\vk|^2 
	%\, + \, |\vk| 
	\, + \, \frac{|\vk|^2}{2} \, - \, E_\vP^\sigma \,,
\eeqn
and we observe that
\eqn
	\, - \, E_\vP^\sigma
	\, \geq \, -\, \frac13 |\vk| \,,
\eeqn
because
\eqn
	E_\vP^\sigma \, \leq \, \frac{1}{\sqrt2}(\frac13+c\alpha) \, (E_\vP^\sigma)^{1/2}
	\, \leq \,  \frac13 \, |\vk|
	%|\vk| \, \geq \, \sqrt2(E_\vP^\sigma)^{1/2} \, \geq \, (\frac13+c\alpha)  E_\vP^\sigma 
\eeqn
for $|\vP|<\frac13$ and $\alpha$ small enough.
This follows from  
\eqn
	0 \, < \, E_\vP^\sigma \, = \, 
	{\rm infspec}H_\vP^\sigma\leq\bra\Omega_f \, , \, H_\vP^\sigma \, \Omega_f\ket
	\, = \, \frac12|\vP|^2+\frac\alpha2\bra (\vec A^\sigma)^2\ket 
\eeqn
by Rayleigh-Ritz, so that $(E_\vP^\sigma)^{1/2}\leq \frac{1}{\sqrt2}(\frac13+c\alpha)$ for $|\vP|<\frac13$ and $\alpha$ small enough. 

If, however,  $\frac{\sqrt2}{2}|\vk|\leq (E_\vP^\sigma)^{1/2}$, so that $x_{min}=(E_\vP^\sigma)^{1/2}$, 
evaluation of the lower bound yields
\eqn
	- \, \sqrt2 |\vk| \,  (E_\vP^\sigma)^{1/2} 
	%\, + \, |k| 
	\, + \, \frac{ |\vk|^2}{2} \, ,
\eeqn
and we observe that
\eqn
	- \, \sqrt2 |\vk| \,  (E_\vP^\sigma)^{1/2} 
	%\, + \, |k| 
	\, + \, \frac{ |\vk|^2}{2} \, 
	\geq \, - \, (|\vP|+c\alpha)|\vk| 
	\, \geq \, - \, (\frac13+ c\alpha)|\vk|
\eeqn
for $|\vP|<\frac13$.  

Therefore, we conclude that
\eqn
	E_{\vP+\vk}^\sigma 
	%\, + \, |\vk| 
	\, > \,   
	E_{\vP}^\sigma  \, - \, C_\alpha \, |\vk|  
\eeqn
for
\eqn
	C_\alpha \, = \, \frac13 \, + \, c\alpha \,,
\eeqn 
and all $\vk\neq0$.

This establishes    statement  $(\mathscr{I}4)$ in Theorem {\ref{thm-cfp-2}}.
\QED

Thus, we have proven our main result, up to auxiliary results proven in the Appendix.

\newpage
%%%%%%%%%%%%%%%%%%%%%%%%%%%%%%%%%%%%%%%%%%%%%%%%%%%%%%%%%%%%%%%%%%%%%%%%%%%%%%%%%%%%%%%%%%%%%%%%%%%%%%%%%%%%%%%%%%%%%%%%%%%%%%%%%%%%%%%%%%%%%%%%%%%%%%%%%%%%%%%%%%%%%%%%%%%%%%%%%%%%%%%%%%%%%%%%%%%%%%%%%%%%%%%%%%%%%%%%%%%%%%%%%%%%%%%%%%%%%%%%%%%%%%%%%%%%%%%%%%%%%%%%%%%%%%%%%%%%%%%%%%%%%%%%%%%%%%%%%%%%%%%%%%%%%%%%%%%%%%%%%%%%%%% A P P E N D I X %%%%%%%%%%%%%%%%%%%%%%%%%%%%%%%%%%%%%%%%%%%%%%%%%%%%%%%%%%%%%%%%%%%%%%%%%%%%%%%%%%%%%%%%%%%%%%%%%%%%%%%%%%%%%%%%%%%%%%%%%%%%%%%%%%%%%%%%%%%%%%%%%%%%%%%%%%%%%%%%%%%%%%%%%%%%%%%%%%%%%%%%%%%%%%%%%%%%%%%%%%%%%%%%%%%%%%%%%%%%%%%%%%%%%%%%%%%%%%%%%%%%%%%%%%%%%%%%%%%%%%%%%%%%%%%%%%%%%%%%%%%%%%%%%%%%%%%%%%%%%%%%%%%%%%%%%%%%%%%%%%%%%%%%%%%%%%%%%%%%%%%%%%%%%%%%%%%%%%%%%%%%%%%%%%%%%%%
\appendix 
\section{ }
\resetequ

\noindent
\subsection{ Well-definedness of the operators $\Hw_{\vP}^{\sigma_j}$ 
and $\widehat{\Hw}_{\vP}^{\sigma_j}$}

We need to verify that the canonical form of the Hamiltonians
$\Hw_{\vP}^{\sigma_j}$ and $\widehat{\Hw}_{\vP}^{\sigma_j}$
in  (\ref{eq-IV.36}) and (\ref{eq-IV.42}) are not only formal. 
This can be achieved by adapting an argument in the work \cite{Nelson} of E. Nelson,
Lemma 3. We shall only outline the proof for $\Hw_{\vP}^{\sigma_j}$; the
case of $\widehat{\Hw}_{\vP}^{\sigma_j}$ is similar.
 
To this end, we write $(\Hw_{\vP}^{\sigma_{j}})'$ for the operator on the right hand side of
(\ref{eq-IV.36}), in order to distinguish it from (\ref{eq:iv-30}). 
We let $\cH_{\vP}(\infty)$ denote the linear span of vectors in
$\cH_{\vP}$ with a finite number of photons. For the values of $\alpha$ and
of $\Lambda$ assumed in Section {\ref{sec-I.2}}, we know that
$H_{\vP}^{\sigma_j}$ is selfadjoint in $D(H_{\vP}^{0})$, where
\begin{equation}
	H_{\vP}^{0} \, := \, \frac{(\vP-\vP^{f})^2}{2}+H^{f}\,.
\end{equation}
Then, we conclude the following:
\\

\begin{itemize}
\item[1)]
The equality (\ref{eq-IV.36}) trivially holds on $\cH_{\vP}(\infty)\bigcap
D(H_{\vP}^{0})$, because vectors in this space are analytic for 
the generator of $W_{\sigma_j}(\vnabla
E_{\vec{P}}^{\sigma_j})$, and since $H_{\vP}^{\sigma_j}$, $H^{0}_{\vP}$ and the
generator of $W_{\sigma_j}(\vnabla
E_{\vec{P}}^{\sigma_j})$ map $\cH_{\vP}(\infty)\bigcap
D(H_{\vP}^{0})$ into itself.
\\

\item[2)]
By standard arguments, one shows that
\begin{equation}\label{eq-B-13}
\|H_{\vP}^{0}W_{\sigma_j}(\vnabla E_{\vec{P}}^{\sigma_j})\psi\|\leq b\,(\|H_{\vP}^{0}\psi\|+\|\psi\|) \, ,
\end{equation}
where $\psi\in \cH_{\vP}(\infty)\bigcap D(H_{\vP}^{0})$, for some $b>0$.

Because $\cH_{\vP}(\infty)\bigcap D(H_{\vP}^{0})$ 
is dense in $D(H_{\vP}^{0})$ with respect to the
norm $\|H_{\vP}^{0}\psi\|+\|\psi\|$,
it follows that $W_{\sigma_j}(\vnabla E_{\vec{P}}^{\sigma_j})$ and 
$W_{\sigma_{j}}^*(\vnabla  E_{\vec{P}}^{\sigma_{j}})$ map $D(H_{\vP}^{0})$ 
into itself.

Consequently,
\begin{equation}\label{eq-B-14}
D(H_{\vP}^{0})\equiv D(\Hw_{\vP}^{\sigma_j})\,.
\end{equation}
$\;$

\item[3)]
The equality (\ref{eq-IV.36}) holds on $D(\Hw_{\vP}^{\sigma_j})$ because
$\cH_{\vP}(\infty)\bigcap D(H_{\vP}^{0})$ is dense in 
$D(H_{\vP}^{0})$
%(\equiv D(\Hw_{\vP}^{\sigma_j}))$ 
in the
norm $\|H_{\vP}^{0}\psi\|+\|\psi\|$, and because of (\ref{eq-B-14}).
Since  $(\Hw_{\vP}^{\sigma_j})'\equiv \Hw_{\vP}^{\sigma_j}$
on the domain of selfadjointness of $\Hw_{\vP}^{\sigma_j}$,
we can therefore conclude that $D((\Hw_{\vP}^{\sigma_j})')\equiv D(\Hw_{\vP}^{\sigma_j})$.
Consequently, we have proven that
%
%. Thus, $D((\Hw_{\vP}^{\sigma_j})')\equiv
%D(\Hw_{\vP}^{\sigma_j})$ and this implies that 
$(\Hw_{\vP}^{\sigma_j})'\equiv \Hw_{\vP}^{\sigma_j}$.
This is what we intended to prove.
\end{itemize}

\newpage
%%%%%%%%%%%%%%%%%%%%%%%%%%%%%%%%%%%%%%%%%%%%%%%%%%%%%%%%%%%%%%%% L E M M A   B.2.  %%%%%%%%%%%%%%%%%%%%%%%%%%%%%%%%%%%%%%%%%%%%%%%%%%%%%%%%%%%%%%%%%%%%%%%%%%%%%%%%
\subsection{Technical lemmata for the proof of $(\mathscr{I}1)$ in Theorem {\ref{thm-cfp-2}}}

\begin{lemma}\label{lm-A-1}
The  Hamiltonian $\widehat{\Hw}_{\vP}^{\sigma_{j+1}}$ has the form (\ref{eq-BogHam}),
with (\ref{eq-III.42.1}), (\ref{eq-III.42.2}), and
(\ref{eq-III.43}).
\end{lemma}

\Proof \\
Recalling the definitions of Section {\ref{sec:iv-2-1}}, we have
\begin{eqnarray}
	\lefteqn{W_{\sigma_{j+1}}(\vnabla
	E_{\vec{P}}^{\sigma_j}) \vec{\beta}^{\sigma_{j+1}}W_{\sigma_{j+1}}^*(\vnabla
	E_{\vec{P}}^{\sigma_j})
	-\langle\vec{\beta}^{\sigma_j}\rangle_{\Psi_{\vec{P}}^{\sigma_j}}
	}
	\\
	&= &W_{\sigma_{j+1}}(\vnabla
	E_{\vec{P}}^{\sigma_j}) \vec{\beta}^{\sigma_{j}}W_{\sigma_{j+1}}^*(\vnabla
	E_{\vec{P}}^{\sigma_j})
	-\langle\vec{\beta}^{\sigma_j}\rangle_{\Psi_{\vec{P}}^{\sigma_j}}\\
	& &- \, \alpha^{\frac{1}{2}}W_{\sigma_{j+1}}(\vnabla
	E_{\vec{P}}^{\sigma_j}) \vec{A}_{\sigma_{j+1}}^{\sigma_j}W_{\sigma_{j+1}}^*(\vnabla
	E_{\vec{P}}^{\sigma_j}) \\
	&=&W_{\sigma_{j}}(\vnabla
	E_{\vec{P}}^{\sigma_j}) \vec{\beta}^{\sigma_{j}}W_{\sigma_{j}}^*(\vnabla
	E_{\vec{P}}^{\sigma_j}) 
	-\langle\vec{\beta}^{\sigma_j}\rangle_{\Psi_{\vec{P}}^{\sigma_j}}\\
	& &+ \, W_{\sigma_{j+1}}^{\sigma_j}(\vnabla
	E_{\vec{P}}^{\sigma_j}) \vec{P}^{f}W_{\sigma_{j+1}}^{\sigma_j\;*}(\vnabla
	E_{\vec{P}}^{\sigma_j}) -\vec{P}^{f}\\
	& &- \,\alpha^{\frac{1}{2}}W_{\sigma_{j+1}}(\vnabla
	E_{\vec{P}}^{\sigma_j}) \vec{A}_{\sigma_{j+1}}^{\sigma_j}W_{\sigma_{j+1}}^*(\vnabla
	E_{\vec{P}}^{\sigma_j}) \\
	&=&\vec{\Pi}_{\vP}^{\sigma_j}-\langle\vec{\Pi}_{\vP}^{\sigma_j}\rangle_{\Phi^{\sigma_j}_{\vec{P}}}\\
	& &- \, \alpha^{\frac{1}{2}}\sum_{\lambda}\int_{\mathcal{B}_{\sigma_j}\setminus
	\mathcal{B}_{\sigma_{j+1}}}\vec{k}\frac{\vnabla
	E_{\vec{P}}^{\sigma_j}\cdot\veps_{\vk,
  	\lambda}^{\;*}b_{\vec{k},\lambda}+h.c.}{|\vec{k}|^{\frac{3}{2}}\delta_{\vP}^{\sigma_j}(\widehat{k})}\,
  	d^3k-\alpha^{\frac{1}{2}}\vec{A}_{\sigma_{j+1}}^{\sigma_j}\\
	& &+ \, \alpha\,
	\sum_{\lambda}\int_{\mathcal{B}_{\sigma_j}\setminus \mathcal{B}_{\sigma_{j+1}}}\,\vec{k}\,\frac{\vnabla
  	E_{\vec{P}}^{\sigma_j}\cdot \veps_{\vk, \lambda}^{\;*}\,
	\vnabla E_{\vec{P}}^{\sigma_j}\cdot 
	\veps_{\vk,\lambda}}{|\vec{k}|^{3}(\delta_{\vP}^{\sigma_j}(\widehat{k}))^2}\,d^3k
	\\
	& &+ \, \alpha\,
	\sum_{\lambda}\int_{\mathcal{B}_{\sigma_j}\setminus \mathcal{B}_{\sigma_{j+1}}}\,\big[\veps_{\vk,
  	\lambda}\,\frac{\vnabla  E_{\vec{P}}^{\sigma_j}\cdot \veps_{\vk,
 	\lambda}^{\;*} }{|\vec{k}|^{\frac{3}{2}}\delta_{\vP}^{\sigma_j}(\widehat{k})}+h.c.\big]
	\,\frac{d^3k}{\sqrt{|\vec{k}|}}\,,
\end{eqnarray}
where
\begin{equation}
	W_{\sigma_{j+1}}^{\sigma_j}(\vnabla E_{\vec{P}}^{\sigma_j}) \, :=\,
	\exp\Big(\, \alpha^{\frac{1}{2}}\sum_{\lambda} 
	\int_{\mathcal{B}_{\sigma_j}\setminus \mathcal{B}_{\sigma_{j+1}}} \, d^3k \,
    \frac{\vnabla E_{\vec{P}}^{\sigma_j}}{|\vec{k}|^{\frac{3}{2}}\delta_{\vP}^{\sigma_j}(\widehat{k})} 
    \cdot (\veps_{\vk,\lambda}b_{\vk,\lambda}^{*} - h.c.)\Big)\,.
\end{equation}
This establishes  (\ref{eq-III.42.1}) and (\ref{eq-III.42.2}).
\QED

\newpage

\begin{lemma}\label{lm:B-2}
For $\vP\in\mathcal{S}$, there exists $C_2>0$ such that, uniformly in $j\in\NN_0$, the inequality  
\begin{equation}\label{eq-B-30}
|\vnabla E^{\sigma_{j+1}}_{\vec{P}}-\vnabla E^{\sigma_{j}}_{\vec{P}}|\leq
C_{2}\Big(\Big\|\frac{\widehat{\Phi}_{\vec{P}}^{\sigma_{j+1}}}{\|\widehat{\Phi}_{\vec{P}}^{\sigma_{j+1}}\|}
-\frac{\Phi_{\vec{P}}^{\sigma_{j}}}{\|\Phi_{\vec{P}}^{\sigma_{j}}\|}\Big\|+\epsilon^{\frac{j+1}{2}}\Big) 
\end{equation}
holds.
\end{lemma}

\Proof
\\
Using (\ref{eq:IV.29}) and (\ref{eq-IV-46bis}),
we write $\vnabla E^{\sigma_{j+1}}_{\vec{P}}\,\text{and}\,\vnabla E^{\sigma_{j}}_{\vec{P}}$ 
in the form
\begin{eqnarray}
\vnabla
E_{\vec{P}}^{\sigma_j}
&= &\vec{P}-\frac{\bra \, \Phi_{\vec{P}}^{\sigma_j}\,,\,\vec{\Pi}_{\vP}^{\sigma_j}
\Phi_{\vec{P}}^{\sigma_j} \, \ket }{\bra \, \Phi_{\vec{P}}^{\sigma_j}\,,\,\Phi_{\vec{P}}^{\sigma_j} \, \ket}
-\langle W_{\sigma_j}(\vnabla
E_{\vec{P}}^{\sigma_j})\,\vec{\beta}^{\sigma_j} \, W_{\sigma_j}^*(\vnabla
E_{\vec{P}}^{\sigma_j}) \rangle_{\Omega_f}
\quad\quad\quad \label{eq-B-31} \\
\vnabla
E_{\vec{P}}^{\sigma_{j+1}}
&= &\vec{P}-\frac{\bra \, \widehat{\Phi}_{\vec{P}}^{\sigma_{j+1}}\,,\,\widehat{\vec{\Pi}}_{\vP}^{\sigma_{j+1}}
\widehat{\Phi}_{\vec{P}}^{\sigma_{j+1}} \, \ket}
{\bra \, \widehat{\Phi}_{\vec{P}}^{\sigma_{j+1}}\,,\,\widehat{\Phi}_{\vec{P}}^{\sigma_{j+1}} \, \ket}
\\
&&\quad\quad\quad\quad-\langle W_{\sigma_{j+1}}(\vnabla
E_{\vec{P}}^{\sigma_{j+1}})\,\vec{\beta}^{\sigma_{j+1}} \, W_{\sigma_{j+1}}^*(\vnabla
E_{\vec{P}}^{\sigma_{j+1}}) \rangle_{\Omega_f} \,.\quad\quad 
\label{eq-B-32}
\nonumber
\end{eqnarray}
By a simple, but slightly lengthy calculation, one can check that 
\begin{eqnarray}
\lefteqn{\langle W_{\sigma_j}(\vnabla
E_{\vec{P}}^{\sigma_j})\,\vec{\beta}^{\sigma_j} \, W_{\sigma_j}^*(\vnabla
E_{\vec{P}}^{\sigma_j})\rangle_{\Omega_f}-}\\
& &-\langle W_{\sigma_{j+1}}(\vnabla
E_{\vec{P}}^{\sigma_{j+1}})\,\vec{\beta}^{\sigma_{j+1}} \, W_{\sigma_{j+1}}^*(\vnabla
E_{\vec{P}}^{\sigma_{j+1}})\rangle_{\Omega_f} \\
&= &\alpha\,
\sum_{\lambda}\int_{\Lambda\setminus\mathcal{B}_{\sigma_j}}\,\vec{k}\,\frac{\vnabla
  E_{\vec{P}}^{\sigma_j}\cdot \veps_{\vk,
  \lambda}^{\;*}\,\vnabla E_{\vec{P}}^{\sigma_j}\cdot \veps_{\vk,
  \lambda}}{|\vec{k}|^{3}(\delta_{\vP}^{\sigma_j}(\widehat{k}))^2}\,d^3k \label{eq-B-34}\\
&&-\alpha\,
\sum_{\lambda}\int_{\Lambda\setminus\mathcal{B}_{\sigma_j}}\,\vec{k}\,\frac{\vnabla
  E_{\vec{P}}^{\sigma_{j+1}}\cdot \veps_{\vk,
  \lambda}^{\;*}\,\vnabla E_{\vec{P}}^{\sigma_{j+1}}\cdot \veps_{\vk,
  \lambda}}{|\vec{k}|^{3}(\delta_{\vP}^{\sigma_{j+1}}(\widehat{k}))^2}\,d^3k \label{eq-B-35}\\
& &+\alpha\,
\sum_{\lambda}\int_{\Lambda\setminus\mathcal{B}_{\sigma_j}}\,\big[\veps_{\vk,
  \lambda}\,\frac{\vnabla  E_{\vec{P}}^{\sigma_j}\cdot \veps_{\vk,
 \lambda}^{\;*} }{|\vec{k}|^{\frac{3}{2}}\delta_{\vP}^{\sigma_j}(\widehat{k})}+h.c.\big]
\,\frac{d^3k}{\sqrt{|\vec{k}|}}\, \label{eq-B-36}\\
& &-\alpha\,
\sum_{\lambda}\int_{\Lambda\setminus\mathcal{B}_{\sigma_{j}}}\,\big[\veps_{\vk,
  \lambda}\,\frac{\vnabla  E_{\vec{P}}^{\sigma_{j+1}}\cdot \veps_{\vk,
 \lambda}^{\;*} }{|\vec{k}|^{\frac{3}{2}}\delta_{\vP}^{\sigma_{j+1}}(\widehat{k})}+h.c.\big]
\,\frac{d^3k}{\sqrt{|\vec{k}|}}\, \label{eq-B-37}\\
&&-\alpha\,
\sum_{\lambda}\int_{\mathcal{B}_{\sigma_j}\setminus \mathcal{B}_{\sigma_{j+1}}}\,\vec{k}\,\frac{\vnabla
  E_{\vec{P}}^{\sigma_{j+1}}\cdot \veps_{\vk,
  \lambda}^{\;*}\,\vnabla E_{\vec{P}}^{\sigma_{j+1}}\cdot \veps_{\vk,
  \lambda}}{|\vec{k}|^{3}(\delta_{\vP}^{\sigma_{j+1}}(\widehat{k}))^2}\,d^3k \label{eq-B-38}\\
& &-\alpha\,
\sum_{\lambda}\int_{\mathcal{B}_{\sigma_j}\setminus \mathcal{B}_{\sigma_{j+1}}}\,\big[\veps_{\vk,
  \lambda}\,\frac{\vnabla  E_{\vec{P}}^{\sigma_{j+1}}\cdot \veps_{\vk,
 \lambda}^{\;*} }{|\vec{k}|^{\frac{3}{2}}\delta_{\vP}^{\sigma_{j+1}}(\widehat{k})}+h.c.\big]
\,\frac{d^3k}{\sqrt{|\vec{k}|}}\,. \label{eq-B-39}
\end{eqnarray}
On the other hand, using definition  (\ref{eq-IV-46bis}), we can calculate
\begin{eqnarray}
\lefteqn{\widehat{\vec{\Pi}}_{\vP}^{\sigma_{j+1}}-\vec{\Pi}_{\vP}^{\sigma_j}}
\label{eq-B-41}\\
&= &
%-\alpha^{\frac{1}{2}}\vec{A}_{\sigma_{j+1}}^{\sigma_j}-
\vec{\cL}_{\sigma_{j+1}}^{\sigma_{j}}\\
& &+\alpha\,
\sum_{\lambda}\int_{\Lambda\setminus\mathcal{B}_{\sigma_{j+1}}}\,\vec{k}\,\frac{\vnabla
  E_{\vec{P}}^{\sigma_j}\cdot \veps_{\vk,
  \lambda}^{\;*}\,\vnabla E_{\vec{P}}^{\sigma_j}\cdot \veps_{\vk,
  \lambda}}{|\vec{k}|^{3}(\delta_{\vP}^{\sigma_j}(\widehat{k}))^2}\,d^3k \label{eq-B-42}\\
&&-\alpha\,
\sum_{\lambda}\int_{\Lambda\setminus\mathcal{B}_{\sigma_{j+1}}}\,\vec{k}\,\frac{\vnabla
  E_{\vec{P}}^{\sigma_{j+1}}\cdot \veps_{\vk,
  \lambda}^{\;*}\,\vnabla E_{\vec{P}}^{\sigma_{j+1}}\cdot \veps_{\vk,
  \lambda}}{|\vec{k}|^{3}(\delta_{\vP}^{\sigma_{j+1}}(\widehat{k}))^2}\,d^3k \label{eq-B-43}\\
& &+\alpha\,
\sum_{\lambda}\int_{\Lambda\setminus\mathcal{B}_{\sigma_{j+1}}}\,\big[\veps_{\vk,
  \lambda}\,\frac{\vnabla  E_{\vec{P}}^{\sigma_j}\cdot \veps_{\vk,
 \lambda}^{\;*} }{|\vec{k}|^{\frac{3}{2}}\delta_{\vP}^{\sigma_j}(\widehat{k})}+h.c.\big]
\,\frac{d^3k}{\sqrt{|\vec{k}|}}\, \label{eq-B-44}\\
& &-\alpha\,
\sum_{\lambda}\int_{\Lambda\setminus\mathcal{B}_{\sigma_{j+1}}}\,\big[\veps_{\vk,
  \lambda}\,\frac{\vnabla  E_{\vec{P}}^{\sigma_{j+1}}\cdot \veps_{\vk,
 \lambda}^{\;*} }{|\vec{k}|^{\frac{3}{2}}\delta_{\vP}^{\sigma_{j+1}}(\widehat{k})}+h.c.\big]
\,\frac{d^3k}{\sqrt{|\vec{k}|}}\,. \label{eq-B-45}
\end{eqnarray}
In order to shorten our notations, we define 
\begin{eqnarray}\label{eq-FG-def-1}
F_{j}%(\mathcal{B}_{\sigma_{j}})
&:=&(\ref{eq-B-34})+(\ref{eq-B-35})+(\ref{eq-B-36})+(\ref{eq-B-37})\\
F_{j+1}%(\mathcal{B}_{\sigma_{j+1}})
&:=&(\ref{eq-B-42})+(\ref{eq-B-43})+(\ref{eq-B-44})+(\ref{eq-B-45})
\label{eq-FG-def-2}\\
G^{j}_{j+1}%(\mathcal{B}_{\sigma_j}\setminus \mathcal{B}_{\sigma_{j+1}})
&:=&(\ref{eq-B-38})+(\ref{eq-B-39})
\label{eq-FG-def-3}
\end{eqnarray}
Returning to (\ref{eq-B-31}), (\ref{eq-B-32}), we can write
\begin{eqnarray}
\lefteqn{\vnabla E_{\vec{P}}^{\sigma_{j+1}}-\vnabla E_{\vec{P}}^{\sigma_j}
-F_{j}
%(\mathcal{B}_{\sigma_{j}})
}
\\
& =&-\frac{1}{\|\widehat{\Phi}_{\vec{P}}^{\sigma_{j+1}}\|}\,
\Bra \, \widehat{\Phi}_{\vec{P}}^{\sigma_{j+1}}\,,\,
\widehat{\vec{\Pi}}_{\vP}^{\sigma_{j+1}}
(\frac{\widehat{\Phi}_{\vec{P}}^{\sigma_{j+1}}}{\|\widehat{\Phi}_{\vec{P}}^{\sigma_{j+1}}\|}
-\frac{\Phi_{\vec{P}}^{\sigma_{j}}}{\|\Phi_{\vec{P}}^{\sigma_{j}}\|}) \, \Ket\\
& &-\frac{\bra \, \widehat{\Phi}_{\vec{P}}^{\sigma_{j+1}}\,,\,\widehat{\vec{\Pi}}_{\vP}^{\sigma_{j+1}}
\Phi_{\vec{P}}^{\sigma_{j}} \, \ket}{\|\widehat{\Phi}_{\vec{P}}^{\sigma_{j+1}}\|\,\|\Phi_{\vec{P}}^{\sigma_{j}}\|}
+\frac{\bra \, \widehat{\Phi}_{\vec{P}}^{\sigma_{j+1}}\,,\,\vec{\Pi}_{\vP}^{\sigma_j}
\Phi_{\vec{P}}^{\sigma_{j}} \, \ket}{\|\widehat{\Phi}_{\vec{P}}^{\sigma_{j+1}}\|\,\|\Phi_{\vec{P}}^{\sigma_{j}}\|}\\
& &-\frac{\bra \, \widehat{\Phi}_{\vec{P}}^{\sigma_{j+1}}\,,\,\vec{\Pi}_{\vP}^{\sigma_j}
\Phi_{\vec{P}}^{\sigma_{j}}\, \ket }{\|\widehat{\Phi}_{\vec{P}}^{\sigma_{j+1}}\|\,\|\Phi_{\vec{P}}^{\sigma_{j}}\|}
+\frac{\bra \, \Phi_{\vec{P}}^{\sigma_{j}}\,,\,\vec{\Pi}_{\vP}^{\sigma_j}
\Phi_{\vec{P}}^{\sigma_{j}} \, \ket}{\|\Phi_{\vec{P}}^{\sigma_{j}}\|^2}
+G^{j}_{j+1}
%(\mathcal{B}_{\sigma_j}\setminus \mathcal{B}_{\sigma_{j+1}})
\,. \quad\quad\quad 
\end{eqnarray}
Using (\ref{eq-B-41}) -- (\ref{eq-B-45}), this can be rewritten into
\begin{eqnarray}
\lefteqn{\vnabla E_{\vec{P}}^{\sigma_{j+1}}-\vnabla E_{\vec{P}}^{\sigma_j}
-F_{j}
%(\mathcal{B}_{\sigma_{j}})
+\frac{\bra \, \widehat{\Phi}_{\vec{P}}^{\sigma_{j+1}}\,,\,
\Phi_{\vec{P}}^{\sigma_{j}} \, \ket }{\|\widehat{\Phi}_{\vec{P}}^{\sigma_{j+1}}\|\,\|\Phi_{\vec{P}}^{\sigma_{j}}\|}\,
F_{j+1}
%(\mathcal{B}_{\sigma_{j+1}})
}
\\
& =&-\frac{1}{\|\widehat{\Phi}_{\vec{P}}^{\sigma_{j+1}}\|}\,
\Bra \, \widehat{\Phi}_{\vec{P}}^{\sigma_{j+1}}\,,\,
\widehat{\vec{\Pi}}_{\vP}^{\sigma_{j+1}}
\big(\frac{\widehat{\Phi}_{\vec{P}}^{\sigma_{j+1}}}{\|\widehat{\Phi}_{\vec{P}}^{\sigma_{j+1}}\|}-
\frac{\Phi_{\vec{P}}^{\sigma_{j}}}{\|\Phi_{\vec{P}}^{\sigma_{j}}\|}\big) \, \Ket\quad\quad\label{eq-B-55}\\
& &-\frac{1}{\|\Phi_{\vec{P}}^{\sigma_{j}}\|}\,
\Bra \, \big(\frac{\widehat{\Phi}_{\vec{P}}^{\sigma_{j+1}}}
{\|\widehat{\Phi}_{\vec{P}}^{\sigma_{j+1}}\|}
-\frac{\Phi_{\vec{P}}^{\sigma_{j}}}{\|\Phi_{\vec{P}}^{\sigma_{j}}\|}\big)\,,\,\vec{\Pi}_{\vP}^{\sigma_j}
\Phi_{\vec{P}}^{\sigma_{j}} \, \Ket\label{eq-B-56}\\
&&-\frac{\bra \, \widehat{\Phi}_{\vec{P}}^{\sigma_{j+1}}\,,\,
%[\alpha^{\frac{1}{2}}\vec{A}_{\sigma_{j+1}}^{\sigma_j}+
\vec{\cL}_{\sigma_{j+1}}^{\sigma_{j}}
%]
\,\Phi_{\vec{P}}^{\sigma_{j}} \, \ket}{\|\widehat{\Phi}_{\vec{P}}^{\sigma_{j+1}}\|\,\|\Phi_{\vec{P}}^{\sigma_{j}}\|}
+G^{j}_{j+1}
%(\mathcal{B}_{\sigma_j}\setminus \mathcal{B}_{\sigma_{j+1}})
\,.\label{eq-B-57}
\end{eqnarray}
We deduce from the definitions (\ref{eq-FG-def-1}) and (\ref{eq-FG-def-2}) that
\begin{equation}
	|F_{j} | \, , \, |F_{j+1} |
	\, < \, c' \, |\vnabla E_{\vP}^{\sigma_{j+1}} - \vnabla E_{\vP}^{\sigma_{j}} |
\end{equation}
where $c'$ is $\cO(\alpha)$ and $j$-independent.
Then, it suffices to check that, for $\alpha$ small enough, there are positive constants $c$, $C$
uniform in $j$, such that
\begin{eqnarray}
\lefteqn{
C \Big(\Big\|\frac{\widehat{\Phi}_{\vec{P}}^{\sigma_{j+1}}}{\|\widehat{\Phi}_{\vec{P}}^{\sigma_{j+1}}\|}
-\frac{\Phi_{\vec{P}}^{\sigma_{j}}}{\|\Phi_{\vec{P}}^{\sigma_{j}}\|}\Big\|+\epsilon^{\frac{j+1}{2}}\Big)
}
    \\
 &\geq& \, \big| \, (\ref{eq-B-55})+(\ref{eq-B-56})+(\ref{eq-B-57}) \, \big| \, \geq \,
 c \, |\vnabla E^{\sigma_{j+1}}_{\vec{P}}-\vnabla E^{\sigma_{j}}_{\vec{P}}| 
\end{eqnarray}
is satisfied.
\QED

\newpage
%%%%%%%%%%%%%%%%%%%%%%%%%%%%%%%%%%%%%%%%%%%%%%%%%%%%%%%%%%%%%%%% L E M M A   B.3.  %%%%%%%%%%%%%%%%%%%%%%%%%%%%%%%%%%%%%%%%%%%%%%%%%%%%%%%%%%%%%%%%%%%%%%%%%%%%%%%%
\begin{lemma}\label{lm:B-3}
Assume $\vP\in\mathcal{S}$, and $\alpha$, $\mu$, and $\epsilon$ small enough. Then, uniformly in $j\in\NN_0$, the bound
\begin{eqnarray}
\lefteqn{\big\|\big(\frac{1}{\Hw_{\vP}^{\sigma_j}-z_{j+1}}\big)^{\frac{1}{2}}
%\big(
\, \mathcal{L}_{\sigma_{j+1}}^{\sigma_j\,(+)\,l}\,(\Gamma_{\vP}^{\sigma_j})^l \,
%\big)
\big(\frac{1}{\Hw_{\vP}^{\sigma_j}-z_{j+1}}\big)^{\frac{1}{2}}\Phi_{\vec{P}}^{\sigma_j}\big\|^2
}\\
& \leq&\,\frac{2}{1-c}\,C_3\,C_4\,Z_{j+1}^{j}\,\frac{1}{|E^{\sigma_{j-1}}_{\vec{P}}-z_{j+1}|}\,\Big|\,
\Bra \, (\Gamma_{\vP}^{\sigma_j})^{l}\Phi_{\vec{P}}^{\sigma_j}\,,\,\frac{1}{\Hw_{\vP}^{\sigma_j}-z_{j+1}}
\,(\Gamma_{\vP}^{\sigma_j})^{l}\Phi_{\vec{P}}^{\sigma_j} \, \Ket \Big| \nonumber
\end{eqnarray}
holds for each $l=1,2,3$, where $\gamma_{\sigma_{j+1}}:=\{z_{j+1}\in\CC \, \big| \,
|z_{j+1}-E_{\vec{P}}^{\sigma_j}| \, =\,\mu\sigma_{j+1}\}$, and $c<1$. $C_3$ and $C_4$ are defined in
(\ref{eq:IV.59}), (\ref{eq:IV.60}) ($(\mathscr{A}3)$ and $(\mathscr{A}4)$ from Section {\ref{ssec-keyingred}}),
and  
\begin{eqnarray}
Z_{j+1}^{j}&:=&\langle \mathcal{L}_{\sigma_{j+1}}^{\sigma_j\,(-)\,l}\,
\mathcal{L}_{\sigma_{j+1}}^{\sigma_j\,(+)\,l}\rangle_{\Omega_f}\\
&= &\alpha\,\sum_{\lambda}\int_{\mathcal{B}_{\sigma_j}\setminus
\mathcal{B}_{\sigma_{j+1}}}\,\,d^3k\,\Big|k^l
\frac{\vnabla E_{\vec{P}}^{\sigma_j}\cdot\veps_{\vk,\lambda}}
{|\vec{k}|^{\frac{3}{2}}\delta_{\vP}^{\sigma_j}(\widehat{k})}
+\frac{(\widehat{l}\cdot\veps_{\vk, \lambda} )}{\sqrt{|\vk| \,})}\Big|^2 \,.\nonumber
\end{eqnarray}
\end{lemma}

\Proof
\\
We first use Eq.~(\ref{eq:IV.60}) to estimate
\begin{eqnarray}
\lefteqn{\big\|\big(\frac{1}{\Hw_{\vP}^{\sigma_j}-z_{j+1}}\big)^{\frac{1}{2}}
%\big(
\mathcal{L}_{\sigma_{j+1}}^{\sigma_j\,(+)\,l}\,(\Gamma_{\vP}^{\sigma_j})^{l}
%\big)
\,\Phi_{\vec{P}}^{\sigma_j}\big\|^2}\\
& \leq&\,\Bra \, \mathcal{L}_{\sigma_{j+1}}^{\sigma_j\,(+)\,l}\,(\Gamma_{\vP}^{\sigma_j})^{l}
\Phi_{\vec{P}}^{\sigma_j}\,,\,\big|\frac{1}{\Hw_{\vP}^{\sigma_j}-z_{j+1}}\big|
\,\mathcal{L}_{\sigma_{j+1}}^{\sigma_j\,(+)\,l}\,(\Gamma_{\vP}^{\sigma_j})^{l}\Phi_{\vec{P}}^{\sigma_j}\, \Ket
\nonumber\\
& \leq&\,C_4\,\Big|\Bra \, \mathcal{L}_{\sigma_{j+1}}^{\sigma_j\,(+)\,l}\,(\Gamma_{\vP}^{\sigma_j})^{l}
\Phi_{\vec{P}}^{\sigma_j}\,,\,\frac{1}{\Hw_{\vP}^{\sigma_j}-z_{j+1}}
\,\mathcal{L}_{\sigma_{j+1}}^{\sigma_j\,(+)\,l}\,(\Gamma_{\vP}^{\sigma_j})^{l}\Phi_{\vec{P}}^{\sigma_j}\, \Ket\Big|\,.
\label{eq-B.62}
\end{eqnarray}
Then we use pull-through formula to derive the following equality which holds in the sense of distributions
for $\vk\in\cB_{\sigma_j}$
\begin{eqnarray}
\lefteqn{\frac{1}{\Hw_{\vP}^{\sigma_j}-z_{j+1}}\,b^*_{\vk,\lambda}\,=\,}\\
&=& b^*_{\vk,\lambda}\,\frac{1}{\frac{(\vec{\Gamma}_{\vP}^{\sigma_j}+\vk)^2}{2}+
\sum_{\lambda}\int_{\RR^3}|\vec{q}|\delta_{\vP}^{\sigma_j}(\widehat{q})\, b^{*}_{\vq, \lambda}
b_{\vq,\lambda} d^3q+\cvar_{\vec{P}}^{\sigma_j}+|\vec{k}|\delta_{\vP}^{\sigma_j}(\widehat{k})-z_{j+1}}\,.\nonumber
\end{eqnarray}
Moreover,  for $\sigma_{j+1}\leq|\vk|\leq\sigma_j$, $j\geq1$, and for $\alpha$, $\mu$, and $\epsilon$ 
small enough but uniform in $j$,
we can control the following series expansion in the space $\cF_{\sigma_j}$
\begin{eqnarray}
\lefteqn{\frac{1}{\frac{(\vec{\Gamma}_{\vP}^{\sigma_j})^2}{2}+H^f_{\delta_{\vP}^{\sigma_j}}
+\cvar_{\vec{P}}^{\sigma_j}+|\vec{k}|\delta_{\vP}^{\sigma_j}(\widehat{k})-z_{j+1}}\times}\\
& &\times\sum_{n=0}^{+\infty}\big[-(\vec{\Gamma}_{\vP}^{\sigma_j}\cdot\vk+\frac{|k|^2}{2})
\frac{1}{\frac{(\vec{\Gamma}_{\vP}^{\sigma_j})^2}{2}+H^f_{\delta_{\vP}^{\sigma_j}}
+\cvar_{\vec{P}}^{\sigma_j}+|\vec{k}|\delta_{\vP}^{\sigma_j}(\widehat{k})-z_{j+1}}\big]^n\nonumber
\end{eqnarray}
where $$H^f_{\delta_{\vP}^{\sigma_j}}\,:=\,\sum_{\lambda}\int_{\RR^3}|\vec{q}|
\delta_{\vP}^{\sigma_j}(\widehat{q})\, b^{*}_{\vq, \lambda}
b_{\vq,\lambda} d^3q\,,$$ the key estimate being
\begin{eqnarray}\label{eq-A-52-1}
\lefteqn{\Big\|\,\Big(\frac{1}{\frac{(\vec{\Gamma}_{\vP}^{\sigma_j})^2}{2}+H^f_{\delta_{\vP}^{\sigma_j}}
+\cvar_{\vec{P}}^{\sigma_j}+|\vec{k}|\delta_{\vP}^{\sigma_j}(\widehat{k})-z_{j+1}}\Big)^{1/2}\times}\\
& \times& (\vec{\Gamma}_{\vP}^{\sigma_j}\cdot\vk+\frac{|k|^2}{2})
\Big(\frac{1}{\frac{(\vec{\Gamma}_{\vP}^{\sigma_j})^2}{2}+H^f_{\delta_{\vP}^{\sigma_j}}
+\cvar_{\vec{P}}^{\sigma_j}+|\vec{k}|\delta_{\vP}^{\sigma_j}(\widehat{k})-z_{j+1}}\Big)^{1/2}
\Big\|_{\cF_{\sigma_j}}\leq\,c \, < \, 1\,.\nonumber
\end{eqnarray}
In order to control the term proportional to $\vec\Gamma_{\vP}^{\sigma_j}\cdot\vk$,
we note that
%for $\alpha$ sufficiently small but uniform in $j$, 
\begin{eqnarray}
	\lefteqn{
	\sum_{i=1}^3\Big\|\,\Big( \frac{1}{ \Hw_{\vP}^{\sigma_j}
	+|\vec{k}|\delta_{\vP}^{\sigma_j}(\widehat{k})-z_{j+1} } \Big)^{1/2}
	\frac{(\vec\Gamma_{\vP}^{\sigma_j})^i}{\sqrt2} \, \Big\|_{\cF_{\sigma_j}}^2
	}
	\nonumber\\
	&& 
	\, \leq \,
	%\frac13 
	3 \, \Big\| \, \big( \, \Hw_\vP^{\sigma_j}+\cO(\alpha) \, \big) \, \Big| \, 
	\frac{1}{\Hw_{\vP}^{\sigma_j}
	+|\vec{k}|\delta_{\vP}^{\sigma_j}(\widehat{k})-z_{j+1} }  
	\, \Big| \, \Big\|_{\cF_{\sigma_j}} \,.
\end{eqnarray}
Then, we observe that 
%for $\alpha$ sufficiently small, 
\eqn
	%\lefteqn{
	&&\lim\sup_{\mu,\alpha,\epsilon\rightarrow0}\Big(\frac{1}{
	|\vec{k}|\delta_{\vP}^{\sigma_j}(\widehat{k}) } \Big)^{1/2}
	\Big\| \, \big( \, \Hw_\vP^{\sigma_j} +\cO(\alpha) \, \big) \, \Big| \, 
	\frac{1}{\Hw_{\vP}^{\sigma_j}
	+|\vec{k}|\delta_{\vP}^{\sigma_j}(\widehat{k})-z_{j+1} } \, \Big| \, \Big\|_{\cF_{\sigma_j}}^{1/2}  
	\sqrt6 \, |\vk| 
	%}
	\nonumber\\
	&&\quad\quad\quad\quad
	\, \leq \, \sup_{\vP\in\mathcal{S}}\frac{|\vP|\sqrt3}{1-|\vP|} \, \leq \, \frac{\sqrt3}{2} \,.
\eeqn
%as $\mu,\alpha\rightarrow0$; 
Therefore, the estimate (\ref{eq-A-52-1}) also holds true for 
the term proportional to $\vec{\Gamma}_\vP^{\sigma_j}\cdot\vk$ if $\mu>0$, $\alpha>0$, and $\epsilon>0$
are small enough, but uniform in $j$. 
%For the last estimate, we used that by assumption, $\vP\in\cS$.
To estimate of the term proportional to $\frac{|\vk|^2}{2}$, we use
\eqn 
	\frac{|\vk|^2}{2} \,  
	\Big\| \, \frac{1}{\Hw_{\vP}^{\sigma_j}
	+|\vec{k}|\delta_{\vP}^{\sigma_j}(\widehat{k})-z_{j+1} }
	\, \Big\|_{\cF_{\sigma_j}}
	\, \leq \, \frac{|\vk|^2}{2(|\vec{k}|\delta_{\vP}^{\sigma_j}(\widehat{k})-\mu\sigma_{j+1}) }
	\, \ll \, 1 \,,
\eeqn
for $\alpha$, $\epsilon$, $\mu$ small enough but unifor in $j$.
Therefore, recalling that $b_{\vk,\lambda} \,\Phi_{\vec{P}}^{\sigma_j}=0$ for $|\vk|\leq\sigma_j$,
we find
\begin{eqnarray}
\lefteqn{(\ref{eq-B.62})\,}\\
& \leq&C_4\,\Big\lbrace\,\alpha\,\sum_{\lambda}\int_{\mathcal{B}_{\sigma_j}\setminus
\mathcal{B}_{\sigma_{j+1}}}\,\,d^3k\,\Big|k^l
\frac{\vnabla E_{\vec{P}}^{\sigma_j}\cdot\veps_{\vk,\lambda}}
{|\vec{k}|^{\frac{3}{2}}\delta_{\vP}^{\sigma_j}(\widehat{k})}
+\frac{(\widehat{l}\cdot\veps_{\vk, \lambda} )}{\sqrt{|\vk| \,})}\Big|^2\times\\
& &\times\big\|\big(\frac{1}{\Hw_{\vP}^{\sigma_j}
+|\vec{k}|\delta_{\vP}^{\sigma_j}(\widehat{k})-z_{j+1}}\big)^\frac{1}{2}
(\Gamma_{\vP}^{\sigma_j})^{l}\Phi_{\vec{P}}^{\sigma_j}\big)\big\|^2\Big\rbrace\,
\sum_{n=0}^{+\infty}c^n\nonumber\\
&\leq &\frac{1}{1-c}\,C_4\,\Big\lbrace\,\alpha\,\sum_{\lambda}\int_{\mathcal{B}_{\sigma_j}\setminus
\mathcal{B}_{\sigma_{j+1}}}\,\,d^3k\,\Big|k^l
\frac{\vnabla E_{\vec{P}}^{\sigma_j}\cdot\veps_{\vk,\lambda}}
{|\vec{k}|^{\frac{3}{2}}\delta_{\vP}^{\sigma_j}(\widehat{k})}
+\frac{(\widehat{l}\cdot\veps_{\vk, \lambda} )}{\sqrt{|\vk| \,})}\Big|^2\times \label{eq-B.68}\\
& &\times\big\|\big(\frac{1}{\Hw_{\vP}^{\sigma_j}
+|\vec{k}|\delta_{\vP}^{\sigma_j}(\widehat{k})-z_{j+1}}\big)^\frac{1}{2}
(\Gamma_{\vP}^{\sigma_j})^{l}\Phi_{\vec{P}}^{\sigma_j}\big)\big\|^2\Big\rbrace\nonumber\\
& \leq&\frac{1}{1-c}\,C_3\,C_4\,\big(\alpha\,\sum_{\lambda}\int_{\mathcal{B}_{\sigma_j}\setminus
\mathcal{B}_{\sigma_{j+1}}}\,\,d^3k\,\Big|k^l
\frac{\vnabla E_{\vec{P}}^{\sigma_j}\cdot\veps_{\vk,\lambda}}
{|\vec{k}|^{\frac{3}{2}}\delta_{\vP}^{\sigma_j}(\widehat{k})}
+\frac{(\widehat{l}\cdot\veps_{\vk, \lambda} )}{\sqrt{|\vk| \,})}\Big|^2\big)\times \label{eq-B.69}\\
& &\times\,\big|\,\Bra \, (\Gamma_{\vP}^{\sigma_j})^{l}\Phi_{\vec{P}}^{\sigma_j}\,,\,
\frac{1}{\Hw_{\vP}^{\sigma_j}-z_{j+1}}
\,(\Gamma_{\vP}^{\sigma_j})^{l}\Phi_{\vec{P}}^{\sigma_j}\, \Ket\big|\nonumber
\end{eqnarray}
where, in passing from (\ref{eq-B.68}) to (\ref{eq-B.69}), we use (\ref{eq:IV.60}),
and property $(\mathscr{A}3)$ from Section {\ref{ssec-keyingred}}.
For $\sigma_1\leq|\vk|\leq\sigma_0$, a similar argument yields (\ref{eq-B.69}).

This proves the lemma.
\QED
 
\newpage
%%%%%%%%%%%%%%%%%%%%%%%%%%%%%%%%%%%%%%%%%%%%%%%%%%%%%%%%%%%%%%%% L E M M A   B.4.  %%%%%%%%%%%%%%%%%%%%%%%%%%%%%%%%%%%%%%%%%%%%%%%%%%%%%%%%%%%%%%%%%%%%%%%%%%%%%%%%
\begin{lemma}\label{lm:B-4}
For $\alpha$ and $\epsilon$ small enough, with $\epsilon>C \, \alpha$, $C$ sufficiently large,
there exist constants $R_1$, $R_2 \leq \cO(\epsilon^{-1})$, uniformly in $j\in\NN$ and $\vP\in\mathcal{S}$,
for which
\begin{eqnarray}
\lefteqn{
8 \, \Big\|\Big(\frac{1}{\Hw_{\vP}^{\sigma_{j-1}}-z_{j+1}}\Big)^{\frac{1}{2}}\,
((\widehat{\Gamma}_{\vP}^{\sigma_j})^{i}\widehat{\Phi}_{\vec{P}}^{\sigma_j}-
(\Gamma_{\vP}^{\sigma_{j-1}})^{i}\widehat{\Phi}_{\vec{P}}^{\sigma_j})\Big\|^2
}
\label{eq:IV.69'}\\
& &\quad\quad\quad\quad\quad\quad
+ \, 8 \, \Big\|\Big(\frac{1}{\Hw_{\vP}^{\sigma_{j-1}}-z_{j+1}}\Big)^{\frac{1}{2}}\,
(\Gamma_{\vP}^{\sigma_{j-1}})^{i}(\widehat{\Phi}_{\vec{P}}^{\sigma_j}-\Phi_{\vec{P}}^{\sigma_{j-1}})\Big\|^2
\nonumber\\
&\leq&\frac{R_1}{\epsilon^{\frac{j}{2}}}
\Big(\frac{\|\widehat{\Phi}_{\vec{P}}^{\sigma_{j}}-\Phi_{\vec{P}}^{\sigma_{j-1}}\|+
\epsilon^{\frac{j}{2}}}{\epsilon^{\frac{j}{4}}}\Big)^2
\label{eq:IV.70'}\\
& &\quad\quad\quad\quad\quad\quad
+\frac{R_2}{\epsilon^{\frac{j}{2}}}
\Big(\frac{\big\|\frac{\widehat{\Phi}_{\vec{P}}^{\sigma_{j}}}{\|\widehat{\Phi}_{\vec{P}}^{\sigma_{j}}\|}-
\frac{\Phi_{\vec{P}}^{\sigma_{j-1}}}{\|\Phi_{\vec{P}}^{\sigma_{j-1}}\|}\big\|+
\epsilon^{\frac{j}{2}}}{4\epsilon^{\frac{j}{4}}}\Big)^2 \,. \nonumber
\end{eqnarray}
\end{lemma}

\Proof
\\
In order to justify the estimate in the statement, it is enough to make the difference
\begin{equation}
(\widehat{\Gamma}_{\vP}^{\sigma_j})^{i}-
(\Gamma_{\vP}^{\sigma_{j-1}})^{i}
\end{equation}
explicit. The definitions are given in (\ref{eq-IV-33}) and (\ref{eq:IV.46}).

>From (\ref{eq-B-31}), (\ref{eq-B-32}), we get
\begin{eqnarray}
\lefteqn{-\frac{\bra \, \widehat{\Phi}_{\vec{P}}^{\sigma_{j}}\,,\,\widehat{\vec{\Pi}}_{\vP}^{\sigma_{j}}
\widehat{\Phi}_{\vec{P}}^{\sigma_{j}} \, \ket}{\bra \, \widehat{\Phi}_{\vec{P}}^{\sigma_{j}}\,,\,
\widehat{\Phi}_{\vec{P}}^{\sigma_{j}} \, \ket}
+\frac{\bra \, \Phi_{\vec{P}}^{\sigma_{j-1}}\,,\,\vec{\Pi}_{\vP}^{\sigma_{j-1}}
\Phi_{\vec{P}}^{\sigma_{j-1}}\, \ket}{\bra \, \Phi_{\vec{P}}^{\sigma_{j-1}}\,,\,\Phi_{\vec{P}}^{\sigma_{j-1}} \, \ket}}\\
& =&\vnabla
E_{\vec{P}}^{\sigma_{j}}-\vnabla
E_{\vec{P}}^{\sigma_{j-1}}\\
& &+\langle W_{\sigma_{j}}(\vnabla
E_{\vec{P}}^{\sigma_{j}})\,\vec{\beta}^{\sigma_{j}} \, W_{\sigma_{j}}^*(\vnabla
E_{\vec{P}}^{\sigma_{j}})\rangle_{\Omega_f}
\\
&&\quad\quad\quad\quad\quad\quad\quad\quad\quad-\langle W_{\sigma_{j-1}}(\vnabla
E_{\vec{P}}^{\sigma_{j-1}})\,\vec{\beta}^{\sigma_{j-1}} \, W_{\sigma_{j-1}}^*(\vnabla
E_{\vec{P}}^{\sigma_{j-1}})\rangle_{\Omega_f} \, .\nonumber
\end{eqnarray}
>From (\ref{eq-B-41}) -- (\ref{eq-B-45}) and (A18) -- (A25), we obtain
\begin{eqnarray}
%\lefteqn{
\widehat{\vec\Gamma}_{\vP}^{\sigma_j}-
\vec\Gamma_{\vP}^{\sigma_{j-1}}
%}\\
&=&\widehat{\vec{\Pi}}_{\vP}^{\sigma_{j}}-\vec{\Pi}_{\vP}^{\sigma_j}\\
& &-\frac{\bra \, \widehat{\Phi}_{\vec{P}}^{\sigma_{j}}\,,\,\widehat{\vec{\Pi}}_{\vP}^{\sigma_{j}}
\widehat{\Phi}_{\vec{P}}^{\sigma_{j}} \, \ket}{\bra \, \widehat{\Phi}_{\vec{P}}^{\sigma_{j}}\,,\,
\widehat{\Phi}_{\vec{P}}^{\sigma_{j}}\, \ket}
+\frac{\bra \, \Phi_{\vec{P}}^{\sigma_{j-1}}\,,\,\vec{\Pi}_{\vP}^{\sigma_{j-1}}
\Phi_{\vec{P}}^{\sigma_{j-1}} \, \ket}{\bra \, \Phi_{\vec{P}}^{\sigma_{j-1}}\,,\,\Phi_{\vec{P}}^{\sigma_{j-1}} \, \ket}
\end{eqnarray}
\begin{eqnarray}
&= &
\vnabla E_{\vec{P}}^{\sigma_{j}} \, - \, 
\vnabla E_{\vec{P}}^{\sigma_{j-1}}
%\\
%& &
%-\alpha^{\frac{1}{2}}\vec{A}_{\sigma_{j}}^{\sigma_{j-1}}-
\, + \, \vec{\cL}_{\sigma_{j}}^{\sigma_{j-1}}\\
& &+\alpha\,
\sum_{\lambda}\int_{\mathcal{B}_{\sigma_{j-1}}\setminus \mathcal{B}_{\sigma_{j}}}\,\vec{k}\,\frac{\vnabla
  E_{\vec{P}}^{\sigma_{j-1}}\cdot \veps_{\vk,
  \lambda}^{\;*}\,\vnabla E_{\vec{P}}^{\sigma_{j-1}}\cdot \veps_{\vk,
  \lambda}}{|\vec{k}|^{3}(\delta_{\vP}^{\sigma_{j-1}}(\widehat{k}))^2}\,d^3k \label{eq-B-42'}\\
%&&+\alpha\,
%\sum_{\lambda}\int_{\mathcal{B}_{\sigma_{j-1}}\setminus \mathcal{B}_{\sigma_{j+1}}}\,\vec{k}\,\frac{\vnabla
%  E_{\vec{P}}^{\sigma_{j+1}}\cdot \veps_{\vk,
%  \lambda}^{\;*}\,\vnabla E_{\vec{P}}^{\sigma_{j+1}}\cdot \veps_{\vk,
%  \lambda}}{|\vec{k}|^{3}(\delta_{\vP}^{\sigma_{j+1}}(\widehat{k}))^2}\,d^3k \label{eq-B-43'}\\
& &+\alpha\,
\sum_{\lambda}\int_{\mathcal{B}_{\sigma_{j-1}}\setminus \mathcal{B}_{\sigma_{j}}}\,\big[\veps_{\vk,
  \lambda}\,\frac{\vnabla  E_{\vec{P}}^{\sigma_{j-1}}\cdot \veps_{\vk,
 \lambda}^{\;*} }{|\vec{k}|^{\frac{3}{2}}\delta_{\vP}^{\sigma_{j-1}}(\widehat{k})}+h.c.\big]
\,\frac{d^3k}{\sqrt{|\vec{k}|}}\, .
\label{eq-B-44'}
%\\
%& &+\alpha\,
%\sum_{\lambda}\int_{\mathcal{B}_{\sigma_{j-1}}\setminus \mathcal{B}_{\sigma_{j+1}}}\,\big[\veps_{\vk,
%  \lambda}\,\frac{\vnabla  E_{\vec{P}}^{\sigma_{j+1}}\cdot \veps_{\vk,
% \lambda}^{\;*} }{|\vec{k}|^{\frac{3}{2}}\delta_{\vP}^{\sigma_{j+1}}(\widehat{k})}+h.c.\big]
%\,\frac{d^3k}{\sqrt{|\vec{k}|}}\, \\
%&&+\alpha\,
%\sum_{\lambda}\int_{\mathcal{B}_{\sigma_{j-1}}\setminus \mathcal{B}_{\sigma_{j+1}}}\,\vec{k}\,\frac{\vnabla
%  E_{\vec{P}}^{\sigma_{j+1}}\cdot \veps_{\vk,
%  \lambda}^{\;*}\,\vnabla E_{\vec{P}}^{\sigma_{j+1}}\cdot \veps_{\vk,
%  \lambda}}{|\vec{k}|^{3}(\delta_{\vP}^{\sigma_{j+1}}(\widehat{k}))^2}\,d^3k \label{eq-B-38'}\\
%& &+\alpha\,
%\sum_{\lambda}\int_{\mathcal{B}_{\sigma_j}\setminus \mathcal{B}_{\sigma_{j+1}}}\,\big[\veps_{\vk,
%  \lambda}\,\frac{\vnabla  E_{\vec{P}}^{\sigma_{j+1}}\cdot \veps_{\vk,
% \lambda}^{\;*} }{|\vec{k}|^{\frac{3}{2}}\delta_{\vP}^{\sigma_{j+1}}(\widehat{k})}+h.c.\big]
%\,\frac{d^3k}{\sqrt{|\vec{k}|}}\,. \label{eq-B-39'}
\end{eqnarray}
Now, we simply combine the result in (\ref{eq-B-30}) with the bounds
\begin{equation}
	\Big\| \, \Big(\frac{1}{\Hw_{\vP}^{\sigma_{j-1}}-z_{j+1}}\Big)^{\frac{1}{2}}\,
	(\Gamma_{\vP}^{\sigma_{j-1}})^{i} \, \Big\|_{\cF_{\sigma_j}} \, \leq \, \cO( \epsilon^{-\frac{j+1}{2}})
\end{equation}
\begin{equation}
	\Big\| \, \Big(\frac{1}{\Hw_{\vP}^{\sigma_{j-1}}-z_{j+1}}\Big)^{\frac{1}{2}}\,
	\vec{A}_{\sigma_{j}}^{\sigma_{j-1}} \, \Big\|_{\cF_{\sigma_j}} \, \leq \, 
	\cO(\epsilon^{\frac{j-1}{2}}) \, ,
\end{equation}
and similarly for $\vec{\cL}_{\sigma_{j}}^{\sigma_{j-1}}$.
The size of all other expressions (\ref{eq-B-42'}) -- (\ref{eq-B-44'}) can
trivially be seen to be of order $\cO(\alpha\,\epsilon^{j-1})$.
The assertion of the lemma follows.
\QED

\subsubsection*{Acknowledgements}

The authors gratefully acknowledge the support and hospitality of the
Erwin Schr\"odinger Institute (ESI) in Vienna in June 2006, 
where this collaboration was initiated.
T.C. was supported by NSF grants DMS-0524909 and DMS-0704031.

%%%%%%%%%%%%%%%%%%%%%%%%%%%%%%%%%%%%%%%%%%%%%%%%%
%\bibliography{/home/vbach/BIB/volle}
%\end{document}
%%%%%%%%%%%%%%%%%%%%%%%%%%%%%%%%%%%%%%%%%%%%%%%%%

\parindent=0pt

\end{document}